\documentclass{ieeeaccess}
\usepackage{cite}
\usepackage{amsmath,amssymb,amsfonts}
\usepackage{algorithmic}
\usepackage{graphicx}
\usepackage{textcomp}
\usepackage[justification=centering]{caption}
\def\BibTeX{{\rm B\kern-.05em{\sc i\kern-.025em b}\kern-.08em
    T\kern-.1667em\lower.7ex\hbox{E}\kern-.125emX}}
\begin{document}
\history{Date of publication xxxx 00, 0000, date of current version xxxx 00, 0000.}
\doi{xx.xxxx/ACCESS.2020.DOI}

\title{A Prospective Look: Key Enabling Technologies, Applications and Open Research Topics in 6G Networks}
\author{\uppercase{Lina Bariah}\authorrefmark{1}, \IEEEmembership{Member, IEEE},
\uppercase{Lina Mohjazi}\authorrefmark{2}, \IEEEmembership{Senior Member, IEEE},
\uppercase{Sami Muhaidat}\authorrefmark{1}, \IEEEmembership{Senior Member, IEEE},
\uppercase{Paschalis C. Sofotasios}\authorrefmark{3}, \IEEEmembership{Senior Member, IEEE},
\uppercase{Gunes Karabulut Kurt}\authorrefmark{4}, \IEEEmembership{Senior Member, IEEE},
\uppercase{Halim Yanikomeroglu}\authorrefmark{5}, \IEEEmembership{Fellow, IEEE}, and
\uppercase{Octavia A. Dobre}\authorrefmark{6}, \IEEEmembership{Fellow, IEEE},
}
\address[1]{KU Center for Cyber-Physical Systems, Department of Electrical and Computer Engineering, Khalifa University, Abu Dhabi, UAE, (e-mails: \{lina.bariah, muhaidat \}@ieee.org.}
\address[2]{KU Center for Cyber-Physical Systems, Department of Electrical and Computer Engineering, Khalifa University, Abu Dhabi, UAE, and with the School of Engineering, University of Glasgow, Glasgow, U.K (e-mail: l.mohjazi@ieee.org)}
\address[3]{P. C. Sofotasios is with the Center for Cyber-Physical Systems, Department of Electrical and Computer Engineering, Khalifa University, Abu Dhabi 127788, UAE, and also with the Department of Electrical Engineering, Tampere University, Tampere 33101, Finland (e-mail: p.sofotasios@ieee.org)}
\address[4]{Electronics and Communication Engineering Department, Istanbul Technical University, 34469 Istanbul, Turkey  (e-mail: gkurt@itu.edu.tr)}
\address[5]{Department of Systems and Computer Engineering, Carleton University, Ottawa, ON K1S 5B6, Canada (e-mail: halim@sce.carleton.ca)}
\address[6]{Department of Electrical and Computer Engineering, Memorial University, St. John's, NL A1C 5S7, Canada (e-mail: odobre@ mun.ca)}

\markboth
{Bariah \headeretal: A Prospective Look: Key Enabling Technologies, Applications and Open Research Topics in 6G Networks}
{Bariah \headeretal: A Prospective Look: Key Enabling Technologies, Applications and Open Research Topics in 6G Networks}

\corresp{Corresponding author: Sami Muhaidat (e-mail: muhaidat@ieee.org).}
 

\begin{abstract}
\boldmath
The fifth generation (5G) mobile networks are envisaged to enable a plethora of breakthrough advancements in wireless technologies, providing support of a diverse set of services over a single platform. While the deployment of 5G systems is scaling up globally, it is time to look ahead for beyond 5G systems. This is mainly driven by the emerging societal trends, calling for fully automated systems and intelligent services supported by extended reality and haptics communications. To accommodate the stringent requirements of their prospective applications, which are data-driven and defined by extremely low-latency, ultra-reliable, fast and seamless wireless connectivity, research initiatives are currently focusing on  a progressive roadmap towards the sixth generation (6G) networks, which are expected to bring transformative changes to this premise. In this article, we shed light on some of the major enabling  technologies for 6G, which are expected to revolutionize the fundamental architectures of cellular networks and provide multiple homogeneous artificial intelligence-empowered services, including distributed communications, control, computing, sensing, and energy, from its core to its end nodes. In particular, the present paper aims to answer several 6G framework related questions: What are the driving forces for the development of 6G? How will the enabling technologies of 6G differ from those in 5G? What kind of applications and interactions will they support which would not be supported by 5G? We address these questions by presenting a comprehensive study of the 6G vision and outlining seven of its disruptive technologies, i.e., mmWave communications, terahertz communications, optical wireless communications, programmable metasurfaces, drone-based communications, backscatter communications and tactile internet, as well as their potential applications. Then, by leveraging the state-of-the-art literature surveyed for each technology, we discuss the associated requirements, key challenges, and open research problems. These discussions are thereafter used to open up the horizon for future research directions. 
 
\end{abstract}


\titlepgskip=-15pt

\maketitle

\IEEEpeerreviewmaketitle


\begin{table}[ht]
\centering
\caption{List of Abbreviations.}
\begin{tabular}{|l l|}
\hline
\textbf{Abbreviation}    & \textbf{Definition}                                                                                                            \\ \hline \hline
AI & Artificial Intelligence  \\ \hline
AR  & Augmented Reality \\ \hline
BackCom & Backscatter Communications \\ \hline
BS  & Base Station \\ \hline
B5G & Beyond 5G \\ \hline
CAV  &  Connected and Autonomous Vehicle \\ \hline
CED &  Cellular-Enabled Drones \\ \hline
CI  & Communication Infrastructure \\ \hline
CR  & Cognitive Radio \\ \hline
EM & Electromagnetic \\ \hline
IoE  & Internet-of-Everything \\ \hline
IoT & Internet-of-Things \\ \hline
LED & Light-Emitting Diode \\ \hline
LiFi & Light Fidelity \\ \hline
LOS & Line-of-Sight \\ \hline
MEC & Multi-Access Edge Computing \\ \hline
MIMO & Multiple-Input Multiple-Output \\ \hline
mmWave & Millimeter Wave \\ \hline
MTC & Machine-type Communication \\ \hline
M2M & Machine-to-Machine \\ \hline
M2P & Machine-to-People \\ \hline
NLOS & non-LOS \\ \hline
NOMA & Non-Orthogonal Multiple Access \\ \hline
OCC & Optical Camera Communications \\ \hline
PU & Primary User \\ \hline
QoS & Quality of Service \\ \hline
RF-EH & Radio Frequency Energy Harvesting \\ \hline
SU & Secondary User \\ \hline
S-I & Signal-Idler \\ \hline
Si & Silicon \\ \hline
TI & Tactile Internet \\ \hline
THz & Terahertz \\ \hline
UAV & Unmanned Aerial Vehicle \\ \hline
UE &  User Equipment \\ \hline
URLLC & {\small Ultra-Reliable Low-Latency Communications} \\ \hline
VLBC & Visible Light  BackCom \\ \hline
VLC & Visible Light Communications \\ \hline
VR & Virtual Reality \\ \hline
WBSN & Wireless Body Sensor Network \\ \hline
WID & Wireless Infrastructure Drone \\ \hline
WiFi & Wireless Fidelity \\ \hline
WLAN & Wireless Local Area Network \\ \hline
WPAN & Wireless Personal Area Network \\ \hline
WPT & Wireless Power Transfer \\ \hline
XR & Extended Reality \\ \hline
5G & Fifth Generation \\ \hline
6G & Sixth Generation \\ \hline
 \hline 
\end{tabular}
\label{tab:Abb}
\end{table}

\section{Introduction} 
\label{sec:intro}

The phenomenal growth of connected devices and the increasing demand for high data rate services have been the main driving forces for the evolution of wireless technologies in the past decades. A forecast study reported by the International Telecommunication Union demonstrates that the volume of mobile data will continue to grow at an exponential rate, reaching up to a remarkable figure of about 5 zettabytes per month in 2030 \cite{ITUreport}. Meanwhile, due to the emergence of the Internet-of-Everything (IoE) paradigm, supporting smart homes, smart cities, and e-health applications seamlessly through connecting billions of people and devices over a single unified communication interface, there is an urgent need to shift the focus from the rate-centric enhanced mobile broadband services to ultra-reliable low-latency communications (URLLC) in order to provide a networked society through massive machine-type communications (MTC) \cite{OD1,OD3}.  Besides generating massive data, the upsurge of IoE will naturally give rise to a myriad of new traffic and data service types, leading to  diverse communication requirements. This grand vision requires a radical departure from the conventional ``one-size-fits-all'' network model of fourth generation systems.

 The fifth generation (5G) of wireless technology represents a technological breakthrough with respect to the previous communication networks. In addition to reducing latency, enhancing connectivity and reliability, and achieving gigabits per second speeds, 5G is set to deliver a variety of service types, often characterized by conflicting requirements and diverse sets of key performance indicators, simultaneously over one platform \cite{OD2}. These features make 5G a key enabler for the Internet-of-Things (IoTs) application environments, where machine-to-people (M2P) communications (e.g., industry automation, smart cities, and intelligent mobility) and machine-to-machine (M2M) communications (e.g.,  autonomous communications between sensors and actuators) are expected to take place alongside people-to-people communications, (e.g., voice over internet protocol (IP), video conferencing, video streaming, and web browsing).

 Delivering a plethora of services, with profound differences in terms of quality of service (QoS) requirements, poses major challenges, such as the need to manage a huge volume of a mixture of human-type and machine-type traffic, which is heterogeneous in nature. To cater to these unique challenges, 5G deployment tends to adopt two main network functionalities, namely softwarization and virtualization \cite{Fernando, Han}. By jointly exploiting softwarization and virtualization, cognition and  programmability of the end-to-end network chain may be achieved by decoupling the network functions from the hardware platform. This yields enhanced flexibility and reliability, as well as fast network auto-reconfiguration, enabling a larger portfolio of  use cases and applications to be supported concurrently.

 In parallel with addressing the aforementioned challenges, 5G has introduced potential disruptive technologies to meet stringent requirements in terms of capacity, connectivity, communication resilience, reliability, deployment costs, power consumption, latency, and data rate. These technologies include, but are not limited to,  millimeter wave (mmWave) communications, massive multiple-input multiple-output (MIMO), non-orthogonal multiple access (NOMA), and network ultra-densification  \cite{7169508,OD4}.

Despite the strong belief that 5G will support the basic MTC and URLLC related applications, it is arguable whether the capabilities of 5G systems will succeed in keeping the pace with the rapid proliferation of ultimately new IoE applications, which are expected to increase by 12\% yearly, and which are enabled by massive connectivity and are based on data-centric and automated processes. Meanwhile, following the revolutionary changes in the individual and societal trends, in addition to the noticeable advancement in human-machine interaction technologies, the market demands by 2030 are envisaged to witness the penetration of a new spectrum of IoE services. These services span from \textit{extended reality} (XR), which comprises \textit{augmented reality} (AR), \textit{virtual reality} (VR), and \textit{mixed reality} services, to flying vehicles, haptics, telemedicine, autonomous systems, and human-machine interfaces. The unprecedented requirements imposed by these services, such as delivering ultra-high reliability, extremely high data rates, and ultra-low latency simultaneously over uplink and downlink, will push the performance of 5G systems to its limits within 10 years of its launch, as speculated in \cite{Tariq}. Moreover, the emergence of such new IoE services necessitates integrating the computing, control, and communication functionalities into a single network design.

In order to deliver future cutting-edge services and accommodate their aforementioned heterogeneous requirements, a new breed of challenges have to be effectively addressed. Examples of these challenges include leveraging sub-terahertz (THz) bands, governing the network performance set by a targeted rate-reliability-latency trade-off, provisioning flexibility in the network architecture and functionalities, and designing an intelligent holistic orchestration platform to coordinate all network resource aspects, including communication, control, computing, and sensing, in an efficient, self-sustainable, and scalable manner, which is tailored to the demands of a specific application scenario or use case \cite{walid, marco,Khaled,Emilio}.

\subsection{6G vision and Requirements}

\begin{figure*}[!t] 
\centering
\includegraphics[width=1\linewidth]{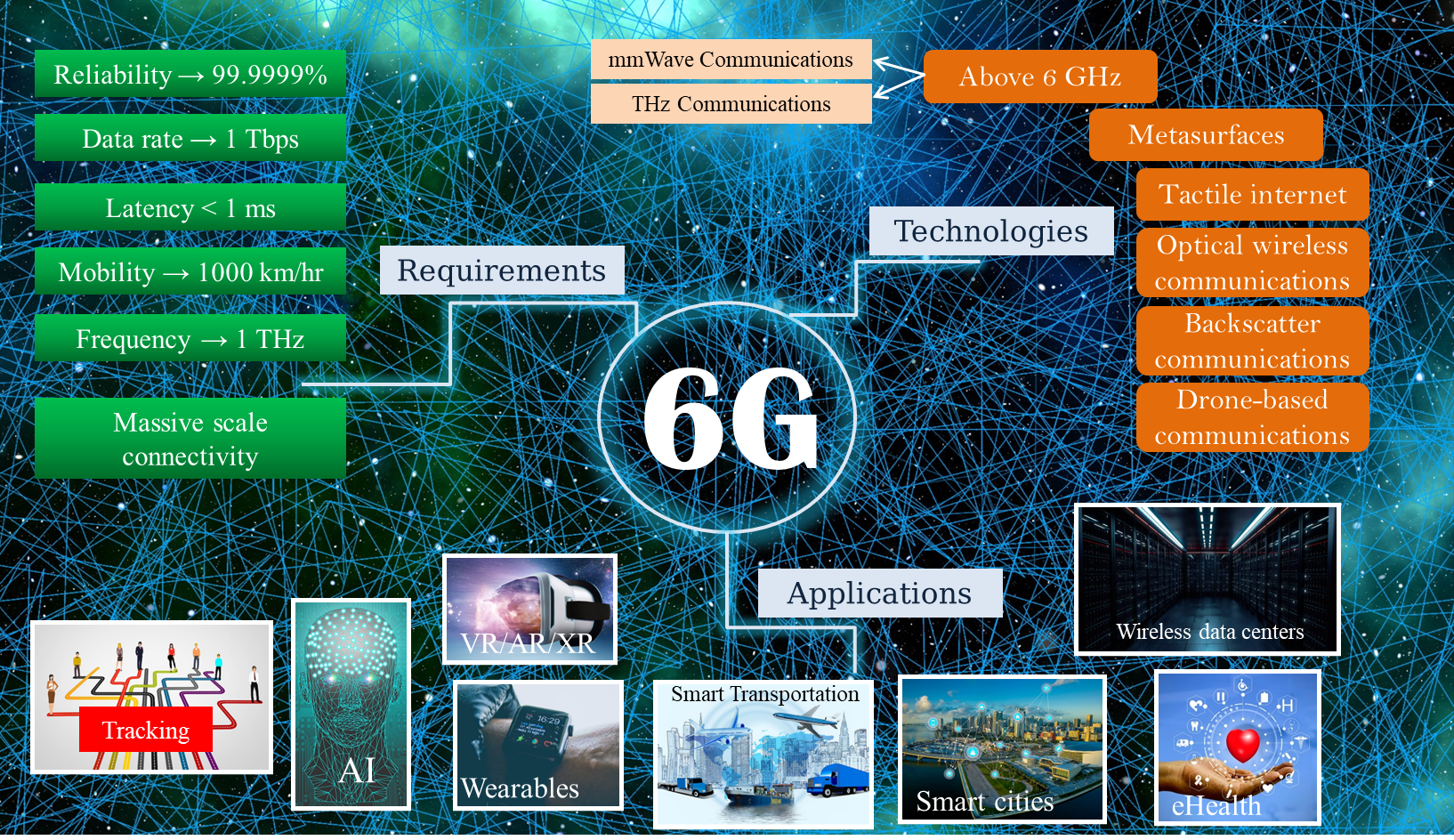}
\caption{\color{black}6G vision: requirements, technologies and applications.}
\label{fig:6G}
\end{figure*}

The evolution of 5G has urged the conceptualization of beyond 5G (B5G) wireless systems, including the sixth generation (6G), which should be capable of unleashing the full potentials of abundant autonomous services comprising past, as well as emerging trends. More precisely, 6G is envisioned to bring novel disruptive wireless technologies and innovative network architecture into perspective. \color{black}It is further envisaged that 6G will ultimately realize the next-generation connectivity, driven by the evolution from \textit{connected everything} to \textit{connected intelligence}, thus enabling ``Human-Thing-Intelligence'' interconnectivity. Additionally, it will support high-precision communications for tactile and haptic applications in order to provide the required sensory experience at different levels, including smell, touch, vision, and hearing \cite{8613209}. The key technical requirements to realize this vision include: 
\begin{itemize}
\item Offering extreme data rates to address the corresponding massive-scale connectivity aspect and to provide ultra-high throughput, even in extreme conditions or emergency scenarios, where varying device densities, spectrum and infrastructure availability, as well as traffic patterns may exist. 
\item Achieving the targeted quality of immersion and per-user capacity and offering a unified quality of experience required by AR and VR applications, which will hit retail, tourism, education, etc.
\item  Delivering real-time tactile feedback with sub-millisecond (ms) latency to fulfill the requirements of haptic applications, such as e-health.
\item Incorporating artificial intelligence (AI) to support seamless data-centric context-aware communications for controlling environments such as smart structures, autonomous transportation systems, and smart industry \cite{2020arXiv200414850P}.  
\item  Meeting the extremely high levels of communication reliability (e.g., more than 99.9999\%) and the low end-to-end latency to support ultra-high mobility scenarios, such as flying vehicles. 
\end{itemize}

The envisioned 6G requirements, technologies and applications are presented in Fig. \ref{fig:6G}.
\color{black}

\subsection{Related Work and Contributions}
While 5G services have begun to roll out across markets, interest in 6G trends has already gained significant momentum  both in academia and industry.  Several research studies have appeared in the recent literature reporting key technological trends and new research directions that would bring 6G into reality, for example, see \cite{David,Tariq,Emilio,walid,Khaled,marco,2020arXiv200400853Z,9040431,9003619,2020arXiv200204929C,Alo1}. In \cite{Tariq}, the authors presented a speculative study on the main use cases that are expected to be brought by 6G and discussed their associated challenges and the potential enabling technologies. The authors in \cite{walid} presented a vision of some potential 6G applications and trends, and discussed the associated service classes and their performance requirements. Additionally, they briefly listed their enabling technologies and pointed out some key open research avenues. In \cite{marco}, the authors presented an overview of a number of potential 6G revolutionary technologies and the associated network architectural innovations that are envisioned to address the shortcomings of 5G systems. Likewise, the authors in \cite{Khaled} delivered a roadmap towards enabling AI in 6G. In particular, they discussed key AI methodologies that can play a central role in the design and optimization of 6G networks.  In \cite{Emilio}, the authors highlighted that in order to support future use cases, the current communication infrastructure has to evolve at both physical and architectural levels. They also discussed the need to develop mechanisms for enabling a holistic resource management platform and described the resulting challenges in terms of privacy and security. \color{black}In \cite{2020arXiv200400853Z}, the authors investigated the 6G vision from the economic standpoint, where they compared the previous wireless generations to the envisioned 6G networks, and predicted that 6G networks will offer a cost reduction by 1000 times, when compared to 5G mobile networks. Moreover, they demonstrated the role of AI towards accomplishing this vision. In the same context, the authors in \cite{9040431} presented the new themes that are likely to emerge and shape the future 6G wireless networks, such as human-machine interface, universal local and cloud computing, multi-sensory data fusion and precision sensing. Furthermore, they emphasized on the potential of AI as the foundation of 6G networks, in addition to transformative 6G technologies, such as spectrum sharing techniques, novel network architectures, and new security mechanisms. The contribution in \cite{9003619} outlines the envisioned new use cases of 6G networks in addition to the revealed paradigms of future wireless networks, including the new radio frontier, micro-service network architecture, deterministic services, and network automation. In \cite{2020arXiv200204929C}, the authors re-stated the 6G vision, trends, requirements and challenges. In particular, they provided an in-depth discussion about future 6G methods, including the integration of terrestrial and satellite communications, new spectrum utilization, modulation schemes, AI, and intelligent mobility management. Dong \textit{et al.} in \cite{Alo1} provided a framework to define the expected future applications and outline the required technologies and anticipated challenges. \color{black}Finally, the authors in \cite{David} presented a vision of 6G and its requirements. With respect to users' perspectives, they also identified innovations that need to be considered towards realizing this vision.  

It is noted that the aforementioned reported contributions mainly take a rather use cases-centric approach to the roadmap of 6G era with a focus on the associated services and technological trends. Conversely, in this survey, we approach the 6G vision from the angle of enabling technologies that manifest themselves as the paradigms needed for the realization of 6G. Specifically, we present an in-depth conceptual overview of the main revolutionary technologies in a holistic manner, taking into explicit account the key drivers, performance metrics, and major ongoing research for every single technology. Apart from the technologies discussed in the previous surveys, which are reviewed here in detail, we shed light on additional innovative technologies, such as \color{black}mmWave communications, THz communications, optical wireless communications, \color{black}metasurfaces (also known as intelligent reflective surfaces), backscatter communications (BackCom), tactile internet (TI), and aerial networks, which are envisaged to ultimately promote the 6G revolution. This survey also delves into the emerging applications of each technology and identifies their associated challenges. This discussion is used to provide a directional guidance towards future research work.

The remainder of the article is organized as follows: In Section \ref{sec:tech}, we present a comprehensive overview of five disruptive 6G technologies. This is followed by outlining some of their potential applications in Section \ref{sec:app}, whereas Section \ref{sec:challenge} highlights the fundamental challenges associated with each technology discussed in Section \ref{sec:tech}. Finally, the article is concluded in Section \ref{sec:conc}.

\section{Key Enabling Technologies for 6G Networks} 
\label{sec:tech}

Future 6G systems will require the support of novel technologies to enable unprecedented functionalities in the network. These technologies are envisioned to introduce a plethora of new applications associated with remarkably stringent requirements in terms of latency, reliability, energy, efficiency, and capacity, compared to their 5G counterparts. In this section, we provide a concrete conceptual background of major disruptive technologies that will shape the future 6G networks, which includes mmWave communications, THz communications, OWC, programmable metasurfaces, drone-based communications, BackCom, and TI. 
\color{black}\subsection{\textbf{Millimeter-Wave and Terahertz Communications}}

One of the key challenges towards realizing 6G networks is the scarcity of spectrum, owing to the unprecedented broadband penetration rate and the emergence of new use cases with rigorous bandwidth requirements. In a recent meeting, 3GPP introduced many new features for the upcoming 3GPP-Rel. 17 towards the evolution of 5G New Radio. Most notably, the enhancement of the physical layer through the support of frequency bands beyond 52.6 GHz, up to 71 GHz. To this effect, it is envisioned that future releases will go beyond this range towards the THz band \cite{2020arXiv200414699S}. Yet, there are several challenges that must be addressed in order to realize this vision. Specifically, the current PHY-layer, which has been mainly optimized for bands below 52.6 GHz, has to be technically redesigned and redeveloped.  

\subsubsection{\textbf{Millimeter-Wave Communications}}
Millimeter-wave (mmWave) communications operate over the 30-300 GHz frequency band, with a corresponding wavelength ranging from 10 to 1 mm \cite{mmw2}. Thanks to the short wavelength, mmWave communications allow the realization of small-sized antenna arrays with a large number of elements in a small physical dimension. Accordingly, narrow directional beam can be achieved, yielding multipath reflection suppression \cite{mmw3}, high immunity against jamming and eavesdropping attacks \cite{mmw5}, as well as robustness to co-user interference, since the involved wireless channels will be largely uncorrelated. Due to these promising potentials, several activities were carried out to standardize the mmWave technology. In particular, mmWave communication was introduced in IEEE 802.11ad and IEEE 802.15.3c standards \cite{mmw4}. The specifications of mmWave communications are summarized in Table \ref{table:mmwave1}.   

Nevertheless, despite the undoubted advantages of mmWave communications, there are still many associated challenges that need to be addressed prior to effective design, and successful deployment and operation. For example, small-sized components manufacturing constitutes a major challenge, due to the increased manufacturing cost. Moreover, severe signal attenuation (as high as 15 dB/km degradation \cite{mmw6}), caused by strong atmospheric absorption, limits the transmission range of mmWave communications to few kilometers.

\begin{table}[ht] \color{black}
\centering
\caption{mmWave Communications Specifications.}
\begin{tabular}{|c|c|}
\hline
\textbf{Parameters}    & \textbf{mmWave}                                                                                                            \\ \hline \hline
 Data rate &  $\geq 10$ \; \textup{Gbps}             \\ \hline
 Latency &  $\leq 1$ \; \textup{ms}             \\ \hline 
Transmission range & Hundreds of meters to a couple of kilometers              \\  \hline
Channel bandwidth &   Up to 2.16 GHz            \\  \hline
 Mobility  &    $\leq 100$ \; \textup{km/hr}         \\
 \hline \hline 
\end{tabular}
\label{table:mmwave1}
\end{table}

\subsubsection{State-of-the-Art} 
Wireless links in mmWave systems are extremely prone to obstacles blockage (including humans), especially when the physical size of the obstacles is greater than that of the wavelength, which is in general short in mmWave communications. In particular, it was shown that a human blockage can cause 20-30 dB degradation in the mmWave link. Motivated by this, several research efforts have been devoted on developing blockage control or scheduling protocols to minimize the effect of blockage in mmWave transmissions. For example, the authors in \cite{6846320} utilized multiple relays with optimum relay selection and scheduling schemes to steer the signal around obstacles, and hence, to minimize incurred signal blockages. In \cite{7417432}, the authors addressed the issue of blockage by proposing a proactive base station (BS) selection scheme based on human blockage prediction, where they utilized RGB depth cameras to estimate the location and velocity of a passing pedestrian, and consequently, estimating the time when the pedestrian blocks the line-of-sight (LOS) component of a mmWave link. In the same context, several other research contributions have investigated the blockage issue in mmWave systems, e.g., \cite{5262296,7959157,6134444}.

In addition to the above, there has recently been a vast attention on the application of non-orthogonal multiple access (NOMA) in the context of mmWave communication scenarios \cite{9050849,2020arXiv200104863Y,8454272,2020arXiv200207452M,9042253,8753467,8485639,8326498,8603758,8844783}. In particular, the research activity in NOMA-based mmWave systems is mainly directed towards investigating the secrecy rate of these systems, in addition to their performance in different realistic communication scenarios, such as drone-based communications, massive MIMO, simultaneous wireless information and power transfer (SWIPT) and M2M communications.

Other relevant research directions are focused on the areas of mmWave channel modeling \cite{8624593,8207426,7400962,7501500,7109864}, transceiver design \cite{8884199,8789649,8115137,7556971,8713847} and antenna design \cite{8731759,8685138,8758798,8316889,8400592,7116515}. The core research directions in mmWave wireless systems are summarized in Table \ref{tab:mmwave2}.

\begin{table}[ht] \color{black}
\centering
\caption{Research Directions in mmWave Wireless Systems.}
\begin{tabular}{|c|c|}
\hline
\textbf{Research direction}    & \textbf{Refs.}                                                                                                            \\ \hline \hline
 Blockage control & \cite{6846320,7417432,5262296,7959157,6134444} \\ \hline
 NOMA-based mmWave                 & \cite{9050849,2020arXiv200104863Y,8454272,2020arXiv200207452M,9042253,8753467,8485639,8326498}                 \\ \hline 
 Channel modeling        & \cite{8624593,8207426,7400962,7501500,7109864}                \\  \hline
Transceiver design        & \cite{8884199,8789649,8115137,7556971,8713847}                \\  \hline
 Antenna design  & \cite{8731759,8685138,8758798,8316889,8400592,7116515}             \\
 \hline \hline 
\end{tabular}
\label{tab:mmwave2}
\end{table}
\color{black}

\subsubsection{\textbf{Terahertz Communications}}

As noted earlier, a promising solution to the current spectrum crunch is to explore the THz-band, which is envisioned to bridge the gap between the mmWave band and infrared light-waves (optical communications), by providing a considerably wider bandwidth and enabling the development of new use cases with high data rates requirements. In addition to extending the bandwidth, THz communications offer an amplified gain due to the shorter wavelength experienced at these bands, allowing for the deployment of a large number of antennas. 

On the other hand, THz based communications require rethinking of existing solutions and investigate novel approaches that offer a seamless operation over the entire THz band. For example, the design of efficient beamforming and tracking techniques that are able to dynamically and precisely track down the location of THz-enabled devices is of great importance, and an open research problem. Other open issues include  hardware architecture design and the integration of massive MIMO and intelligent surfaces. An overview of the opportunities and challenges associated with THz communications is given in \cite{gunes4}.

\subsubsection{State-of-the-Art} Motivated by the important role of THz modulators in enabling THz technology in future wireless systems, the performance of several amplitude and phase modulators was examined for various materials and fabrication processes \cite{TH1,TH2,TH3,TH4,8409966,8306983,8691541,8616810}. For example, silicon (Si) substrates, coated with effective materials such as gold \cite{TH1} and manganese iron oxide \cite{TH2}, as well as graphene-based modulators \cite{GR1,GR2} are proven to offer a performance enhancement to the THz modulators by extending the transmission range and pumping power density, in addition to their tunability and high-speed characteristics. Nevertheless, the main drawbacks of these nano-particle-based THz modulators are the high cost and the increased complexity. On the contrary, graphene-based modulators suffer from a low modulation depth and high energy consumption. 

The implementation of THz communications in outdoor environments is rather challenging, which is particularly due to the inevitable loss caused by molecular absorption and other atmospheric conditions, such as rain \cite{6005345}. Accordingly, THz transceiver and antenna designs have to be thoroughly investigated prior to the effective design and deployment of these systems. To this end, antenna and transceiver designs have attracted great interests in the research community \cite{7967651,8737689,7859330,7930523,8345575,8601401,7956223,8239849,8340871,7565507,6709822,6594881,5325122,7175082}. Specifically, the development of Si-germanium signal generators, quantum cascade laser photonic sources, compact graphene antennas and graphene/liquid crystal based phase shifters are some of the reported research work in the field of THz transceiver and antenna designs \cite{7036065}.

Moreover, channel modeling is vital in THz communications in order to ensure reliability and high spectral efficiency. Existing research is focusing on the characterization of LOS and non-LOS (NLOS) components, with emphasis on scattering properties for the NLOS component and free space loss, molecular absorption and harsh weather conditions for the LOS component. In particular, research efforts are mainly centered around characterizing channel coefficients for deterministic and statistical conditions in indoor and outdoor environments. Ray-tracing is a reliable method for modeling LOS and NLOS components and is utilized extensively to characterize the deterministic and stochastic channel coefficients. For instance, in \cite{THC3,THC4,THC5,THC6,THC7,THC8,THC9}, the authors proposed efficient propagation deterministic models for THz nano-communications while incorporating the LOS and NLOS components for 2D and 3D scenarios. Although deterministic models provide  higher accuracy in describing channel coefficients compared to stochastic models, the underlying high computational complexity and the required geometrical information of the propagation environment are critical drawbacks of such models. On the other hand, statistical characterization of THz channels is rather challenging, especially when taking into account channel mobility, channel state information estimation, and channel correlation. The recent advancements in the design of THz communications systems are summarized in Table \ref{tab:THz}.

\begin{table}[ht]
\centering
\caption{Advancements in the Design of THz Wireless Systems.}
\begin{tabular}{|c|c|}
\hline
\textbf{Design aspects }    & \textbf{Refs.}                                                                                                            \\ \hline \hline
 {\small THz modulator} & \cite{TH1,TH2,TH3,TH4,8409966,8306983,8691541,8616810} \\ \hline
{\small Antenna design}                   & \cite{7967651,8737689,7859330,7930523,8345575,8601401,7956223,8239849}                 \\ \hline 
{\small Transceiver design}                  & \cite{8340871,7565507,6709822,6594881,5325122,7175082}                \\  \hline
{\small Channel modeling}   & \cite{THC1,THC2,8651537,8684885,7582545,8123513,7539586,8093757,THC3,THC4,THC5,THC6,THC7,THC8,THC9}             \\
 \hline \hline 
\end{tabular}
\label{tab:THz}
\end{table}

\color{black}\subsection{\textbf{Optical Wireless Communications}} 

Optical wireless communication (OWC) systems has emerged as a key technology for 6G networks and beyond,  enabling broadband connectivity. There has been an increasing interest in OWC with terrestrial, space and underwater applications. This interest is stimulated by the advancement in solid state optical technology, in addition to the promising features of OWC. These features include ultra-high bandwidth, inherent physical layer security, spatial reuse, ultra-low latency, high data rates, immunity to interference, and low cost, hence, fulfilling the demanding requirements of beyond 5G wireless networks \cite{murat}. It is recalled that information in OWC is carried over optical links, whose wavelengths vary between infrared and ultraviolet, including the visible light. OWC systems in the infrared frequency range enables long-range data transmission over high-speed wireless links, which are often encountered in wireless backhaul networks \cite{Chowdhury_2019}. In the following, we  summarize the most common OWC technologies.

Visible light communication (VLC) has emerged as a prominent technology that is anticipated to offer high-speed indoor connectivity. In VLC systems, light-emitting diodes (LEDs)/laser diodes (LDs) are used as transmitters while the receivers consist of photodetectors (PDs) \cite{7239528,8970387}. The transmission range of VLC can reach up to 20 m, with data rates of 10 Gbps and 100 Gbps, when using LEDs and LDs, respectively \cite{vlc2}. Furthermore, low-complexity implementation in VLC scenarios can be realized by using LEDs, where extra power supplies are not required. In this case, LEDs can be utilized to perform illumination, communication and localization simultaneously \cite{vlc2,8970387}. 

\color{black}
Light-Fidelity (LiFi), a promising optical solution that is envisioned to complement wireless fidelity (WiFi), is a subset of OWCs that realizes bidirectional and high-speed wireless communication \cite{Haas}. LiFi leverages visible light in downlink to realize illumination as well as  wireless communication, and infrared or RF in uplink. Similar to VLC, LiFi communication systems depend on LEDs and PDs as transmitters and receivers, respectively \cite{Haas2}. 

Optical camera communication (OCC), is another promising OWC technology, which is mainly used for positioning and navigation in indoor environments. An OCC receiver consists of embedded cameras or image sensors, while transmitter is a typical commercial LED \cite{OCC}. Moreover, OCC spectrum spans between the infrared and ultraviolet bands, with wavelength in the range of 10,000 nm \cite{7890427}. Due to the wide spread of smartphone devices with sophisticated embedded cameras, OCC can be easily implemented in these smart devices, rendering it as the practical version of VLC. 

\color{black}

Free space optical (FSO) communication, which takes place in the near infrared, is considered an effective approach in realizing high data rate communications over several kilometers \cite{8259465}. For example, for reasonable distances (around 1 km), FSO can achieve data rates in the order of 10 Gbps \cite{5771213}. High frequency reuse factor, physical security, and robustness against electromagnetic interference are other advantages exhibited by FSO systems, when restricting the use of a very narrow laser beam at the transmitter side \cite{6844864}.

\subsubsection{State-of-the-Art} 
Despite their superior features, OWC systems are impaired by several factors that have detrimental effects on their performance, such as ambient light noise, nonlinearity of LEDs, etc. The authors in \cite{533654,581112,6779317,8309378,7903587,a11,7804058,pat1} quantified the effect of ambient light noise on the performance of different optical systems, and presented efficient solutions to enhance the performance of OWC in the presence of ambient light noise. Atmospheric loss represents another major challenge in OWC, and severely degrades the performance of OWC systems in indoor and outdoor environments. Particularly, free space loss is dominating in indoor scenarios, while path loss and atmospheric turbulence are the main affecting factors on the performance of outdoor OWC. Research works in \cite{a12,6313871,6608639,8732427,8703708,7031362,1215849,ATA2019108,LATAL2019184,BAYKAL201729} investigated the effect of atmospheric loss on the performance of OWC systems. Moreover, OWC performance is vulnerable to pointing errors, which is caused by the horizontal movement of high buildings due to thermal expansion, weak earthquakes and wind \cite{1299334}. Due to pointing errors, transmitters and receivers in OWC may experience misalignment, resulting in the degradation of the system performance. The consequences of pointing errors on OWC systems under different scenarios have been thoroughly investigated in the literature, e.g., \cite{5062297,8754864,8445766,7882670,7192727,6932439,7076656}. A summary of the research efforts in OWC impairments is provided in Table \ref{tab:OWC}.

\begin{table}[ht] \color{black}
\centering
\caption{Research in OWC Impairments.}
\begin{tabular}{|c|c|}
\hline
\textbf{Impairment}    & \textbf{Refs.} \\ \hline \hline
Ambient light & \cite{533654,581112,6779317,8309378,7903587,a11,7804058,pat1} \\ \hline
Atmospheric loss & \cite{a12,6313871,6608639,8732427,8703708,7031362,1215849,ATA2019108,LATAL2019184,BAYKAL201729} \\ \hline
Pointing errors & \cite{5062297,8754864,8445766,7882670,7192727,6932439,7076656} \\ \hline 
\hline 
\end{tabular}
\label{tab:OWC}
\end{table}
\color{black}

\subsection{\textbf{Programmable Metasurfaces for Wireless Communications}} 

The mmWave and THz communications are envisioned as key enablers for 6G systems. They are expected to satisfy the stringent requirements of various potential 6G use cases by exploiting higher frequency bands. However, owing to the severe attenuation and scattering properties, the detrimental effects on communication efficiency remains the grand challenge in wireless communications. For example, small and large objects in indoor environments, e.g., walls and furniture, typically scatter rays in all directions, leading to severe multipath propagation environments. The Doppler effect is another key challenge, which can limit the realization of ultra-broadband communications, particularly in the mmWave and THz bands \cite{MSs3}. Existing solutions mainly rely on device-side approaches, which consider the wireless environment to be uncontrollable and hence, it remains unaware of the on-going communication processes.

\subsubsection{Metasurfaces} 
Metasurfaces have recently emerged as an innovative technology, which is envisioned to revolutionize  wireless communications by allowing  wireless system designers  to fully manipulate the propagation of electromagnetic (EM) waves in a wireless link. The building block of a metasurface is a meta-atom, which is an artificial conductive structure with engineered EM properties that is repeated periodically across a rectangular surface (also called tile).  At the macroscopic level,  metasurfaces exhibit unique EM properties such as customized permittivity and permeability levels, and negative refraction \cite{8466374}. As a consequence, metasurfaces enable unprecedented capabilities when interacting with impinging EM waves, which include wave focusing, absorption, imaging, scattering, polarization, to name but a few \cite{8466374}.  It is worth mentioning that metasurfaces leverage these unique abilities without any limitation on their operating frequency. 
\par Recently, there has been a steadily growing interest in both industry and academia on tunable metasurfaces, also called programmable metasurfaces. In this context, the meta-atom design can be dynamically altered through a simple external stimuli, such as a binary switch, empowering metasurfaces with unique adaptivity. More specifically, dynamic meta-atoms are fitted with tunable switching components, such as micro-electro-mechanical switches or CMOS transistors, which can alter the structure of the meta-atoms. This allows metasurface tiles to receive commands from an external programming interface, where parameters of the incident and reflected waves, e.g.,  phase, amplitude, frequency, and polarization, are carefully manipulated in order to enable the EM behavior of interest \cite{MSs3}. Moreover, the discovery of communicating nodes in the surrounding wireless environment can be realized by equipping the metasurface tile with efficient sensing and reporting features. 

 Tunable mechanisms of metasurfaces facilitate massive connectivity,  interference mitigation, and enhanced diversity by introducing an additional degree of freedom. These tunable features are essential in order to realize the flexibility needed for future wireless communications. The authors in \cite{8449754} presented the first model to describe a programmable wireless indoor environment using programmable metasurfaces. The introduction of programmatically controlled wireless environments has undeniably opened the door for a broad range of functionalities to be ultimately achieved even at the mmWave and THz bands. Some metasurface functionalities are summarized as follows:
\begin{itemize}
\item[$-$] \textit{Beam steering:} This function can be achieved by allowing a metasurface to change the direction of the impinging wave towards the desired direction through manipulating either the refraction or reflection index which can override the outgoing directions defined by the Snell's Law \cite{MSs3}. 
\item[$-$] \textit{Beam splitting:} In this function, a metasurface tile splits an incident wave into customized orthogonal  multiple beams steered towards multiple directions to serve multiple users simultaneously \cite{8449754}.      
\item[$-$] \textit{Wave absorption:} Blocking the access of an unauthorized wireless device can be accomplished by adjusting the properties of the metasurfaces to ensure no or minimal reflection or refraction of the incident wave. This functionality can be utilized to prevent eavesdropping and optimize the network physical-layer security. As demonstrated in \cite{MS}, for a given incident wave, metasurfaces are able to reduce the wave power by 35 dB. 
\item[$-$] \textit{Wave polarization:} This function allows a metasurface to fully control the polarization of  impinging waves and manipulate their oscillation orientation \cite{8449754}.      
\item[$-$] \textit{Phase control:} This functionality of metasurfaces allows for the alteration of the carrier phase \cite{8449754}.           
\end{itemize}

\color{black}\subsubsection{Artificial intelligence (AI)-empowered Metasurfaces}
To facilitate simultaneous functionalities within an uninterrupted connectivity, AI tools are envisioned to be indispensable in programmable metasurfaces as they enable the identification of the best operation policy based on data driven techniques \cite{9007666,mohjazi2020outlook}. Leveraging AI is particularly attractive in heterogeneous wireless applications, with diverse networks, and QoS user requirements, as it can potentially provide an efficient and dynamic means to adapt network parameters, such as coding rate, route selection, frequency band, and symbol modulation. 

Machine learning (ML)-enabled solutions, which are a subfield of AI, are expected to be a core component in smart programmable metasurfaces, allowing them to achieve a complex level of coordination, and thereby maintaining a desired global behavior while ensuring scalability, energy and overhead reduction \cite{mohjazi2020outlook}. In this setup, a metasurface interface can be empowered with intelligence by applying novel data processing paradigms that can learn from data and perform functionalities to complete complex tasks efficiently. This would support self-organization and automation of all metasurface functions, including maintenance, management, and operational tasks. A recent research study proposed a deep learning (DL)-based ML approach to achieve signal focusing through learning the mapping between the estimated channel state information (CSI) at a user location and the optimal configuration of the metasurface's unit cell \cite{huang2019indoor}. Furthermore, adaptive control and coordination of multiple metasurfaces in programmable wireless environments was demonstrated for a set of users through the application of neural networks \cite{liaskos2019interpretable}. More recently, convolutional neural network approaches were proven to exhibit their merit for beamforming by learning the physics of the beamforming from computed data to make online prediction of the coding matrices, to fulfill the network requirements \cite{8988246}. Additionally, the principle of ML-enabled imager was proposed for programmable metasurfaces to produce high-quality EM imaging and high-accuracy object recognition \cite{li2019machine}. The results are promising in real-time compressed imaging applications in the microwave, millimeter-wave, and THz frequencies \cite{li2019machine}.
\color{black}

 \subsubsection{State-of-the-Art} 
Programmable metasurfaces have recently attracted a large attention of the international scientific community.  In particular,  a number of research studies examined the potentials of programmable metasurfaces as modulators  \cite{zhao, QPSK,zhang2,wang,basar}. Furthermore, the research works in \cite{guan,yu,shen,chen} investigated the design of smart beamforming in metasurface-based wireless secrecy communication systems. Researchers also explored the use of metasurfaces in wireless power transfer (WPT) \cite{wu,li,7354885}. The role of machine learning in controlling the functionalities of metasurfaces to actively improve the coverage of the highly dynamic indoor environments is analyzed in \cite{taha,del,Huang}. The aforementioned state-of-the-art is summarized in Table \ref{tab:meta}.

\begin{table}[ht]
\centering
\caption{\footnotesize{Advancements in Metasurfaces.}}
\begin{tabular}{|c|c|}
\hline
\textbf{Addressed Schemes}     & \textbf{Refs.                                                                                                       } \\ \hline \hline
{\small Modulation} & \cite{zhao, QPSK,zhang2,wang,basar} \\ \hline
{\small Beamforming}    & \cite{guan,yu,shen,chen}             \\ \hline
{\small Wireless power transfer (WPT) }                  & \cite{wu,li,7354885}                 \\ \hline 
{\small Machine learning}                  & \cite{taha,del,Huang}                \\ \hline \hline 
\end{tabular}
\label{tab:meta}
\end{table}

\subsection{\textbf{Drone-Based Communications and Autonomous Systems}} 

\begin{figure*}[!t] 
\centering
\includegraphics[width=1\linewidth]{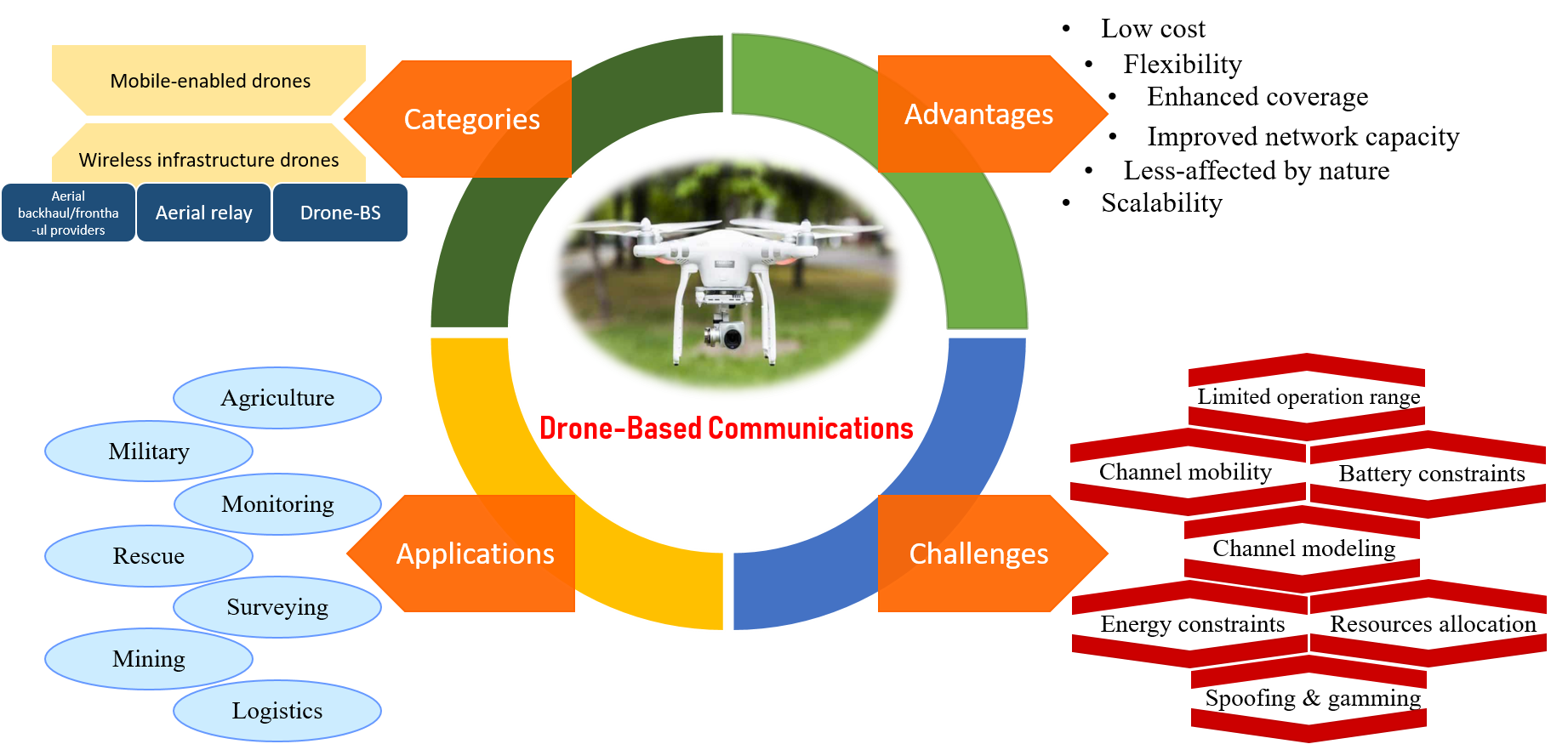}
\caption{Advantages, categories, applications, and challenges of drone-based systems.}
\label{fig:drones2}
\end{figure*}

A key driving force behind the vision of 6G is the deployment of connected and autonomous vehicle (CAV) systems and drone-based communications. The research efforts in CAV and drone (also known as unmanned aerial vehicles (UAVs))-based communication systems, have been steadily growing in both academia and industry, targeting strict requirements, particularly ultra-low latency and unprecedented communication reliability. The advantages, categories, applications, and challenges of drone-based systems are depicted in Fig. \ref{fig:drones2}. In the following, we focus our attention on the current and futuristic application scenarios of UAVs, as well as the state-of-the-art. An in-depth discussion of the underlying challenges is then provided in the following section.  

\subsubsection{Drone-assisted Wireless Communications}

Drones constitute the basic building block of aerial networks, whose inherent features, such as mobility and flexibility, enable several  imminent and futuristic applications in wireless networks \cite{8598647}. It is shown that the use of drones can significantly improve the coverage and transmission rates \cite{gunes1}. Furthermore, standardization activities led by 3GPP are currently ongoing to adapt the necessary changes in order to integrate drones in the future wireless networks \cite{gunes2}. 

Drone-assisted wireless communications can be categorized as follows:

\textit{\underline{Cellular-enabled Drones (CED)}:} CEDs are operated as user equipments (UEs) (i.e., drone-UEs) in order to enable several key applications such as  mining, oil and gas, transportation, surveying and monitoring, with a velocity of 160 km/h in urban and rural environments \cite{dronesO}. To ensure connectivity with  cellular networks, several essential requirements, which include reliable and low-latency communications between the drone-UEs and ground BSs, have to be met. User-centric cell-free massive MIMO, also known as distributed massive MIMO, was recently proposed as a prominent solution to efficiently increase the system coverage and energy efficiency of aerial networks \cite{8756714,7827017}. By efficiently utilizing cell-free massive MIMO, in which massive number of antennas are distributed over a wide area to serve multiple drone-UEs, the effect of large scale fading becomes less as all users approximately have equal distances to the allocated access point. This is beneficial for cell-edge users who experience severe large scale fading. Recent results showed that the utilization of cell-free massive MIMO architecture brings substantial benefits to the performance of drone-UEs, compared to multi-cell massive MIMO, as inter-cell interference is eliminated in the former \cite{8952782}. In some applications, drone-UEs will require high-speed connectivity from ground BSs or from drone-BSs (i.e., drones operating as BSs). 
It should be pointed out that nowadays cellular networks are designed for ground users with unique mobility and  traffic characteristics  that are considerably distinct from those experienced with drone-UEs. Therefore, the integration of drone-UEs into cellular networks in a single wireless network presents a set of new key challenges and design considerations that must be addressed for the efficient realization and successful deployment of CEDs \cite{8660516}.  

\textit{\underline{Wireless Infrastructure Drones (WIDs)}:}  WIDs are intended to extend the network capabilities by enhancing network coverage or capacity. WIDs can be further classified based on their functionalities into: 
\begin{itemize}
\item  \textit{Drone-BSs:} 
Drone-BSs are aerial nodes with some   BS features and functionalities, that are envisioned to provide capacity and coverage enhancements for 6G networks. They are cost-effective solutions that render wireless connectivity to hard-to-reach areas, as well as geographical areas with limited cellular infrastructure. Drone-BSs are also attractive solutions for  delivering reliable, broadband and wide-scale temporary wireless connectivity in special events or harsh scenarios, such as sport events and natural disasters. Furthermore, high altitude drone-BSs are expected to provide a long-term and cost effective connectivity for rural areas.  The integration of drone-BSs with other physical layer techniques such as mmWave and massive MIMO, cognitive radios, etc., is a promising solution to provide data-hungry services  and is expected to create a new set of challenges for the next generation of flying BSs \cite{8660516}. Optimal positioning of the drone-BSs is one of the critical challenges that needs to be addressed in dense deployment scenarios \cite{gunes3}.

\item \textit{Aerial Relays:} Relaying has been extensively investigated in the context of terrestrial communications aiming to enhance network reliability, throughput, and coverage. However, such relays are subject to limited mobility and often are constrained by wired backhauling \cite{7470933}.  On the contrary, drones acting as wireless relays, are versatile and offer high mobility. This feature makes them a promising candidate for providing enhanced wireless connectivity beyond LOS. Moreover, aerial relays can play a significant role in extending the battery life of drones \cite{7463007}.  

\item  \textit{Aerial Backhaul for Cellular Networks:} Wireless back-hauling has been shown to provide a cost-effective solution compared to wired backhauling. However, it is subject to interference, blockage and path loss, which can significantly degrade the performance and reduce the data rate \cite{7306534}. In this respect, drone-based networks are foreseen to play a fundamental role in achieving robust and high-speed backhaul connectivity for cellular networks \cite{8254715}. Such networks are expected to provide flexible drone-based backhaul communications that will enhance the network capacity, reliability, as well as the operation cost \cite{8660516}. 

\end{itemize}

\subsubsection{{Air-Ground-Space Integrated Networks}}

There is a strong belief that existing terrestrial, aerial and satellite networks  will not be able to cope with the massive volume of generated data, which will continue to grow at an exponential rate together with the rapid proliferation of new IoE applications. On the other hand, the integration of these networks is viewed as the next evolution of  wireless infrastructure, that  is envisioned to cater to diverse use cases with different QoS requirements, particularly in realistic scenarios such as urban, rural, and lightly dense areas.
Yet, despite their indispensable benefits, the envisioned integrated architecture will introduce unprecedented challenges that include, but are not limited to, heterogeneity, security, resource management, self-organization, energy consumption, and backhauling \cite{8533634}.

\subsubsection{{Drone-based Multi-access Edge Computing}}

Multi-access edge computing (MEC), enables cloud computing capabilities at the edge of cellular networks, and has recently emerged as one of the potential technologies for 5G networks.  In particular, MEC enables mobile devices with limited resources to offload their computation tasks to the edge of the network.

In  drone-enabled networks, mobile devices can offload their computationally demanding tasks to drones with MEC capabilities, typically at the edge of the network, thus reducing the network congestion and allowing for the rapid deployment of new applications. Additionally, the drone-based network can provide an effective mobility management without the necessity of handover, as well as uninterrupted MEC services for high mobility users, due to their large-scale coverage and LOS connection \cite{N1,N2}.  
  
Within the same context, the  limited on-board processing capabilities of drones, which are mainly due to their storage and battery constraints, bring about several concerns towards the efficient execution of complex tasks \cite{7423671}. In particular, heavy computation demanding applications, such as real-time image processing, may not be supported by the anticipated vision of drones. Recent research efforts proposed  efficient techniques to tackle these limitations by leveraging cloud computing to offload the computation-intensive  tasks from the drones to remote cloud servers \cite{7545004}. The role of these servers can be summarized as follows \cite{7964096}: 
\begin{itemize}

\item[$-$] {\textit{Storage:}} Storage services can be offered by the cloud to store  drones data streams that include environment and mission-related parameters, captured images and sensed data. 

\item[$-$] {\textit{Computation:}} Intensive computations are executed in the cloud in order to minimize the processing time and energy consumption at the drone. Moreover, the large amount of stored data from the drones can be exploited to perform data analytic tasks in order to enhance the performance of drones-enabled networks, in terms of trajectory adaptation, altitude optimization, and energy consumption customization, in an intelligent manner.
\end{itemize}

\color{black}\subsubsection{SWIPT-enabled Drones}
Short built-in battery lifetime restricts the utilization of drones in numerous applications, due to the limited size and weight of drones, which results in a limited energy storage, and consequently, short cruising duration. Aiming to prolong the system's lifetime, trajectory design, location adjustment, and power allocation optimization approaches have been proposed in the literature as promising mechanisms to overcome the energy shortage issue in drones \cite{7888557,8316986,8119562}. However, these schemes are not always feasible in practice and do not provide fundamental solutions to the involved energy inefficiency problem \cite{9075988}. In this context, SWIPT has been foreseen as an emerging energy-replenishment solution, in which drones harvest energy from received RF signals to extend the cruising duration \cite{8915758,8876848,8821282}. Despite their remarkable advantages, SWIPT-enabled drones are more vulnerable to physical layer attacks, such as eavesdropping, spoofing and jamming attacks, due to the LOS and broadcast features of drone channels. There have been several attempts to propose efficient solutions in order to enhance the secrecy rate performance of SWIPT-enabled drones, including conventional physical layer security mechanisms, such as cooperative jamming, artificial noise, and multiple antenna techniques, in addition to position optimization and resource allocation \cite{9075988,8667026,8611204}. An interesting work reported in \cite{8653332} investigated the resource allocation optimization in UAV-assisted SWIPT systems, in which drones are exploited to send data to specific ground receivers, in addition to transmitting the same RF signals to ground energy receivers, which are equipped with wireless power harvesting devices. These energy receivers may be possible eavesdroppers. Therefore, the authors in \cite{8653332} formulated an optimization problem to obtain the optimum resource allocation scheme that maximizes the secrecy rate of the information receivers. \color{black}

\subsubsection{State-of-the-Art} 
The research efforts have focused thus far on basic and more advanced design aspects of drones, including CEDs and WIDs. Specifically, extensive research efforts are directed towards proposing encryption and authentication mechanisms to ensure a robust secure communication infrastructure that also maintains privacy \cite{8581510,8691741,8486625,8795473,8255736,8255739,8337900,8337908,8337902,8325268,8675384}. 

The energy efficiency and battery properties are  key practical design aspects, in which the battery lifetime, charging mechanism and energy consumption must be optimized to enable seamless and uninterrupted wireless communications. In this context, the optimization of energy consumption and  charging time of drones have lately received significant attention \cite{8301585,8255733,8701666,8653315,8708295,8613833,8469090,8846189}.

Various design issues, that allow for the realization of the full potential of drone networks, are tackled in the literature, such as network architecture \cite{8253543,8316776,8533634,7744808}, image processing \cite{8701689,7900406,8418072}, interference management \cite{8755983,8555533,8287974,8851647,7756327,8654727} and storage \cite{8733128,8734799}. Recent advances in aerial networks are summarized in Table \ref{tab:AN}.
\begin{table}[ht]
\centering
\caption{Recent Advancements in Aerial Networks.}
\begin{tabular}{|c|c|}
\hline
\textbf{Addressed issue}     & \textbf{Refs.}                                                                                                            \\ \hline \hline
Security \& privacy & \cite{8581510,8691741,8486625,8795473,8255736,8255739,8337900,8337908,8337902,8325268,8675384} \\ \hline
Energy and battery efficiency   & \cite{8301585,8255733,8701666,8653315,8708295,8613833,8469090,8846189}             \\ \hline
Network architecture                   & \cite{8253543,8316776,8533634,7744808}                 \\ \hline 
Image processing                  & \cite{8701689,7900406,8418072}                 \\ \hline 
Interference management                   & \cite{8755983,8555533,8287974,8851647,7756327,8654727}                 \\ \hline
Storage                   & \cite{8733128,8734799}                   \\ \hline \hline 
\end{tabular}
\label{tab:AN}
\end{table}

\subsection{\textbf{The Next Frontier for IoE: Backscatter Communications and Wireless Powered Networks}} 

The exponential growth of connected devices, constituting the emerging IoE, is a major driving force towards the development of energy-efficient solutions to sustain wireless communication among connected nodes \cite{7120024}. Nonetheless, despite the notable advancements, the short battery lifetime of the deployed battery-operated devices still constitutes a major design challenge, which requires a paradigm shift towards the development of the next generation green communication architecture. Accordingly, ambient BackCom have emerged as a new communication paradigm for low power communications in 5G networks. This approach is based on the concept that a transmitter sends data to its receiver by backscattering ambient signals, e.g., TV or Wi-Fi signals. Compared to conventional systems, backscatter transceivers consume significantly less power (orders of magnitude), rendering it a strong candidate for low power networks and IoE applications \cite{8368232}. Owing to its promising features, several new and disruptive technologies can be integrated with BackCom.

\subsubsection{Radio-frequency (RF)-Powered BackCom Networks}
 
RF energy harvesting (RF-EH) has been recently proposed as a promising solution to provide perpetual energy replenishment for such networks. RF-EH is realized by allowing wireless devices, equipped with dedicated EH circuits, to harvest energy from either ambient RF signals or dedicated RF sources. It can be divided into two main categories, namely wireless-powered communications \cite{7984754} and simultaneous wireless information and power transfer, which have been shown to provide noticeable gains in terms of power and spectral efficiencies by enabling simultaneous information process. Despite the remarkable advantages, RF-EH techniques still suffer from particular limitations, especially in the context of low power wireless networks. Specifically, wireless-powered devices are not able to communicate perpetually, as they require dedicated time for energy harvesting. Additionally, these devices depend on active RF signals for communication; as a consequence, they suffer from relatively high power consumption, which can pose major issues, particularly in large-scale low power wireless networks \cite{8368232}. 
Motivated by this concern, a new trend is to integrate BackCom systems with various RF-EH techniques in a single network. This promising paradigm is
envisioned to address some of these challenges and catalyze the deployment of new technologies and services \cite{BS-EH}. 

\subsubsection{{RF-Powered Cognitive Radio Networks and Ambient BackCom}}

The integration of RF-EH techniques with cognitive radio (CR) networks has led to the development of a new communication paradigm, called RF-powered CR networks \cite{RFID}. In such networks, a CR transmitter harvests RF energy when a primary (licensed) user (PU) is active, which is subsequently utilized for data transmissions between secondary (unlicensed) users (SUs) \cite{RFID,RFID2}. Evidently, the performance of these networks greatly depends on the availability of PU signals. In this context, BackCom is envisioned as a potential solution to address this challenge by allowing SUs to harvest energy from PU signals in addition to transmitting data by backscattering the PU signals. Therefore, it is evident that although BackCom and energy harvesting have not played a major role yet in 5G, they are envisaged to be a core part of 6G with full potentials. 

\subsubsection{{Visible Light BackCom}}

VLC is  a new  paradigm that is foreseen to provide ubiquitous connectivity while addressing some of the limitations and challenges of RF communications. It is based on intensity modulation and direct detection, where the intensity of LEDs is modulated to convey information, and then demodulated/detected directly using a photodiode. There are several  key advantages  of VLC that include inherent communication security, high degree of spatial reuse, and its immunity to RF interference, which makes it safe to be used in critical places with high electromagnetic interference, e.g., hospitals and industrial plants. The principle of visible light  BackCom (VLBC) systems is similar to its RF counterpart, in which VLBC leverages ambient light to harvest energy and then modulates  VLC signals to transmit its data to backscatter receivers \cite{AL,AL2}.

\subsubsection{{Quantum BackCom}}
 Quantum backscatter communications is another promising technology which is anticipated to contribute towards the development of 6G and the next generation IoT, particularly in terms of performance and security \cite{8491095}. In this new paradigm, a transmitter produces entangled signal-idler (S-I) photon pairs. The S-photon is transmitted and backscattered from a backscatter transmitter, while the I-photon is kept at the receiver. This quantum setting provides a significant gain in the error exponent for the communication link and facilitates secure communication by exploiting quantum cryptography.
 
\subsubsection{{State-of-the-Art}}

BackCom networks are subject to critical security threats, such as eavesdropping and jamming. This stems from the typical simplicity and low-complexity of BackCom transceivers. As a result, existing security solutions, including encryption and digital signatures, may not be applicable due to the power and complexity constraints of BackCom devices. This has motivated the international research community to investigate new security mechanisms that can guarantee fully secure and private wireless communications \cite{7432027,8000668,8681643,6987335,7122464,7556997,7867859,6836141}.  Self-interference is another major limitation in BackCom systems. The sources of self-interference include: (i) signals from ambient RF sources,  and (ii) multipath propagation. Based on this, several self-interference cancellation techniques have been recently reported in the open technical literature \cite{8219367,7913737,8353418,6942226,8666220,8302845,8735851,8700258}.

It is noted that BackCom networks are not optimized and/or designed for large-scale low power networks comprising a massive number of IoT devices, e.g., sensors in environmental monitoring, sensors in smart roads to collect data about the pavement conditions and traffic, etc. Furthermore, such systems are different from human-centric communications with diverse and unique traffic characteristics as well as QoS requirements, which requires the development of  efficient physical layer and media access control schemes to prevent access congestion. Within this context, multiple access techniques in BackCom systems are regarded to be instrumental in improving the efficiency of backscatter networks. In particular,  conventional orthogonal multiple access  and non-orthogonal schemes (e.g., NOMA \cite{8501953} and rate splitting multiple access), are recognized as promising candidates for enabling massive connectivity, while maintaining high energy and spectral efficiency \cite{8756283,8851217,8847703,8636518,8439079,8412618,8017383,7995033,8399824}.

In addition to the aforementioned studies, major research efforts have focused on channel modeling and estimation \cite{8320359,8618337,8746230,7982802,8523801,7820135}, resource allocation \cite{8851217,8730429,8476159,8756283,8700623} and wireless energy harvesting \cite{8360017,8116748,8340034,7981380,8413073,8434224,8588295} in BackCom systems.
Table \ref{tab:BS} summarizes the timely open research topics in the field of BackCom.

\begin{table}[ht]
\centering
\caption{Open Research Topics in BackCom Systems.}
\begin{tabular}{|c|c|}
\hline
\textbf{Research topics}     & \textbf{Refs.}                                                                                                            \\ \hline \hline
Security  &   \cite{7432027,8000668,8681643,6987335,7122464,7556997,7867859,6836141} \\ \hline
Interference control                    & \cite{8219367,7913737,8353418,6942226,8666220,8302845,8735851,8700258}                \\ \hline 
Multiple access   & \cite{8756283,8851217,8847703,8636518,8439079,8412618,8017383,7995033,8399824}            \\ \hline
Channel modeling \& estimation                  & \cite{8320359,8618337,8746230,7982802,8523801,7820135}               \\ \hline 
Resource allocation                    & \cite{8851217,8730429,8476159,8700623}                \\ \hline 
Wireless energy harvesting                    & \cite{8360017,8116748,8340034,7981380,8413073,8434224,8588295}                \\ \hline \hline 
\end{tabular}
\label{tab:BS}
\end{table}



\subsection{\textbf{Tactile Internet}}

Tactile Internet (TI) is seen as the next frontier of IoE, focusing on M2P and M2M interactions. With the recent advances in tactile/haptic devices, it is predicted that TI will catalyze the deployment of a plethora of new applications ranging from health care to education and smart manufacturing. Therefore, it is expected to reshape our daily lives and ultimately realize the full potential of the next industrial revolution, also known as Industry 4.0. 
 
To fully realize TI, the communication infrastructure (CI) has to meet  strict design guidelines, as it is currently unable to address the stringent requirements of the use cases envisioned for TI. In particular, the CI has to support extremely low end-to-end latency with high-reliability \cite{Berg}. Furthermore, it must ensure data security without jeopardizing the latency requirements imposed by the computationally demanding encryption techniques.  

To address these requirements and catalyze the deployment of new use cases with unique requirements, the development of unique and disruptive B5G wireless communication technologies is of paramount importance. To this end, we envision the development of: 1) communication technologies in the THz band; 2) novel wireless network architectures; and 3) AI-enabled communication networks.

\subsubsection{State-of-the-Art} 

The use cases  of TI have recently drawn  significant research and industrial attention, as they are envisioned to have great potential to advance all aspects of our daily lives. As reported in the literature, TI has been adopted in numerous applications, such as VR and AR \cite{8197491,8344795,7786938,8689144,8026164,8733996,8337839,7924239,8291482,8019876}, healthcare \cite{8197481,5665813,7707355,7829437,8755882,8673005,8371236}, education \cite{7036100,8765378,7845618,7891552,7123626}, intelligent transportation \cite{8377984,8624569,7995874,5648353}, industry \cite{8718538,8542940,8281493,7962162} and robotics \cite{7980645,8736514,8063891,8556371,8491368,8241709}. Table \ref{tab:TI} summarizes the envisioned applications of TI.

\begin{table}[ht]
\centering
\caption{Applications of Tactile Internet.}
\begin{tabular}{|c|c|}
\hline
\textbf{Applications}     & \textbf{Refs.}                                                                                                            \\ \hline \hline
VR and AR  &   \cite{8197491,8344795,7786938,8689144,8026164,8733996,8337839,7924239,8291482,8019876} \\ \hline
Healthcare                    & \cite{8197481,5665813,7707355,7829437,8755882,8673005,8371236}               \\ \hline 
Education                    & \cite{7036100,8765378,7845618,7891552,7123626}               \\ \hline 
Intelligent transportation   & \cite{8377984,8624569,7995874,5648353}            \\ \hline
Industry                  & \cite{8718538,8542940,8281493,7962162}               \\ \hline  
Robotics                  & \cite{7980645,8736514,8063891,8556371,8491368,8241709}              \\ \hline \hline 
\end{tabular}
\label{tab:TI}
\end{table}

\section{Driving Applications of 6G Technologies} 
\label{sec:app}

In this section, we highlight the potential applications associated with the aforementioned technologies, which are expected to realize the vision of 6G systems.
\color{black}\subsection{\textbf{Millimeter-wave Communications}} 


\begin{itemize}

\item \textit{Wireless Backhaul:} Due to ultra-dense deployment of heterogeneous multi-tier small cells in future wireless communications, fiber-based backhauling will no longer be an economical option to connect multiple BSs to each others and to the gateway, due to several installation restrictions. Therefore, wireless backhaul represents a scalable promising alternative to connect multiple BSs in small cells. This can be achieved by utilizing the wide and underutilized mmWave bandwidth, such as the unlicensed 60 GHz band, as well as the 70-80 GHz band. Specifically, leveraging the mmWave band to realize wireless backhaul solutions is expected to provide increased flexibility, high speed transmission, cost efficiency and enhanced data rates \cite{7306533}. Another key advantage of adopting the mmWave band in wireless backhauling is the controlled level of inter-cell interference, due to the LOS nature of the mmWave links \cite{6600706}.

\item \textit{Wearable devices:} Recent advancements in miniature electronics fabrication technology prompts the spread of smart high-end wearable devices, which require higher data rates and longer battery lifetime, such as smart watches, smart AR/VR glasses and helmets, health-care gadget, and motion trackers \cite{mmw6}. Given that the transmission range constraint is relaxed in these applications, wireless communication between wearable devices and the smart receiver (which can be a smartphone) can be efficiently realized by incorporating mmWave communications as a perfect candidate to establish high data rate, low interference and reliable communication between the device and the receiver, especially in densely populated indoor environments \cite{7445132}. 

\item \textit{Imaging and Tracking:} Given that signals in the 60 GHz band will be mostly reflected from objects larger than their short wavelength, mmWave communications is considered as a promising candidate in imaging and tracking systems. Particularly, objects' dimensions can be accurately measured by relying on the highly directional beams of the 60 GHz links, which also helps reducing the interference, and subsequently, providing a precise imaging and tracking systems. Furthermore, the miniaturized antenna arrays facilitate the integration of these antennas in small receivers (such as smart phones or tablets).
\end{itemize}
\color{black}

\subsection{\textbf{Terahertz Communications}}

The wide bandwidth in the THz band is envisioned to drive the deployment of a large array of new use cases, as outlined next.

\begin{figure*}[!t] 
\centering
\includegraphics[width=1\linewidth]{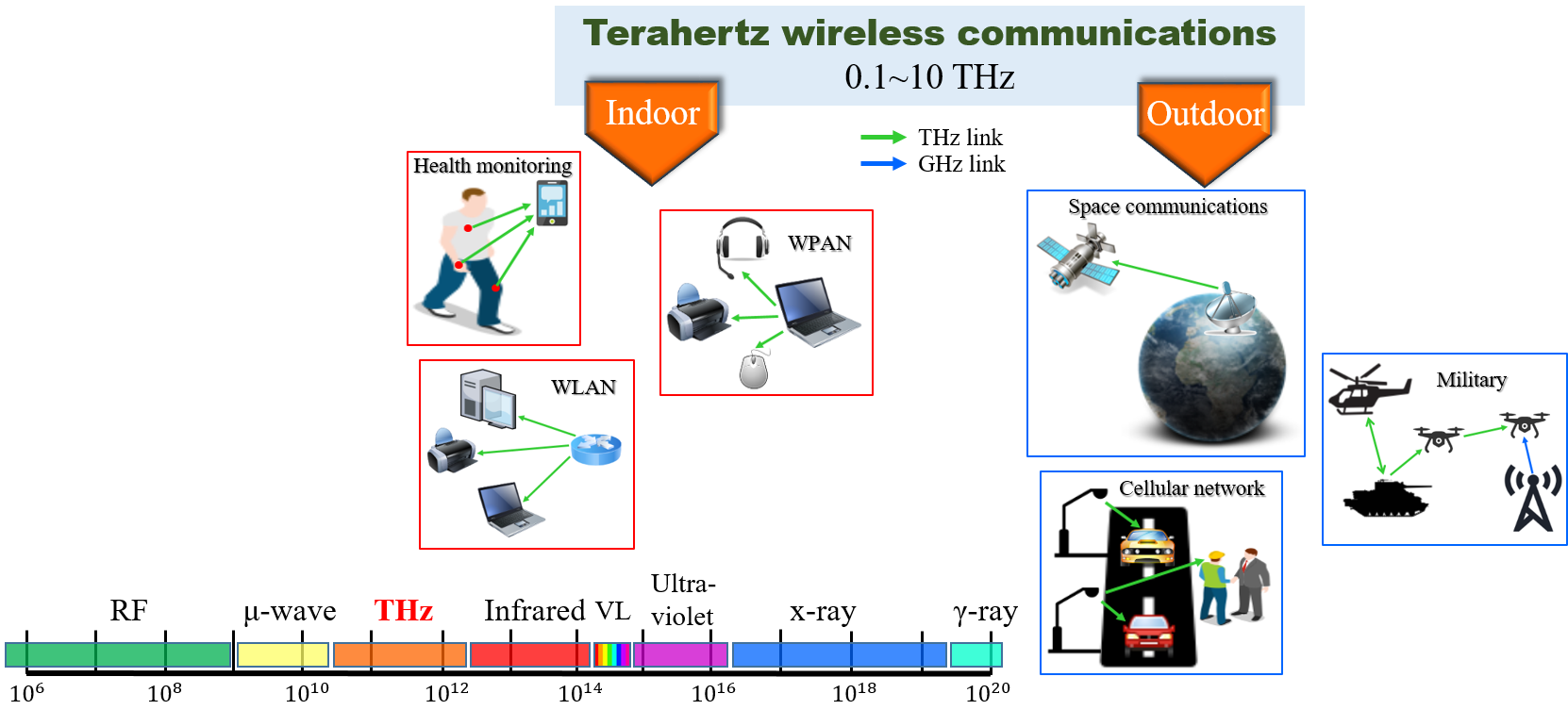}
\caption{Indoor and outdoor applications of THz communications.}
\label{fig:THz}
\end{figure*}

\begin{itemize}

\item \textit{Wireless Data Centers:} Today's data centers suffer from high complexity, power consumption, maintenance cost, and wasted spaces occupied by large cables. Therefore, there have been attempts to address these issues in order to enable fast and reliable access to cloud-based services. According to  \cite{8663550}, the power consumption of all data centers will reach 73 billion KWh by 2020. Therefore, THz communications could be a promising candidate for the next generation of data centers, satisfying the peak data rate of 10-20 Gbps required by 5G, and even higher \cite{THz3}. Although still in infancy, there have been recent research investigations on channel modeling for indoor environments, which have paved the way for utilizing THz communications in indoor wireless data centers \cite{THZC,THZC2}. Initial results showed that the cabling cost can be reduced without compromising the bandwidth. 

\item \textit{Secure Drone Communications:} Drone communications in the THz band is one of the envisioned applications of THz communications that are expected to achieve higher capacity gains and support increased mobility \cite{THz4}. Moreover,  the deployment of large antenna arrays for coverage extension enables extremely narrow beams, which inherently limits the probability of eavesdropping, and therefore, it achieves secure communications \cite{THz5}.

\item \textit{Health Monitoring:} THz communications is a promising candidate in the field of health care. Specifically, several nano-sensors can be utilized to monitor different ions in the human blood, such as glucose and sodium, in addition to cholesterol levels, infections and cancer bio-markers. The collected data by the sensors are forwarded to a micro interface, e.g., a cellular phone or a medical device, using THz communications \cite{8663550}. It is noted that the THz radiations are considered safe for the human-being bodies compared to the gamma rays \cite{gamma}. 

\item \textit{Wireless local area networks (WLANs)/Wireless personal area networks (WPANs):} THz band communications are envisioned to enable bandwidth-intensive applications such as high definition holographic video conferencing and ultra-high speed data transfer. This stems from the fact that a seamless interconnection may be facilitated between ultra-high wired networks (e.g., fiber optical links) and wireless devices (e.g., laptops or tablets) in WLANs or between personal wireless devices in WPANs \cite{THz5}. 

\end{itemize}

Potential applications of THz communications are outlined in Fig. \ref{fig:THz}.

\color{black}
\subsection{\textbf{Optical Wireless Communications}}  
\begin{itemize}

\item \textit{Smart Transportation Systems:} The wide spread of LEDs in current transportation lighting systems, such as vehicles lights, street lamps and traffic signals, facilitates the utilization of these LEDs to perform wireless communications besides their original role of illumination. LEDs can be exploited to implement OWC to realize safe and smart transportation systems, by allowing vehicles to communicate road-related information, including vehicles speed, navigation, and traffic status, while maintaining low-complexity and low-cost implementation \cite{6852088}.
\item \textit{Airplane Passenger Lights:} OWC can be interestingly applied to travelers lighting in airplanes, in which LEDs can realize wireless communications for in-flight passengers audio and video files transmission, in addition to instant messaging and data exchange \cite{5722675}.
\item \textit{Underground Mining:} Serious accidents, due to cave collapsing, chemical leakage, and gas explosions, in underground mining raise a critical issue in the mining industries. In such events, it is of paramount importance to detect the location of the miners in order to provide the proper assistance. Recently, VLC has emerged as a promising technology for indoor positioning, due to its suitability in enclosed places, in addition to its low cost, low interference and high data rates features \cite{6924006}.
\item \textit{Healthcare:} Electromagnetic interference caused by RF signals is considered as a threatening factor for expensive medical machines. Moreover, intensive care units pose restrictions on the use of mobile phones operating over the RF band \cite{med}. Intra-clinical data transmission is considered an attractive application to OWC, which is reported as safe for human health. OWC can be implemented in healthcare buildings to provide safe and high data rate transmission over short distances, in addition to lighting, which minimizes the implementation cost and provides a health-friendly alternative to RF communications \cite{7151783}. 
\end{itemize}
\color{black}
\subsection{\textbf{Metasurfaces for Wireless Communications}}

\begin{itemize}
\item \textit{Metasurfaces in WPT Applications:} WPT is foreseen as a game-changing technology, in which  future networks are envisioned to provide perpetual energy replenishment, particularly for low power devices/sensors.  A major concern, however, is the ability of devices to harvest enough energy in wireless channels. 
The unique properties of metasurfaces, that include their abilities to steer and concentrate electromagnetic waves, enable efficient power transfer and energy harvesting.  The work in \cite{song} discussed the potentials of integrating smart tables with metasurfaces in order to enable multiple wireless devices to be powered/charged simultaneously. The integration of WPT in metasurfaces for biological applications was studied in \cite{li}, where a metasurface-based wearable device was placed over the human skin surface to improve the efficiency of an implanted WPT system.

\item \textit{Metasurface-based Textiles for Wireless Body Sensor Networks (WBSNs)}: Very recently, metasurface-based textiles were developed for energy-efficient and secure WBSN applications \cite{tian}. In this approach, regular clothing is fitted with conductive metasurface textiles, where wireless signals can glide around the surface of the body on the clothes to interconnect wireless wearable devices with each other forming a WBSN. In this application scenario, wearable devices are located in close proximity to the body. This results in a significant reduction in the power dissipated by the wireless devices, leading to an improvement in the battery life and data rates of these devices. In fact, this innovative WBSN is foreseen to boost the received signal compared to conventional technologies. Furthermore, metasurface-based textiles may enable personal sensor networks, which are highly efficient, immune to interference, and inherently secure \cite{tian}. Looking ahead, they are envisioned to have future applications in high-tech athletic wear, health monitoring, and human-machine interfaces.  
\end{itemize}

\subsection{\textbf{Drone-Based Communications and Autonomous Systems}}

\begin{itemize}

\item \textit{Search and Rescue Missions:} Search and rescue missions are some of the critical driving applications of drone networks. This is primarily due to the flexibility of drones compared to manned vehicles, which take a longer time to deploy \cite{droneApp}. 

\item \textit{Mailing and Delivery:} Package delivery is one of the attractive civil applications of drones, adopted by  major courier companies around the world in order to accomplish fast, cost-effective and reliable delivery. This is motivated by the fact that most of the packages' weights are below the maximum tolerable load of a single drone \cite{7423671}. For example, Amazon reported that 83\% of their packages weights fall below the 2.5 kg \cite{Amazon}, while FedEx average package weight is less than 5 kg \cite{7423671}.

\item \textit{Marine Drones:} Underwater drones can accomplish several military and civil underwater missions, such as oil spills exploration and identification, in addition to performing intensive studies relating to marine organisms and ecosystem \cite{marine}.

\end{itemize}

\subsection{\textbf{BackCom and Energy Harvesting}}
\begin{itemize}
\item \textit{Smart Homes:} Low power battery-less backscatter sensors equipped with energy harvesting devices can be efficiently embedded in homes to perform a wide range of tasks, such as gas leak detection, smoke and carbon oxide detection, and movement monitoring.  Another driving application of BackCom is smart dustbins, in which backscatter devices keep track of the garbage level and report it to garbage collecting trucks. 
\item \textit{Smart Cities:} Backscatter enabled sensors can be flexibly placed in street lamps, parking lots, buildings, and bridges to realize the envisioned energy-efficient low-cost smart cities. BackCom can be utilized in smart cities to enhance  air quality by monitoring the pollution and noise level in the air. Additionally, it can be used to manage traffic in closed parking areas and ease the process of finding an available parking place by indicating the available slots. 
\item \textit{Biomedical Applications:} Wearable and implantable human medical devices, in addition to plants and animals monitoring, are some of the key drivers of BackCom technology. For example, Smart Google contact lenses, which are equipped with miniaturized BackCom devices, are designed to continuously measure the glucose levels in the tears for diabetes patients and backscatter the reported results to a wireless controller. Other serious diseases, such as epilepsy and Parkinson's, are envisioned to be diagnosed and treated by the assistance of BackCom technology. In particular, it is envisaged that brain-implantable BackCom neural devices will play the role of the brain-computer interface needed for studying and diagnosing diseases of interest.
\end{itemize}

\begin{figure*}[!t] 
\centering
\includegraphics[width=0.8\linewidth]{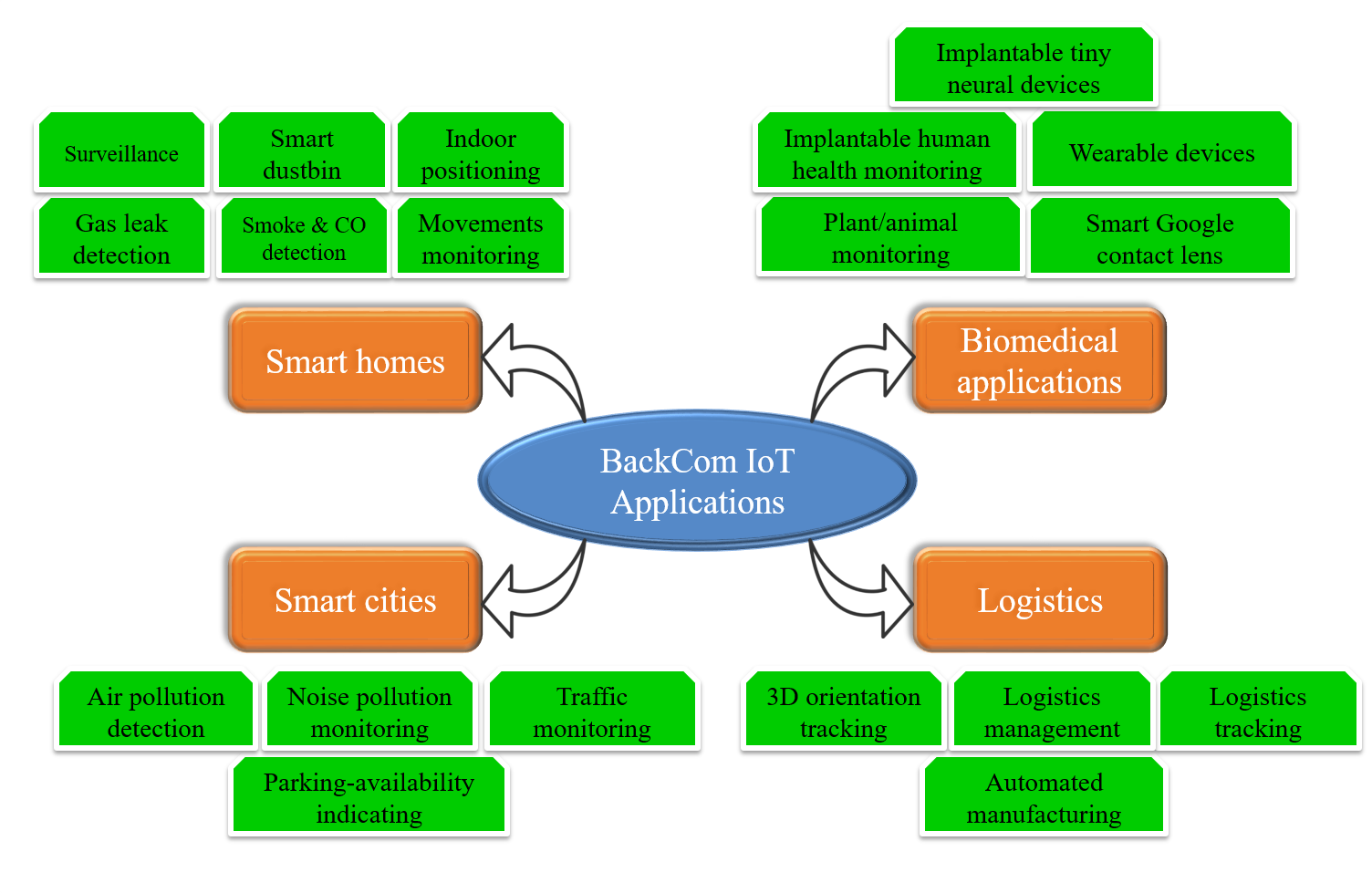}
\caption{Applications of BackCom systems.}
\label{fig:BSC}
\end{figure*}

Potential applications of BackCom systems are presented in Fig. \ref{fig:BSC}. 

\subsection{\textbf{Tactile Internet}}

\begin{itemize}
 \item \textit{Industry:} Automation in industry realizes the control of machinery and processes through a large network of sensors and actuators in order to improve  productivity and reduce  labor costs. Industrial automation is steadily growing in the context of TI, enabling the full control of rapidly moving devices with high sensitivity while  meeting the end-to-end latency requirements. However, the ever-growing need for control processes with different latency, reliability, data rate, and security demands is envisioned to catalyze the development of new wireless solutions tailored to these requirements. 
 
 \item  {\textit{Virtual Reality} (VR):} VR enables users to physically interact with each other by applying various motor skills over a VR simulation platform. 
In this context, TI is anticipated to provide the low latency required to facilitate shared virtual environments. High-fidelity interaction requires haptic feedback to allow users to touch objects in a VR environment and enable users to feel one another's actions on the same touched object. This requires a stable and seamless user communication coordination, which is not supported by today's VR systems. Hence, TI is foreseen as a key enabler for haptic communications with ultra-low delay communication  and reliability requirements.

 \item  {\textit{Augmented Reality} (AR):} AR applications are fast growing, owing to the availability of AR glasses and powerful smart devices equipped with small sensors and cameras. However, the present AR systems are restricted to deliver pre-processed content due to the limited computational capabilities of the small wireless devices and the inherent delays in the communication network. TI, on the other hand, is perceived to enable the augmentation of dynamic and real-time information to the contents.

 \item \textit{Healthcare:} Potential applications of TI in healthcare include  tele-surgery, tele-rehabilitation, and tele-diagnosis. Different from healthcare services provided by current communication networks, which are location-dependent, medical expertise provided by the TI will not be bounded by time and/or a physical location. For example, a physician can diagnose patients at their locations by remotely controlling a robot while receiving haptic feedback as well as audio-visual information. Tele-surgery is another example, which  has the  potential to revolutionize healthcare delivery in the next decade. 

\item \textit{Education:} Improved learning experiences over distances can be achieved via  TI by allowing teachers and learners to exchange haptic information. Identical multi-modal human-machine interfaces are required to enable auditory, visual, and haptic interactions, which can be realized by enabling ultra-low latency communication systems. For example, TI may allow a remote music instructor to apply instant actions over the haptic overlay to correct the hand moves of a student learning a musical instrument.
 \end{itemize}

\section{Challenges and Future Directions}
\label{sec:challenge}

In this section, we discuss the open research issues associated with the previously presented potential 6G technologies and highlight their research challenges.

\color{black}\subsection{\textbf{Millimeter-Wave Communications}}

\begin{itemize}

\item \textit{User Mobility:} User mobility constitutes a major challenge in the implementation of mmWave networks, that severely affects the system's capacity. Therefore, to realize the full potential of mmWave communications, there is a need to develop novel, efficient and adaptive modulation and coding schemes in order to compensate for channel variations. Moreover, in indoor small cell scenarios, which comprise sets of access points serving multiple devices (each set called basic service set), user mobility causes severe and rapid load fluctuations in each set, in addition to recurrent handovers between multiple access points \cite{bss}. This calls for the development of sophisticated user association and handovers mechanisms between multiple access points, which are capable of providing a guaranteed QoS, balanced load, and improved system capacity for the realization of efficient mmWave communications in future wireless networks. 

\item \textit{Interference Management:} To overcome the short range limitation in mmWave communications, a large number of access points are employed to extend the link coverage in small cell environments. In several indoor scenarios, such as office cubicles and conference rooms, networks experience interference due to the deployment of a large number of access points (i.e., large number of basic service sets). This interference can be detrimental if the device is close to the interfering access point, which is a highly probable event. Therefore, research interests should be directed towards developing novel interference management mechanisms to prevent significant deterioration in the performance of mmWave communications.

\item \textit{Blockage and Shadowing Control:} Sensitivity to blockage represents a fundamental challenge for wireless mmWave communications. Specifically, a sudden blockage for the LOS transmission between the BS and the user causes a disconnected session, yielding a significant degradation in the system's reliability. Additionally, re-establishing a new connection between the user and another BS increases the network overhead, affecting the system's latency, which is a major issue in the envisioned 6G mobile networks. Signal steering, to avoid human blocking, requires a very large number of access points, which augments the level of interference in addition to the increased complexity. Therefore, the design of reliable anti-blockage schemes is necessary before implementing efficient mmWave communications in future wireless networks.
\end{itemize}\color{black}
\subsection{\textbf{Terahertz Communications}}

\begin{itemize}

\item \textit{Transceiver Architecture:} Transceiver architectures in the THz band is one of the critical aspects to be considered due to the unique characteristics of the propagation environment of THz links. In order to realize the full potentials of the THz-band, there is a growing interest in the development of novel transceiver architectures that can operate across the entire THz-band. The developed architectures are expected to combat the severe path loss, thus, enabling high sensitivity and high power gains. Moreover, the co-existence of different frequency bands, such as THz, mmWave, and microwave cells, requires thorough investigations over different layers.

\item \textit{THz Modulator:}
The characteristics of THz modulators, including amplitude and phase modulators, play a central role in quantifying the efficiency of THz communication systems. These characteristics include, but not limited to, modulation speed and depth in amplitude modulators and phase shift amount in phase modulators. Current modulators designs, with the adopted architecture and utilized materials, limit the modulator ability to achieve ultra-high speed and consequently to realize efficient THz wireless systems. This stems from the fact that the existing modulators do not allow EM radiation manipulation in the THz band, which is required in order to facilitate high-speed control of the modulator characteristics. Therefore, this calls for research intervention to develop intelligent and tunable ultra-high speed modulators, with approximately 1 picosecond response time, to enable efficient and reliable THz wireless communications \cite{8663550}.

\item \textit{Channel Modeling for THz Communications:} Existing low frequency channel models can not accurately capture the entire behavior of high frequency THz links, which experience severe attenuation due to molecular absorption and antenna aperture, in addition to the free space loss. Note that the multi-path channel of THz communications compromises LOS and NLOS components. On the contrary, LOS attenuation, represented by path loss, is measured by the addition of the spreading and molecular absorption losses, which are encountered due to wave expansion and molecular absorption, respectively. The severity of molecular absorption is determined based on the density of molecules experienced along with the transmission link, distance, weather conditions (e.g., heavy rain), and frequency window in the THz band. Accordingly, LOS channel component in the THz band is described as severely frequency selective. Therefore, it is essential to develop an accurate model to represent the LOS component in the THz wireless system, which is necessary to identify the performance limits of THz communications and propose enhancement schemes for such technology \cite{8387218}. 

On the other hand, due to the unavailability of the LOS components in some scenarios, the THz link might be limited to the NLOS component, which can be classified into specular reflected, diffusely scattered and diffracted EM waves. Therefore, for precise channel characterization, it is required to accurately trace the reflection, scattering and diffraction coefficients of the incident beam in the THz system \cite{6998944}, which depend on the incident angle and surface material and geometry.

Hence, the development of realistic and accurate channel models for THz links is still an open research problem, which requires thorough investigation to enable the implementation of an efficient THz wireless system.
\end{itemize}
\color{black}\subsection{\textbf{Optical Wireless Communications}}

\begin{itemize}
\item \textit{Physical Layer Security:} Heterogeneous ultra-dense networks are envisioned to shape the future 6G wireless networks, in which hybrid RF/optical wireless communications are widely deployed. Although physical layer security in OWC is thoroughly investigated in the literature, implementing secure hybrid RF/optical networks, which require the development of efficient physical layer security mechanisms, represents a major challenge in such systems. By noting that legitimate users and eavesdroppers in hybrid networks are connected to either RF or VLC sources, it is essential to propose physical layer security mechanisms to associate users to the appropriate source in a way that maximizes the system secrecy rate \cite{8622294}.  

\item \textit{Multiple Access Networks:} The limited modulation bandwidth and peak optical power represents a performance limiting factor in the realization of efficient OWC systems, as it directly affects the spectral efficiency in such systems, particularly in VLC scenarios. Several multiple access schemes have been proposed in the literature to accommodate multiple users, and hence improve the spectral efficiency in OWC networks, such as NOMA, rate splitting multiple access (RSMA) and space-division multiple access (SDMA) \cite{2020arXiv200207583N}. It is worth noting that existing research in the field of optical multiple access is primarily relying on perfect impractical scenarios. In specific, in the reported work, perfect CSI, Gaussian noise, and availability of LOS assumptions are considered. Therefore, it is essential to examine the performance of OWC systems under imperfect scenarios, such as investigating the effect of ambient light. Additionally, considering that VLC links may experience fading and shadowing, it is essential to examine the performance of different optical multiple access schemes under imperfect channel conditions. Moreover, physical layer security of NOMA, RSMA, and SDMA systems, in the context of OWC, remains an open research problem.

\item \textit{Reconfigurable Intelligent Surfaces for FSO:} The existence of a LOS link constraint constitutes a major challenge in the implementation and generic deployment of FSO networks. This is due to the fact that optical links in FSO systems are usually impaired by several factors, such as atmospheric turbulence and geometric and misalignment losses \cite{schober}. Consequently, optical reconfigurable intelligent surfaces (RIS) have emerged as an efficient solution to relax the LOS constraint in FSO networks. Different than relay, RIS is considered an energy efficient technology to extend the coverage area of FSO wireless networks at low implementation cost and complexity. Motivated by the promising potentials of RIS in FSO networks, the research community has recently started to actively investigate the integration of RIS in FSO scenarios \cite{schober,9057633,2020arXiv200105715W}.

\end{itemize}
\color{black}

\subsection{\textbf{Metasurfaces for Wireless Communications}}

In spite of the promising prospects of metasurfaces in 6G, several design aspects should be further investigated in order to realize the full potential of this promising technology.

\begin{itemize}

\item {\textit {Dynamic Structure Design:}}  The ability to manipulate the configurations of meta-atoms constitutes a key design challenge for the efficient operation of reconfigurable metasurfaces, whose deployment is needed to support a wide range of functionalities in highly dynamic wireless environments. Although there exist some research studies that have successfully demonstrated that multiple functionalities can be achieved by multiple metasurfaces, only a few have presented the capability of a metasurface to perform different functionalities simultaneously \cite{yang2016}. In this case, each unit cell of the metasurface has to be controlled independently, raising the need to develop effective distributed meta-atom control mechanisms and to examine the performance of the variety of functions supported by each metasurface. Additionally, since metasurfaces are envisioned to be deployed in application scenarios involving the operation over a wide frequency range (varying from 1 to 60 GHz), designing efficient metasurface structures that are capable of dynamically switching the operation frequency poses an essential research goal \cite{8466374}.

\item \textit{Efficient Programmable Interface:} Apart from the need to develop metasurface structures capable of realizing different functions in real-time, there is a compelling need to investigate advanced multi-functional metasurfaces that can switch from one EM behavior to another in a fast manner to cater for the increasingly diverse user demands, especially in high mobility scenarios where the system convergence rate may not be within the coherence time of the surrounding wireless environment.  As a result, research efforts should be directed towards developing control software that incorporates low-complexity and fast configuration optimizers to  facilitate the optimization and adaptation of metasurfaces functionalities to the surrounding environment. Also, advanced signal processing and machine learning algorithms may be developed to leverage the sensing capabilities of metasurfaces for enabling intelligent system performance optimization, which can converge within the coherence time of the environment and can be aligned with the network requirements of 6G systems, such as massive connectivity, ultra-low latency, and high reliability \cite{8466374}. 

\item \textit{High-order Modulation:} The design of high-order modulation  and  novel waveform designs for metasurface-based wireless communication systems constitute promising solutions for enabling high data rate transmissions. This is of paramount importance since current metasurface-based transmitters are limited to single-carrier low-order modulation schemes, such as binary/quadrature phase-shift-keying \cite{QPSK,zhang2}.

\item \textit{WPT in {Metasurfaces}:} It is recalled that the last years witnessed remarkable advancements in battery design. Nonetheless, the short battery life of wireless devices still constitutes a major design challenge and requires a paradigm shift towards the development of the next generation green communication architectures. WPT was proposed recently as a promising solution to provide perpetual energy replenishment for such networks. It is realized by allowing wireless devices, equipped with dedicated energy harvesting circuits, to harvest energy from either ambient RF signals or dedicated RF sources. Given that metasurfaces have the ability to steer, absorb and collimate EM waves, particular research efforts should be dedicated to exploit the unique functionalities of metasurfaces to wirelessly charge the wireless devices from long distances.

\end{itemize}

\subsection{\textbf{Drone-Based Communications and Autonomous Systems}}

\begin{itemize}

\item \textit{Network Architecture and Analysis:} Network planning, performance evaluation and resource allocation are some of the challenges that drone-BSs encounters in aerial networks. While terrestrial mobile networks are designed to meet the requirements of ground users, they  are not optimized to support aerial networks. Specifically, terrestrial BS antennas are not designed to support the ultra-low latency requirements of high elevation angle users in aerial networks. Therefore, there is a need to develop a novel and efficient system architecture that can efficiently integrate terrestrial BSs  with drone-based UEs.

\item \textit{Energy and Storage Efficiency:}  Energy constraint is a limiting factor in mobile-enabled drones, particularly since solar energy and the limited size of built-in batteries are the only sources of power.  This is a crucial issue, especially in power-hungry monitoring missions, where continuous monitoring  and transmission  are inevitable.  Various energy-aware mechanisms have been reported in the literature  to address the energy efficiency problem in drones. For example, an approach is to utilize multiple cooperative drones to allow a single drone to temporarily leave the network for energy replenishment \cite{7470933}. Storage constraint is another major  concern, e.g., in monitoring missions, where drones must store a large amount of data. This motivates the investigation of novel forwarding and compression schemes to efficiently handle this huge amount of data.

\item \textit{Collision Avoidance:} Buildings and large obstacles represent a major hazard to drones, so they must be addressed thoroughly in order to avoid collisions to objects in the surrounding environment. A way to address this problem is to restrict the drone flying zones to limited areas. However, this will increase the interference between multiple drones and lead to higher collision probability \cite{7465674}. Therefore, there is a need for  efficient collision avoidance schemes to enable drones dynamically adjust their trajectories to minimize collision probability. 

\item \textit{Channel Modeling:} Efficient implementation of cooperative aerial networks requires accurate characterization of  communication links to ensure reliable and safe operation of air-to-air and air-to-ground links. In flying ad-hoc network architectures, drone communications require the development of robust theoretical framework to model air-to-air and air-to-ground links. While there have been reported works on link characteristics of aerial networks in different frequency bands \cite{8288376,7407385,7842372}, there are still not enough results to characterize the channel models, particularly for cooperative (relaying) scenarios \cite{8353365}. 

\par Although the communication link characteristics of drone-based systems are unique, some terrestrial channel models, such as two-ray and Rician models, were shown to be a good fit for drone environments; however, more experimental and real-time tests are required in order to verify the accuracy of such models and properly select their parameters. More importantly, further research efforts must be dedicated to verify the validity of these models in different frequency bands, such as 433 MHz, 1575.42 MHz, and 2.4 GHz bands \cite{8353365}.

\end{itemize}

\subsection{\textbf{Backscatter Communications and Energy Harvesting}}

\begin{itemize}

\item \textit{Security and Jamming:} BackCom systems typically suffer from potential security and jamming attacks, owing to their simple modulation and coding schemes. The key issue is that the limited  resources in backscatter systems are not able to support the implementation of conventional security solutions that include encryption and  digital signatures \cite{8219367}. This calls for the development of simple, yet highly efficient security solutions to realize secure BackCom systems.

\item \textit{Interference to Licensed Systems:} Data transmission in ambient BackCom is based on reflecting ambient signals received from licensed sources. Therefore,  interference  imposed on licensed users is inevitable, which  calls for the need to develop communication protocols that guarantee no or minimal interference. Recent research efforts have focused on interference modeling and development of compensation schemes  \cite{8368232}.   

\item \textit{Full-Duplex Ambient Backscatter:} Full-duplex BackCom systems are proposed to enable simultaneous communication between multiple ambient backscatter nodes. In such cases, the same antenna is used by a backscatter receiver to transmit and receive signals. As a result, a significant amount of self-interference exists between different components of the BackCom transceiver. This calls for the development of self-interference mitigation schemes and constitutes an open research issue towards addressing this challenge \cite{AL3,7913737}.

\end{itemize}

\subsection{\textbf{Tactile Internet}}
Although TI is considered as a new paradigm envisioned to generate a plethora of new applications, several open research challenges exist and need to be fully addressed for the successful realization of this enabling technology.

\begin{itemize}

\item \textit{Haptic Devices:} Haptic devices, such as sensors and actuators, enable users to feel, touch, and manipulate objects in real or virtual environments. Although haptic devices have already been commercialized, they still fall short in terms of degrees of freedom as well as the cost effectiveness. Additionally, in order to realize the envisioned applications, haptic devices have to offer kinesthetic and tactile control simultaneously \cite{Aijaz}.

\item \textit{Data Compression:} Bandwidth-limited networks represent a major challenge for haptic communications, which requires a paradigm shift towards the development of innovative solutions that would enhance system reliability and user experience. In this regard, several haptic data compression techniques have been thoroughly investigated in the literature, to realize the full potential of TI. However, further investigations towards haptic codec design for TI are required. This might include the proposal of a new set of kinesthetic and tactile codec solutions that will lead to highly efficient data compression techniques.

\item \textit{Integration of Multi-Modal Sensory:}  One of the key challenging aspects in enabling haptic feedback is  multi-modal sensory, where visual, haptic, and auditory feedback are integrated simultaneously. However, these different modalities vary in terms of their latency, sampling, and transmission rate. Subsequently, novel multiplexing schemes have to be studied in order to temporally integrate multiple modalities with different priorities. 

%

\item \textit{Ultra-Reliability:} Since TI is expected to disrupt major attributes in the society, ultra-reliable network connectivity is necessary to minimize the packet losses and reduce the outage to $10^{-7}$ \cite{Aijaz}. A highly lossy environment in haptic communications leads to erroneous sensations and directly interrupts the user's activity. There are several factors that impact the reliability of TI applications. This includes uncontrollable interference, lack of resources, equipment failure, and reduced signal strength. This will require the investigation of efficient reliability enhancement mechanisms to achieve ultra-high reliability in real-time operations \cite{Antonakoglou}.   

\item \textit{Ultra-Low Latency:}  As stated earlier, TI requires sub-ms end-to-end latency. Therefore, it is essential to understand the latency budget between  sensors and actuators in order to investigate the impact of each contributing factor in the chain. In general, the end-to-end latency is dominated by air-interface, backhaul, and core latencies. To cater to the critical latency requirements, innovative latency optimization mechanisms are necessary in addition to effective protocol stack and hardware designs.  
\end{itemize}

\section{Conclusion}
\label{sec:conc}

Although the glory of 5G networks is at its peak, initial implementation and testing phase of 5G networks along with the emergence of a plethora of new applications, such as bio-interface applications, are revealing new challenges and limitations of the upcoming wireless networks, including but not limited to, ultra-high reliability, extremely high data rates and ultra-low latency. Accordingly, this spots the lights on the fundamental question: \textit{Will the forthcoming 5G wireless networks be able to accommodate the newly emerged applications with the concurrent stringent requirements, necessary for realizing fully autonomous and intelligent systems?} To answer this question, in this paper, we sketched out the roadmap into the future hypothetical vision of B5G networks. Particularly, we focused on exploring expected new technologies for 6G networks, such as mmWave communications, THz communications, OWC, metasurfaces, aerial networks, BackCom, and TI, along with their potential applications and inherent challenges. The technical challenges associated with these technologies call for a deeper investigation, which will potentially accelerate the development of innovative solutions as well as standardization efforts for 6G.



\bibliographystyle{IEEEtran}
\bibliography{references-gk}

\begin{thebibliography}{100}
\providecommand{\url}[1]{#1}
\csname url@samestyle\endcsname
\providecommand{\newblock}{\relax}
\providecommand{\bibinfo}[2]{#2}
\providecommand{\BIBentrySTDinterwordspacing}{\spaceskip=0pt\relax}
\providecommand{\BIBentryALTinterwordstretchfactor}{4}
\providecommand{\BIBentryALTinterwordspacing}{\spaceskip=\fontdimen2\font plus
\BIBentryALTinterwordstretchfactor\fontdimen3\font minus
  \fontdimen4\font\relax}
\providecommand{\BIBforeignlanguage}[2]{{%
\expandafter\ifx\csname l@#1\endcsname\relax
\typeout{** WARNING: IEEEtran.bst: No hyphenation pattern has been}%
\typeout{** loaded for the language `#1'. Using the pattern for}%
\typeout{** the default language instead.}%
\else
\language=\csname l@#1\endcsname
\fi
#2}}
\providecommand{\BIBdecl}{\relax}
\BIBdecl

\bibitem{ITUreport}
{R}ep. {ITU}-{R} {M}.2370-0, ``{IMT} traffic estimates for the years 2020 to
  2030,'' 2015.

\bibitem{OD1}
A.~{Yadav} and O.~A. {Dobre}, ``All technologies work together for good: A
  glance at future mobile networks,'' \emph{IEEE Wireless Commun.}, vol.~25,
  no.~4, pp. 10--16, Aug. 2018.

\bibitem{OD3}
M.~{Mohammadkarimi}, M.~A. {Raza}, and O.~A. {Dobre}, ``Signature-based
  nonorthogonal massive multiple access for future wireless networks: Uplink
  massive connectivity for machine-type communications,'' \emph{IEEE Veh.
  Technol. Mag.}, vol.~13, no.~4, pp. 40--50, Dec. 2018.

\bibitem{OD2}
G.~I. {Tsiropoulos}, A.~{Yadav}, M.~{Zeng}, and O.~A. {Dobre}, ``Cooperation in
  {5G} {HetNets}: Advanced spectrum access and {D2D} assisted communications,''
  \emph{IEEE Wireless Commun.}, vol.~24, no.~5, pp. 110--117, Oct. 2017.

\bibitem{Fernando}
D.~{Kreutz}, F.~M.~V. {Ramos}, P.~E. Verissimo, C.~E. {Rothenberg},
  S.~{Azodolmolky}, and S.~{Uhlig}, ``Software-defined networking: {A}
  comprehensive survey,'' \emph{Proc. IEEE}, vol. 103, no.~1, pp. 14--76, Jan.
  2015.

\bibitem{Han}
B.~Han, V.~Gopalakrishnan, L.~Ji, and S.~Lee, ``Network function
  virtualization: {C}hallenges and opportunities for innovations,'' \emph{IEEE
  Commun. Mag.}, vol.~53, no.~2, pp. 90--97, Feb. 2015.

\bibitem{7169508}
A.~{Gupta} and R.~K. {Jha}, ``A survey of {5G} network: {A}rchitecture and
  emerging technologies,'' \emph{IEEE Access}, vol.~3, pp. 1206--1232, Jul.
  2015.

\bibitem{OD4}
S.~M.~R. {Islam}, N.~{Avazov}, O.~A. {Dobre}, and K.~{Kwak}, ``Power-domain
  non-orthogonal multiple access ({NOMA}) in {5G} systems: Potentials and
  challenges,'' \emph{IEEE Commun. Surveys Tuts.}, vol.~19, no.~2, pp.
  721--742, Second quarter 2017.

\bibitem{Tariq}
F.~Tariq, M.~Khandaker, K.-K. Wong, M.~Imran, M.~Bennis, and M.~Debbah, ``A
  speculative study on 6{G},'' \emph{arXiv preprint, arXiv:1902.06700}, Aug.
  2019.

\bibitem{walid}
W.~{Saad}, M.~{Bennis}, and M.~{Chen}, ``A vision of {6G} wireless systems:
  {A}pplications, trends, technologies, and open research problems,''
  \emph{IEEE Network}, vol.~PP, pp. 1--1, Oct. 2019.

\bibitem{marco}
M.~Giordani, M.~Polese, M.~Mezzavilla, S.~Rangan, and M.~Zorzi, ``Towards {6G}
  networks: {U}se cases and technologies,'' \emph{arXiv preprint,
  arXiv:1903.12216}, Mar. 2019.

\bibitem{Khaled}
K.~B. {Letaief}, W.~{Chen}, Y.~{Shi}, J.~{Zhang}, and Y.~A. {Zhang}, ``The
  roadmap to 6{G}-{AI} empowered wireless networks,'' \emph{IEEE Commun. Mag.},
  vol.~57, no.~8, pp. 84--90, Aug. 2019.

\bibitem{Emilio}
E.~{Calvanese Strinati}, S.~{Barbarossa}, J.~L. {Gonzalez-Jimenez},
  D.~{Ktenas}, N.~{Cassiau}, L.~{Maret}, and C.~{Dehos}, ``6{G}: The next
  frontier: {F}rom holographic messaging to artificial intelligence using
  subterahertz and visible light communication,'' \emph{IEEE Veh. Technol.
  Mag.}, vol.~14, no.~3, pp. 42--50, Sep. 2019.

\bibitem{8613209}
M.~{Katz}, M.~{Matinmikko-Blue}, and M.~{Latva-Aho}, ``{6Genesis} flagship
  program: Building the bridges towards {6G}-enabled wireless smart society and
  ecosystem,'' in \emph{IEEE 10th Latin-American Conference on Communications
  (LATINCOM)}, 2018, pp. 1--9.

\bibitem{2020arXiv200414850P}
E.~{Peltonen}, M.~{Bennis}, M.~{Capobianco}, M.~{Debbah}, A.~{Ding},
  F.~{Gil-Casti{\~n}eira}, M.~{Jurmu}, T.~{Karvonen}, M.~{Kelanti}, A.~{Kliks},
  T.~{Lepp{\"a}nen}, L.~{Lov{\'e}n}, T.~{Mikkonen}, A.~{Rao}, S.~{Samarakoon},
  K.~{Sepp{\"a}nen}, P.~{Sroka}, S.~{Tarkoma}, and T.~{Yang}, ``{6G} white
  paper on edge intelligence,'' \emph{arXiv e-prints, arXiv:2004.14850}, Apr.
  2020.

\bibitem{David}
K.~{David} and H.~{Berndt}, ``6{G} vision and requirements: {I}s there any need
  for beyond 5{G}?'' \emph{IEEE Veh. Technol. Mag.}, vol.~13, no.~3, pp.
  72--80, Sep. 2018.

\bibitem{2020arXiv200400853Z}
S.~{Zhang}, C.~{Xiang}, and S.~{Xu}, ``{6G}: Connecting everything by 1000
  times price reduction,'' \emph{arXiv preprint, arXiv:2004.00853}, Apr. 2020.

\bibitem{9040431}
H.~{Viswanathan} and P.~E. {Mogensen}, ``Communications in the {6G} era,''
  \emph{IEEE Access}, vol.~8, pp. 57\,063--57\,074, Mar. 2020.

\bibitem{9003619}
K.~{Samdanis} and T.~{Taleb}, ``The road beyond {5G}: A vision and insight of
  the key technologies,'' \emph{IEEE Network}, vol.~34, no.~2, pp. 135--141,
  Mar. 2020.

\bibitem{2020arXiv200204929C}
S.~{Chen}, Y.-C. {Liang}, S.~{Sun}, S.~{Kang}, W.~{Cheng}, and M.~{Peng},
  ``Vision, requirements, and technology trend of {6G}: How to tackle the
  challenges of system coverage, capacity,user data-rate and movement speed,''
  \emph{arXiv preprint, arXiv:2002.04929}, Feb. 2020.

\bibitem{Alo1}
S.~Dang, O.~Amin, B.~Shihada, and M.-S. Alouini, ``What should {6G} be?''
  \emph{Nat. Electron.}, vol.~3, p. 20–29, Jan. 2020.

\bibitem{2020arXiv200414699S}
H.~{Saarnisaari}, S.~{Dixit}, M.-S. {Alouini}, A.~{Chaoub}, M.~{Giordani},
  A.~{Kliks}, M.~{Matinmikko-Blue}, N.~{Zhang}, A.~{Agrawal}, M.~{Andersson},
  V.~{Bhatia}, W.~{Cao}, Y.~{Chen}, W.~{Feng}, M.~{Heikkil{\"a}}, J.~M.
  {Jornet}, L.~{Mendes}, H.~{Karvonen}, B.~{Lall}, M.~{Latva-aho}, X.~{Li},
  K.~{L{\"a}hetkangas}, M.~T. {Masonta}, A.~{Pandey}, P.~{Pirinen}, K.~{Rabie},
  T.~M. {Ramoroka}, H.~{Saarela}, A.~{Singhal}, K.~{Tian}, J.~{Wang},
  C.~{Zhang}, Y.~{Zhen}, and H.~{Zhou}, ``A {6G} white paper on connectivity
  for remote areas,'' \emph{arXiv preprints, arXiv:2004.14699}, Apr. 2020.

\bibitem{mmw2}
M.~{Xiao}, S.~{Mumtaz}, Y.~{Huang}, L.~{Dai}, Y.~{Li}, M.~{Matthaiou}, G.~K.
  {Karagiannidis}, E.~{Bjornson}, K.~{Yang}, C.~{I}, and A.~{Ghosh},
  ``Millimeter wave communications for future mobile networks,'' \emph{IEEE J.
  Sel. Areas Commun.}, vol.~35, no.~9, pp. 1909--1935, Sep. 2017.

\bibitem{mmw3}
O.~{Semiari}, W.~{Saad}, M.~{Bennis}, and M.~{Debbah}, ``Integrated millimeter
  wave and sub-6 {GHz} wireless networks: A roadmap for joint mobile broadband
  and ultra-reliable low-latency communications,'' \emph{IEEE Wireless
  Commun.}, vol.~26, no.~2, pp. 109--115, Apr. 2019.

\bibitem{mmw5}
L.~{Zhang}, H.~{Zhao}, S.~{Hou}, Z.~{Zhao}, H.~{Xu}, X.~{Wu}, Q.~{Wu}, and
  R.~{Zhang}, ``A survey on {5G} millimeter wave communications for
  {UAV}-assisted wireless networks,'' \emph{IEEE Access}, vol.~7, pp.
  117\,460--117\,504, Jul. 2019.

\bibitem{mmw4}
Y.~{Niu}, Y.~{Li}, D.~{Jin}, L.~{Su}, and A.~V. {Vasilakos}, ``A survey of
  millimeter wave ({mmWave}) communications for {5G}: Opportunities and
  challenges,'' \emph{arXiv preprint, arXiv:1502.07228}, Feb. 2015.

\bibitem{mmw6}
X.~{Wang}, L.~{Kong}, F.~{Kong}, F.~{Qiu}, M.~{Xia}, S.~{Arnon}, and G.~{Chen},
  ``Millimeter wave communication: A comprehensive survey,'' \emph{IEEE Commun.
  Surveys Tuts.}, vol.~20, no.~3, pp. 1616--1653, Third quarter 2018.

\bibitem{6846320}
Y.~{Niu}, Y.~{Li}, D.~{Jin}, L.~{Su}, and D.~{Wu}, ``Blockage robust and
  efficient scheduling for directional {mmWave WPANs},'' \emph{IEEE Trans. Veh.
  Technol.}, vol.~64, no.~2, pp. 728--742, Feb. 2015.

\bibitem{7417432}
Y.~{Oguma}, R.~{Arai}, T.~{Nishio}, K.~{Yamamoto}, and M.~{Morikura},
  ``Proactive base station selection based on human blockage prediction using
  {RGB-D} cameras for {mmWave} communications,'' in \emph{Proc. IEEE Global
  Communications Conference (GLOBECOM)}, San Diego, CA, Dec. 2015, pp. 1--6.

\bibitem{5262296}
S.~{Singh}, F.~{Ziliotto}, U.~{Madhow}, E.~{Belding}, and M.~{Rodwell},
  ``Blockage and directivity in 60 {GHz} wireless personal area networks: from
  cross-layer model to multihop {MAC} design,'' \emph{IEEE J. Sel. Areas
  Commun.}, vol.~27, no.~8, pp. 1400--1413, Oct. 2009.

\bibitem{7959157}
D.~{Ramirez}, L.~{Huang}, Y.~{Wang}, and B.~{Aazhang}, ``On opportunistic
  {mmWave} networks with blockage,'' \emph{IEEE J. Sel. Areas Commun.},
  vol.~35, no.~9, pp. 2137--2147, Sep. 2017.

\bibitem{6134444}
Y.~M. {Tsang} and A.~S.~Y. {Poon}, ``Detecting human blockage and device
  movement in {mmWave} communication system,'' in \emph{Proc. IEEE Global
  Telecommunications Conference - GLOBECOM}, Houston, TX, Dec. 2011, pp. 1--6.

\bibitem{9050849}
F.~{Zhao}, W.~{Hao}, L.~{Shen}, G.~{Sun}, Y.~{Zhou}, and Y.~{Wang}, ``Secure
  energy efficiency transmission for {mmWave-NOMA} system,'' \emph{IEEE Systems
  J.}, pp. 1--4, Mar. 2020.

\bibitem{2020arXiv200104863Y}
Y.~{Yapici}, N.~{Rupasinghe}, I.~{Guvenc}, H.~{Dai}, and A.~{Bhuyan},
  ``Physical layer security for {NOMA} transmission in {mmWave} drone
  networks,'' \emph{arXiv preprint, arXiv:2001.04863}, Jan. 2020.

\bibitem{8454272}
J.~{Cui}, Z.~{Ding}, P.~{Fan}, and N.~{Al-Dhahir}, ``Unsupervised machine
  learning-based user clustering in {Millimeter-Wave-NOMA} systems,''
  \emph{IEEE Trans. Wireless Commun.}, vol.~17, no.~11, pp. 7425--7440, Nov.
  2018.

\bibitem{2020arXiv200207452M}
D.~{Marasinghe}, N.~{Jayaweera}, N.~{Rajatheva}, and M.~{Latva-aho},
  ``Hierarchical user clustering for {mmWave-NOMA} systems,'' \emph{arXiv
  preprint, arXiv:2002.07452}, Feb. 2020.

\bibitem{9042253}
B.~{Wang}, R.~{Shi}, F.~{Shi}, and J.~{Hu}, ``{mmWave-NOMA}-based low-latency
  and high-reliable communications for enhancement of {V2X} services,''
  \emph{IEEE Access}, vol.~8, pp. 57\,049--57\,062, Mar. 2020.

\bibitem{8753467}
J.~{Kaur} and M.~L. {Singh}, ``User assisted cooperative relaying in beamspace
  massive {MIMO NOMA} based systems for millimeter wave communications,''
  \emph{China Communications}, vol.~16, no.~6, pp. 103--113, Jun. 2019.

\bibitem{8485639}
L.~{Dai}, B.~{Wang}, M.~{Peng}, and S.~{Chen}, ``Hybrid precoding-based
  millimeter-wave massive {MIMO-NOMA} with simultaneous wireless information
  and power transfer,'' \emph{IEEE J. Sel. Areas Commun.}, vol.~37, no.~1, pp.
  131--141, Jan. 2019.

\bibitem{8326498}
T.~{Lv}, Y.~{Ma}, J.~{Zeng}, and P.~T. {Mathiopoulos}, ``Millimeter-wave {NOMA}
  transmission in cellular {M2M} communications for internet of things,''
  \emph{EEE Internet Things J.}, vol.~5, no.~3, pp. 1989--2000, Jun. 2018.

\bibitem{8603758}
M.~{Zeng}, W.~{Hao}, O.~A. {Dobre}, and H.~V. {Poor}, ``Energy-efficient power
  allocation in uplink {mmWave} massive {MIMO} with {NOMA},'' \emph{IEEE Trans.
  Veh. Technol.}, vol.~68, no.~3, pp. 3000--3004, 2019.

\bibitem{8844783}
W.~{Hao}, M.~{Zeng}, G.~{Sun}, O.~{Muta}, O.~A. {Dobre}, S.~{Yang}, and
  H.~{Gacanin}, ``Codebook-based {Max}–{Min} energy-efficient resource
  allocation for uplink {mmWave} {MIMO}-{NOMA} systems,'' \emph{IEEE Trans.
  Commun.}, vol.~67, no.~12, pp. 8303--8314, 2019.

\bibitem{8624593}
N.~{Iqbal}, J.~{Luo}, R.~{Muller}, G.~{Steinbock}, C.~{Schneider}, D.~A.
  {Dupleich}, S.~{Hafner}, and R.~S. {Thoma}, ``Multipath cluster fading
  statistics and modeling in millimeter-wave radio channels,'' \emph{IEEE
  Trans. Antennas Propag.}, vol.~67, no.~4, pp. 2622--2632, Apr. 2019.

\bibitem{8207426}
I.~A. {Hemadeh}, K.~{Satyanarayana}, M.~{El-Hajjar}, and L.~{Hanzo},
  ``Millimeter-wave communications: Physical channel models, design
  considerations, antenna constructions, and link-budget,'' \emph{IEEE Commun.
  Surveys Tuts.}, vol.~20, no.~2, pp. 870--913, Second quarter 2018.

\bibitem{7400962}
S.~{Hur}, S.~{Baek}, B.~{Kim}, Y.~{Chang}, A.~F. {Molisch}, T.~S. {Rappaport},
  K.~{Haneda}, and J.~{Park}, ``Proposal on millimeter-wave channel modeling
  for {5G} cellular system,'' \emph{IEEE J. Sel. Topics Signal Process.},
  vol.~10, no.~3, pp. 454--469, Apr. 2016.

\bibitem{7501500}
M.~K. {Samimi} and T.~S. {Rappaport}, ``{3-D} millimeter-wave statistical
  channel model for {5G} wireless system design,'' \emph{IEEE Trans. Microw.
  Theory Techn.}, vol.~64, no.~7, pp. 2207--2225, Jul. 2016.

\bibitem{7109864}
T.~S. {Rappaport}, G.~R. {MacCartney}, M.~K. {Samimi}, and S.~{Sun}, ``Wideband
  millimeter-wave propagation measurements and channel models for future
  wireless communication system design,'' \emph{IEEE Trans. Commun.}, vol.~63,
  no.~9, pp. 3029--3056, Sep. 2015.

\bibitem{8884199}
L.~N. {Ribeiro}, S.~{Schwarz}, and A.~L.~F. {De Almeida}, ``Double-sided
  massive {MIMO} transceivers for {mmWave} communications,'' \emph{IEEE
  Access}, vol.~7, pp. 157\,667--157\,679, Oct. 2019.

\bibitem{8789649}
C.~G. {Tsinos}, S.~{Chatzinotas}, and B.~{Ottersten}, ``Hybrid analog-digital
  transceiver designs for multi-user {MIMO mmWave} cognitive radio systems,''
  \emph{IEEE Trans. Cogn. Commun. Netw.}, vol.~6, no.~1, pp. 310--324, Mar.
  2020.

\bibitem{8115137}
K.~{Satyanarayana}, M.~{El-Hajjar}, P.~{Kuo}, A.~A.~M. {Mourad}, and
  L.~{Hanzo}, ``Adaptive transceiver design for {C-RAN in mmWave}
  communications,'' \emph{IEEE Access}, vol.~6, pp. 16\,770--16\,782, Nov.
  2018.

\bibitem{7556971}
P.~{Xia}, R.~W. {Heath}, and N.~{Gonzalez-Prelcic}, ``Robust analog precoding
  designs for millimeter wave {MIMO} transceivers with frequency and time
  division duplexing,'' \emph{IEEE Trans. Commun.}, vol.~64, no.~11, pp.
  4622--4634, Nov. 2016.

\bibitem{8713847}
K.~{Satyanarayana}, M.~{El-Hajjar}, A.~A.~M. {Mourad}, and L.~{Hanzo},
  ``Multi-user full duplex transceiver design for {mmWave} systems using
  learning-aided channel prediction,'' \emph{IEEE Access}, vol.~7, pp.
  66\,068--66\,083, May 2019.

\bibitem{8731759}
H.~{Chen}, Y.~{Shao}, Y.~{Zhang}, C.~{Zhang}, and Z.~{Zhang}, ``A low-profile
  broadband circularly polarized {mmWave} antenna with special-shaped ring
  slot,'' \emph{IEEE Antennas Wireless Propag. Lett.}, vol.~18, no.~7, pp.
  1492--1496, Jul. 2019.

\bibitem{8685138}
K.~{Gulur Sadananda}, M.~P. {Abegaonkar}, and S.~K. {Koul}, ``Gain equalized
  shared-aperture antenna using dual-polarized {ZIM} for {mmWave 5G} base
  stations,'' \emph{IEEE Antennas Wireless Propag. Lett.}, vol.~18, no.~6, pp.
  1100--1104, Jun. 2019.

\bibitem{8758798}
C.~{Shu}, J.~{Wang}, S.~{Hu}, Y.~{Yao}, J.~{Yu}, Y.~{Alfadhl}, and X.~{Chen},
  ``A wideband dual-circular-polarization horn antenna for {mmWave} wireless
  communications,'' \emph{IEEE Antennas Wireless Propag. Lett.}, vol.~18,
  no.~9, pp. 1726--1730, Sep. 2019.

\bibitem{8316889}
S.~{Zhu}, H.~{Liu}, Z.~{Chen}, and P.~{Wen}, ``A compact gain-enhanced vivaldi
  antenna array with suppressed mutual coupling for {5G mmWave} application,''
  \emph{IEEE Antennas Wireless Propag. Lett.}, vol.~17, no.~5, pp. 776--779,
  May 2018.

\bibitem{8400592}
Q.~{Wu}, J.~{Hirokawa}, J.~{Yin}, C.~{Yu}, H.~{Wang}, and W.~{Hong},
  ``Millimeter-wave multibeam endfire dual-circularly polarized antenna array
  for {5G} wireless applications,'' \emph{IEEE Trans. Antennas Propag.},
  vol.~66, no.~9, pp. 4930--4935, Sep. 2018.

\bibitem{7116515}
H.~{Kaouach} and A.~{Kabashi}, ``Simple tri-layer linearly polarized discrete
  lens antenna with high-efficiency for {mmWave} applications,'' \emph{IEEE
  Antennas Wireless Propag. Lett.}, vol.~15, pp. 259--262, Jun. 2016.

\bibitem{gunes4}
K.~Tekbiyik, A.~Ekti, G.~Kurt, and A.~Gorcin, ``Terahertz band communication
  systems: {C}hallenges, novelties and standardization efforts,''
  \emph{Physical Commun.}, vol.~35, pp. 1--18, Aug. 2019.

\bibitem{TH1}
T.~L. Wen \emph{et~al.}, ``Enhanced optical modulation depth of {T}erahertz
  waves by self-assembled monolayer of plasmonic gold nanoparticles,''
  \emph{Adv. Opt. Mater.}, vol.~4, no.~12, pp. 1974--1980, Sep. 2016.

\bibitem{TH2}
W.~E. Lai, P.~Huang, B.~Pelaz, P.~D. Pino, and Q.~Zhang, ``Enhanced all-optical
  modulation of {T}erahertz waves on the basis of manganese ferrite
  nanoparticles,'' \emph{J. Phys. Chem. C.}, vol. 121, no.~39, pp.
  21\,634--21\,640, Sep. 2017.

\bibitem{TH3}
T.~A.~P. Tran and P.~H. {Bolivar}, ``Terahertz modulator based on vertically
  coupled {F}ano metamaterial,'' \emph{IEEE Trans. THz Sci. Technol.}, vol.~8,
  no.~5, pp. 502--508, Sep. 2018.

\bibitem{TH4}
S.~Y. et~al., ``Optically driven {T}erahertz wave modulator using ring-shaped
  microstripline with {GaInAs} photoconductive mesa structure,'' \emph{IEEE J.
  Sel. Topics Quantum Electron}, vol.~23, no.~4, pp. 1--8, Jul. 2017.

\bibitem{8409966}
L.~{Xiong}, B.~{Zhang}, H.~{Ji}, W.~{Wang}, X.~{Liu}, S.~{He}, and J.~{Shen},
  ``Active optically controlled broadband {T}erahertz modulator based on
  {Fe3O4Nanoparticles},'' \emph{IEEE Trans. THz Sci. Technol.}, vol.~8, no.~5,
  pp. 535--540, Sep. 2018.

\bibitem{8306983}
L.~{Jiu-Sheng}, L.~{Shao-he}, and Z.~{Le}, ``Terahertz modulator using
  {4-N,N-Dimethylamino-4'-N'-Methyl-Stilbazolium Tosylate (DAST)/Si} hybrid
  structure,'' \emph{IEEE Photon. J.}, vol.~10, no.~2, pp. 1--6, Apr. 2018.

\bibitem{8691541}
C.~{Yi}, S.~H. {Choi}, M.~{Urteaga}, and M.~{Kim}, ``20-{Gb/s ON-OFF-Keying
  Modulators} using 0.25- $\mu$ m {InP DHBT} switches at 290 {GHz},''
  \emph{IEEE Microw. Wireless Compon. Lett.}, vol.~29, no.~5, pp. 360--362, May
  2019.

\bibitem{8616810}
G.~{Isic}, G.~{Sinatkas}, D.~C. {Zografopoulos}, B.~{Vasic}, A.~{Ferraro},
  R.~{Beccherelli}, E.~E. {Kriezis}, and M.~{Belic}, ``Electrically tunable
  metal-semiconductor-metal {T}erahertz metasurface modulators,'' \emph{IEEE J.
  Sel. Topics Quantum Electron}, vol.~25, no.~3, pp. 1--8, May 2019.

\bibitem{GR1}
Y.~Wu \emph{et~al.}, ``Tunable {T}erahertz reflection of graphene via ionic
  liquid gating,'' \emph{Nanotechnology}, vol.~28, no.~9, pp. 1--5, Jan. 2017.

\bibitem{GR2}
O.~Ozdemir, A.~M. Aygar, O.~Balci, C.~Kocabas, H.~Caglayan, and E.~Ozbay,
  ``Enhanced tunability of {V}-shaped plasmonic structures using ionic liquid
  gating and graphene,'' \emph{Carbon}, vol. 108, pp. 515--520, Nov. 2016.

\bibitem{6005345}
H.~{Song} and T.~{Nagatsuma}, ``Present and future of {T}erahertz
  communications,'' \emph{IEEE Trans. THz Sci. Technol.}, vol.~1, no.~1, pp.
  256--263, Sep. 2011.

\bibitem{7967651}
Q.~{Yu}, J.~{Gu}, Q.~{Yang}, Y.~{Zhang}, Y.~{Li}, Z.~{Tian}, C.~{Ouyang},
  J.~{Han}, J.~F. {O'Hara}, and W.~{Zhang}, ``All-dielectric meta-lens designed
  for photoconductive {T}erahertz antennas,'' \emph{IEEE Photon. J.}, vol.~9,
  no.~4, pp. 1--9, Aug. 2017.

\bibitem{8737689}
Z.~{Shen}, X.~{Ji}, Y.~{Liao}, K.~{Wang}, B.~{Jin}, and F.~{Yan}, ``Resonant
  polysilicon antenna for {T}erahertz detection,'' \emph{IEEE Photon. J.},
  vol.~11, no.~4, pp. 1--8, Aug. 2019.

\bibitem{7859330}
Z.~{Hao}, J.~{Wang}, Q.~{Yuan}, and W.~{Hong}, ``Development of a low-cost
  {THz} metallic lens antenna,'' \emph{IEEE Antennas Wireless Propag. Lett.},
  vol.~16, pp. 1751--1754, Feb. 2017.

\bibitem{7930523}
K.~{Fan}, Z.~{Hao}, Q.~{Yuan}, and W.~{Hong}, ``Development of a high gain
  325-500 {GHz} antenna using quasi-planar reflectors,'' \emph{IEEE Trans.
  Antennas Propag.}, vol.~65, no.~7, pp. 3384--3391, Jul. 2017.

\bibitem{8345575}
S.~{Bhardwaj} and J.~L. {Volakis}, ``Hexagonal waveguide based circularly
  polarized horn antennas for sub-mm-{W}ave/{T}erahertz band,'' \emph{IEEE
  Trans. Antennas Propag.}, vol.~66, no.~7, pp. 3366--3374, Jul. 2018.

\bibitem{8601401}
G.~{Rana}, A.~{Bhattacharya}, A.~{Gupta}, D.~{Ghindani}, R.~{Jain}, S.~P.
  {Duttagupta}, and S.~S. {Prabhu}, ``A polarization-resolved study of
  nanopatterned photoconductive antenna for enhanced {T}erahertz emission,''
  \emph{IEEE Trans. THz Sci. Technol.}, vol.~9, no.~2, pp. 193--199, Mar. 2019.

\bibitem{7956223}
G.~P. {Szakmany}, A.~O. {Orlov}, G.~H. {Bernstein}, and W.~{Porod}, ``{THz}
  wave detection by antenna-coupled nanoscale thermoelectric converters,''
  \emph{IEEE Trans. THz Sci. Technol.}, vol.~7, no.~5, pp. 582--585, Sep. 2017.

\bibitem{8239849}
N.~M. {Burford}, M.~J. {Evans}, and M.~O. {El-Shenawee}, ``Plasmonic nanodisk
  thin-film {T}erahertz photoconductive antenna,'' \emph{IEEE Trans. THz Sci.
  Technol.}, vol.~8, no.~2, pp. 237--247, Mar. 2018.

\bibitem{8340871}
J.~{Al-Eryani}, H.~{Knapp}, J.~{Kammerer}, K.~{Aufinger}, H.~{Li}, and
  L.~{Maurer}, ``Fully integrated single-chip 305-375-{GHz} transceiver with
  on-chip antennas in {SiGe BiCMOS},'' \emph{IEEE Trans. THz Sci. Technol.},
  vol.~8, no.~3, pp. 329--339, May 2018.

\bibitem{7565507}
C.~{Jiang}, A.~{Mostajeran}, R.~{Han}, M.~{Emadi}, H.~{Sherry}, A.~{Cathelin},
  and E.~{Afshari}, ``A fully integrated 320 {GHz} coherent imaging transceiver
  in 130 nm {SiGe BiCMOS},'' \emph{IEEE J. Solid-State Circuits}, vol.~51,
  no.~11, pp. 2596--2609, Nov. 2016.

\bibitem{6709822}
B.~{Blázquez}, K.~B. {Cooper}, and N.~{Llombart}, ``Time-delay multiplexing
  with linear arrays of {THz} radar transceivers,'' \emph{IEEE Trans. THz Sci.
  Technol.}, vol.~4, no.~2, pp. 232--239, Mar. 2014.

\bibitem{6594881}
T.~{Bryllert}, V.~{Drakinskiy}, K.~B. {Cooper}, and J.~{Stake}, ``Integrated
  200-240-{GHz} {FMCW} radar transceiver module,'' \emph{IEEE Trans. Microw.
  Theory Techn.}, vol.~61, no.~10, pp. 3808--3815, Oct. 2013.

\bibitem{5325122}
T.~{Gobel}, D.~{Schoenherr}, C.~{Sydlo}, M.~{Feiginov}, P.~{Meissner}, and
  H.~L. {Hartnagel}, ``Continuous-wave single-sampling-point characterisation
  of optoelectronic on-chip {T}erahertz transceiver,'' \emph{Electron. Lett.},
  vol.~45, no.~23, pp. 1171--1172, Nov. 2009.

\bibitem{7175082}
S.~{Moghadami}, F.~{Hajilou}, P.~{Agrawal}, and S.~{Ardalan}, ``A 210 {GHz}
  fully-integrated {OOK} transceiver for short-range wireless chip-to-chip
  communication in 40 nm {CMOS} technology,'' \emph{IEEE Trans. THz Sci.
  Technol.}, vol.~5, no.~5, pp. 737--741, Sep. 2015.

\bibitem{7036065}
C.~{Lin} and G.~Y. {Li}, ``Indoor {T}erahertz communications: {H}ow many
  antenna arrays are needed?'' \emph{IEEE Trans. Wireless Commun.}, vol.~14,
  no.~6, pp. 3097--3107, Jun. 2015.

\bibitem{THC3}
C.~Han, A.~O. Bicen, and I.~F. Akyildiz, ``Multiray channel modeling and
  wideband characterization for wireless communications in the {T}erahertz
  band,'' \emph{IEEE Trans. Wireless Commun.}, vol.~14, no.~5, pp. 2402--2412,
  May 2015.

\bibitem{THC4}
D.~He \emph{et~al.}, ``Channel modeling for kiosk downloading communication
  system at 300 {GHz},'' in \emph{Proc. European Conf. on Antennas and
  Propagation (EUCAP)}, Paris, Mar. 2017, pp. 1331--1335.

\bibitem{THC5}
C.~Han and I.~F. Akyildiz, ``Three-dimensional end-to-end modeling and analysis
  for graphene-enabled {T}erahertz band communications,'' \emph{IEEE Trans.
  Veh. Technol.}, vol.~66, no.~7, pp. 5626-- 5634, Jul. 2017.

\bibitem{THC6}
C.~Zhang, C.~Han, and I.~F. Akyildiz, ``Three dimensional end-to-end modeling
  and directivity analysis for graphene-based antennas in the {T}erahertz
  band,'' in \emph{Proc. IEEE Global Commun. Conf. (GLOBECOM)}, San Diego, CA,
  Dec. 2015, pp. 1--6.

\bibitem{THC7}
D.~He \emph{et~al.}, ``Stochastic channel modeling for kiosk applications in
  the {T}erahertz band,'' \emph{IEEE Trans. THz Sci. Technol.}, vol.~7, no.~5,
  pp. 502-- 513, Sep. 2017.

\bibitem{THC8}
S.~Kim and A.~Zaji, ``Statistical modeling and simulation of short-range
  device-to-device communication channels at sub-{THz} frequencies,''
  \emph{IEEE Trans. Wireless Commun.}, vol.~15, no.~9, pp. 6423--6433, Sep.
  2016.

\bibitem{THC9}
K.~{Tekbiyik}, E.~{Ulusoy}, A.~R. {Ekti}, S.~{Yarkan}, T.~{Baykas},
  A.~{Gorcin}, and G.~K. {Kurt}, ``Statistical channel modeling for short range
  line–of–sight {T}erahertz communication,'' in \emph{Proc. Int. Symp. on
  Personal, Indoor and Mobile Radio Communications (PIMRC)}, Sep. 2019, pp.
  1--5.

\bibitem{THC1}
R.~Piesiewicz, T.~Kleine-Ostmann, N.~Krumbholz, D.~Mittleman, M.~Koch, and
  T.~Krner, ``Terahertz characterisation of building materials,''
  \emph{Electronics Letters}, vol.~41, no.~18, pp. 1002--1004, Sep. 2005.

\bibitem{THC2}
C.~Jansen, R.~Piesiewicz, D.~Mittleman, T.~Krner, and M.~Koch, ``The impact of
  reflections from stratified building materials on the wave propagation in
  future indoor {T}erahertz communication systems,'' \emph{IEEE Trans. Antennas
  Propag.}, vol.~56, no.~5, pp. 1413--1419, May 2008.

\bibitem{8651537}
H.~{Zhao}, L.~{Wei}, M.~{Jarrahi}, and G.~J. {Pottie}, ``Extending spatial and
  temporal characterization of indoor wireless channels from 350 to 650
  {GHz},'' \emph{IEEE Trans. THz Sci. Technol.}, vol.~9, no.~3, pp. 243--252,
  May 2019.

\bibitem{8684885}
K.~{Guan}, B.~{Peng}, D.~{He}, J.~M. {Eckhardt}, S.~{Rey}, B.~{Ai}, Z.~{Zhong},
  and T.~{Kurner}, ``Measurement, simulation, and characterization of
  train-to-infrastructure inside-station channel at the {T}erahertz band,''
  \emph{IEEE Trans. THz Sci. Technol.}, vol.~9, no.~3, pp. 291--306, May 2019.

\bibitem{7582545}
X.~{Gao}, L.~{Dai}, Y.~{Zhang}, T.~{Xie}, X.~{Dai}, and Z.~{Wang}, ``Fast
  channel tracking for {T}erahertz beamspace massive {MIMO} systems,''
  \emph{IEEE Trans. Veh. Technol.}, vol.~66, no.~7, pp. 5689--5696, Jul. 2017.

\bibitem{8123513}
K.~{Tsujimura}, K.~{Umebayashi}, J.~{Kokkoniemi}, J.~{Lehtomaki}, and
  Y.~{Suzuki}, ``A causal channel model for the {T}erahertz band,'' \emph{IEEE
  Trans. THz Sci. Technol.}, vol.~8, no.~1, pp. 52--62, Jan. 2018.

\bibitem{7539586}
B.~{Peng} and T.~{Kurner}, ``Three-dimensional angle of arrival estimation in
  dynamic indoor {T}erahertz channels using a forward-backward algorithm,''
  \emph{IEEE Trans. Veh. Technol.}, vol.~66, no.~5, pp. 3798--3811, May 2017.

\bibitem{8093757}
J.~{Park}, S.~{Choi}, J.~J. {Kim}, Y.~{Kim}, M.~{Lee}, H.~{Kim}, B.~{Bae},
  H.~{Song}, K.~{Cho}, S.~{Lee}, H.~{Lee}, and J.~{Kim}, ``A novel stochastic
  model-based eye-diagram estimation method for {8B/10B and TMDS}-encoded
  high-speed channels,'' \emph{IEEE Trans. Electromagn. Compat.}, vol.~60,
  no.~5, pp. 1510--1519, Oct. 2018.

\bibitem{murat}
M.~{Uysal} and H.~{Nouri}, ``Optical wireless communications — an emerging
  technology,'' in \emph{International Conference on Transparent Optical
  Networks (ICTON)}, Graz, Jul. 2014, pp. 1--7.

\bibitem{Chowdhury_2019}
\BIBentryALTinterwordspacing
M.~Z. Chowdhury, M.~Shahjalal, M.~K. Hasan, and Y.~M. Jang, ``The role of
  optical wireless communication technologies in {5G/6G} and {IoT} solutions:
  Prospects, directions, and challenges,'' \emph{Appl. Sci.}, vol.~9, no.~20,
  p. 4367, Oct. 2019. [Online]. Available:
  \url{http://dx.doi.org/10.3390/app9204367}
\BIBentrySTDinterwordspacing

\bibitem{7239528}
P.~H. {Pathak}, X.~{Feng}, P.~{Hu}, and P.~{Mohapatra}, ``Visible light
  communication, networking, and sensing: A survey, potential and challenges,''
  \emph{IEEE Commun. Surveys Tuts.}, vol.~17, no.~4, pp. 2047--2077, Fourth
  quarter 2015.

\bibitem{8970387}
A.~R. {Ndjiongue}, T.~M.~N. {Ngatched}, O.~A. {Dobre}, and A.~G. {Armada},
  ``{VLC}-based networking: Feasibility and challenges,'' \emph{IEEE Netw.},
  pp. 1--8, 2020.

\bibitem{vlc2}
P.~S. S.~Kumar, ``A comprehensive survey of visible light communication:
  Potential and challenges,'' \emph{Wireless Pers. Commun.}, vol. 109, p.
  1357–1375, Nov. 2019.

\bibitem{Haas}
H.~{Haas}, L.~{Yin}, C.~{Chen}, S.~{Videv}, D.~{Parol}, E.~{Poves},
  H.~{Alshaer}, and M.~S. {Islim}, ``Introduction to indoor networking concepts
  and challenges in {LiFi},'' \emph{IEEE/OSA J. Opt. Commun. Netw.}, vol.~12,
  no.~2, pp. A190--A203, Feb. 2020.

\bibitem{Haas2}
H.~{Haas}, L.~{Yin}, Y.~{Wang}, and C.~{Chen}, ``What is {LiFi}?'' \emph{J.
  Lightw. Technol.}, vol.~34, no.~6, pp. 1533--1544, Mar. 2016.

\bibitem{OCC}
N.~{Saha}, M.~S. {Ifthekhar}, N.~T. {Le}, and Y.~M. {Jang}, ``Survey on optical
  camera communications: challenges and opportunities,'' \emph{IET
  Optoelectron.}, vol.~9, no.~5, pp. 172--183, Oct. 2015.

\bibitem{7890427}
T.~{Nguyen}, A.~{Islam}, T.~{Hossan}, and Y.~M. {Jang}, ``Current status and
  performance analysis of optical camera communication technologies for {5G}
  networks,'' \emph{IEEE Access}, vol.~5, pp. 4574--4594, Mar. 2017.

\bibitem{8259465}
M.~Z. {Chowdhury}, M.~T. {Hossan}, A.~{Islam}, and Y.~M. {Jang}, ``A
  comparative survey of optical wireless technologies: Architectures and
  applications,'' \emph{IEEE Access}, vol.~6, pp. 9819--9840, Jan. 2018.

\bibitem{5771213}
F.~{Demers}, H.~{Yanikomeroglu}, and M.~{St-Hilaire}, ``A survey of
  opportunities for free space optics in next generation cellular networks,''
  in \emph{Proc. Ninth Annual Communication Networks and Services Research
  Conference}, 2011, pp. 210--216.

\bibitem{6844864}
M.~A. {Khalighi} and M.~{Uysal}, ``Survey on free space optical communication:
  A communication theory perspective,'' \emph{IEEE Commun. Surveys Tuts.},
  vol.~16, no.~4, pp. 2231--2258, Fourth quarter 2014.

\bibitem{533654}
A.~C. {Boucouvalas}, ``Ambient light noise and its effect on receiver design
  for indoor wireless optical links,'' in \emph{Proc. of ICC/SUPERCOMM -
  International Conference on Communications}, vol.~3, Dallas, TX, Jun. 1996,
  pp. 1472--1476.

\bibitem{581112}
------, ``Indoor ambient light noise and its effect on wireless optical
  links,'' \emph{IEE Proc. - Optoelectron.}, vol. 143, no.~6, pp. 334--338,
  Dec. 1996.

\bibitem{6779317}
S.~{Verma}, A.~{Shandilya}, and A.~{Singh}, ``A model for reducing the effect
  of ambient light source in {VLC} system,'' in \emph{Proc. IEEE International
  Advance Computing Conference (IACC)}, Gurgaon, Feb. 2014, pp. 186--188.

\bibitem{8309378}
X.~{Li}, B.~{Hussain}, L.~{Wang}, J.~{Jiang}, and C.~P. {Yue}, ``Design of a
  {2.2-mW 24-Mb/s CMOS VLC Receiver SoC} with ambient light rejection and
  post-equalization for {Li-Fi} applications,'' \emph{J. Lightw. Technol.},
  vol.~36, no.~12, pp. 2366--2375, Jun. 2018.

\bibitem{7903587}
C.~{Lin}, P.~{Chen}, C.~{Chang}, J.~{Yu}, C.~{Chang}, and Y.~{Tseng}, ``A
  hydrogenated amorphous silicon thin-film transistor optical pixel sensor for
  ameliorating influences of ambient light and reflected light,'' \emph{IEEE J.
  Electron Devices Soc.}, vol.~5, no.~4, pp. 262--265, Jul. 2017.

\bibitem{a11}
S.~Lee, ``Reducing the effects of ambient noise light in an indoor optical
  wireless system using polarizers,'' \emph{Microwave Opt. Technol. Lett.},
  vol.~40, no.~3, pp. 228--231, Dec. 2003.

\bibitem{7804058}
T.~{Adiono}, A.~{Pradana}, R.~V.~W. {Putra}, and S.~{Fuada}, ``Analog filters
  design in {VLC} analog front-end receiver for reducing indoor ambient light
  noise,'' in \emph{Proc. IEEE Asia Pacific Conference on Circuits and Systems
  (APCCAS)}, Jeju, Oct. 2016, pp. 581--584.

\bibitem{pat1}
``Ambient light interference reduction for optical input devices,'' Patent
  US20\,100\,001\,978A1.

\bibitem{a12}
S.~Arnon, ``Effects of atmospheric turbulence and building sway on optical
  wireless-communication systems,'' \emph{Opt. Lett.}, vol.~28, no.~2, pp.
  129--131, Dec. 2003.

\bibitem{6313871}
X.~{Song} and J.~{Cheng}, ``Optical communication using subcarrier intensity
  modulation in strong atmospheric turbulence,'' \emph{J. Lightw. Technol.},
  vol.~30, no.~22, pp. 3484--3493, Nov. 2012.

\bibitem{6608639}
------, ``Subcarrier intensity modulated {MIMO} optical communications in
  atmospheric turbulence,'' \emph{IEEE/OSA J. Opt. Commun. Netw.}, vol.~5,
  no.~9, pp. 1001--1009, Sep. 2013.

\bibitem{8732427}
A.~{Jaiswal}, M.~R. {Bhatnagar}, P.~{Soni}, and V.~K. {Jain}, ``Differential
  optical spatial modulation over atmospheric turbulence,'' \emph{IEEE J. Sel.
  Topics Signal Process.}, vol.~13, no.~6, pp. 1417--1432, Oct. 2019.

\bibitem{8703708}
S.~{Arya} and Y.~H. {Chung}, ``A novel blind spectrum sensing technique for
  multi-user ultraviolet communications in atmospheric turbulence channel,''
  \emph{IEEE Access}, vol.~7, pp. 58\,314--58\,323, May 2019.

\bibitem{7031362}
M.~{Aggarwal}, P.~{Garg}, and P.~{Puri}, ``Exact capacity of
  amplify-and-forward relayed optical wireless communication systems,''
  \emph{IEEE Photon. Technol. Lett.}, vol.~27, no.~8, pp. 903--906, Apr. 2015.

\bibitem{1215849}
M.~{Gebhart}, E.~{Leitgeb}, and J.~{Bregenzer}, ``Atmospheric effects on
  optical wireless links,'' in \emph{Proc. International Conference on
  Telecommunications (ConTEL)}, vol.~2, Zagreb, Jun. 2003, pp. 395--401.

\bibitem{ATA2019108}
Y.~Ata, Y.~Baykal, and M.~C. Gokce, ``Error performance of optical wireless
  communication systems exercising {BPSK} subcarrier intensity modulation in
  non-kolmogorov turbulent atmosphere,'' \emph{Opt. Commun.}, vol. 436, pp. 108
  -- 112, Apr. 2019.

\bibitem{LATAL2019184}
J.~Latal, J.~Vitasek, L.~Hajek, A.~Vanderka, R.~Martinek, and V.~Vasinek,
  ``Influence of simulated atmospheric effect combined with modulation formats
  on {FSO} systems,'' \emph{Optical Switching and Networking}, vol.~33, pp. 184
  -- 193, Jul. 2019.

\bibitem{BAYKAL201729}
Y.~Baykal, ``{BER} of asymmetrical optical beams in oceanic and marine
  atmospheric media,'' \emph{Optics Commun.}, vol. 393, pp. 29 -- 33, Jun.
  2017.

\bibitem{1299334}
D.~{Kedar} and S.~{Arnon}, ``Urban optical wireless communication networks: the
  main challenges and possible solutions,'' \emph{IEEE Commun. Mag.}, vol.~42,
  no.~5, pp. S2--S7, May 2004.

\bibitem{5062297}
H.~G. {Sandalidis}, T.~A. {Tsiftsis}, and G.~K. {Karagiannidis}, ``Optical
  wireless communications with heterodyne detection over turbulence channels
  with pointing errors,'' \emph{J. Lightw. Technol.}, vol.~27, no.~20, pp.
  4440--4445, Oct. 2009.

\bibitem{8754864}
M.~{Smilic}, Z.~{Nikolic}, D.~{Milic}, P.~{Spalevic}, and S.~{Panic},
  ``Comparison of adaptive algorithms for free space optical transmission in
  {Malaga} atmospheric turbulence channel with pointing errors,'' \emph{IET
  Commun.}, vol.~13, no.~11, pp. 1578--1585, Jul. 2019.

\bibitem{8445766}
S.~E. {Trevlakis}, A.~A. {Boulogeorgos}, and G.~K. {Karagiannidis}, ``Outage
  performance of transdermal optical wireless links in the presence of pointing
  errors,'' in \emph{Proc. IEEE 19th International Workshop on Signal
  Processing Advances in Wireless Communications (SPAWC)}, Kalamata, Jun. 2018,
  pp. 1--5.

\bibitem{7882670}
A.~{Jaiswal}, M.~R. {Bhatnagar}, and V.~K. {Jain}, ``Performance of optical
  space shift keying over {Gamma}-{Gamma} fading with pointing error,''
  \emph{IEEE Photon. J.}, vol.~9, no.~2, pp. 1--16, Apr. 2017.

\bibitem{7192727}
I.~S. {Ansari}, F.~{Yilmaz}, and M.~{Alouini}, ``Performance analysis of
  free-space optical links over {Malaga} ($\mathcal{M} $) turbulence channels
  with pointing errors,'' \emph{IEEE Trans. Wireless Commun.}, vol.~15, no.~1,
  pp. 91--102, Jan. 2016.

\bibitem{6932439}
L.~{Yang}, X.~{Gao}, and M.~{Alouini}, ``Performance analysis of relay-assisted
  all-optical {FSO} networks over strong atmospheric turbulence channels with
  pointing errors,'' \emph{J. Lightw. Technol.}, vol.~32, no.~23, pp.
  4613--4620, Dec. 2014.

\bibitem{7076656}
I.~S. {Ansari}, M.~{Alouini}, and J.~{Cheng}, ``Ergodic capacity analysis of
  free-space optical links with nonzero boresight pointing errors,'' \emph{IEEE
  Tran. Wireless Commun.}, vol.~14, no.~8, pp. 4248--4264, Aug. 2015.

\bibitem{MSs3}
M.~Di~Renzo, M.~Debbah, D.-T. Phan-Huy, A.~Zappone, M.-S. Alouini, C.~Yuen,
  V.~Sciancalepore, G.~C. Alexandropoulos, J.~Hoydis, H.~Gacanin \emph{et~al.},
  ``Smart radio environments empowered by reconfigurable {AI} meta-surfaces:
  {A}n idea whose time has come,'' \emph{EURASIP J Wirel Commun Netw}, vol.
  2019, no.~1, pp. 1--20, May 2019.

\bibitem{8466374}
C.~{Liaskos}, S.~{Nie}, A.~{Tsioliaridou}, A.~{Pitsillides}, S.~{Ioannidis},
  and I.~{Akyildiz}, ``A new wireless communication paradigm through
  software-controlled metasurfaces,'' \emph{IEEE Commun. Mag.}, vol.~56, no.~9,
  pp. 162--169, Sep. 2018.

\bibitem{8449754}
\emph{Realizing Wireless Communication Through Software-Defined HyperSurface
  Environments}, Chania, Jun. 2018.

\bibitem{MS}
C.~Liaskos, S.~Nie, A.~Tsioliaridou, A.~Pitsillides, S.~Ioannidis, and
  I.~Akyildiz, ``A novel communication paradigm for high capacity and security
  via programmable indoor wireless environments in next generation wireless
  systems,'' \emph{Ad Hoc Netw.}, vol.~87, pp. 1--16, May 2019.

\bibitem{9007666}
S.~{Abadal}, T.~{Cui}, T.~{Low}, and J.~{Georgiou}, ``Programmable
  metamaterials for software-defined electromagnetic control: Circuits,
  systems, and architectures,'' \emph{IEEE Trans. Emerg. Sel. Topics Circuits
  Syst.}, vol.~10, no.~1, pp. 6--19, Feb. 2020.

\bibitem{mohjazi2020outlook}
L.~Mohjazi, A.~Zoha, L.~Bariah, S.~Muhaidat, P.~C. Sofotasios, M.~A. Imran, and
  O.~A. Dobre, ``An outlook on the interplay of {AI} and software-defined
  metasurfaces,'' \emph{arXiv preprint, arXiv:2004.00365}, 2020.

\bibitem{huang2019indoor}
C.~Huang, G.~C. Alexandropoulos, C.~Yuen, and M.~Debbah, ``Indoor signal
  focusing improvement via deep learning configured intelligent metasurfaces,''
  in \emph{Proc. IEEE Int. Workshop. Signal Process. Advances in Wireless
  Commun. Cannes, France}, Jul. 2019.

\bibitem{liaskos2019interpretable}
C.~Liaskos, A.~Tsioliaridou, S.~Nie, A.~Pitsillides, S.~Ioannidis, and
  I.~Akyildiz, ``An interpretable neural network for configuring programmable
  wireless environments,'' in \emph{Proc. IEEE 20th International Workshop on
  Signal Processing Advances in Wireless Communications (SPAWC)}.\hskip 1em
  plus 0.5em minus 0.4em\relax IEEE, Jul. 2019, pp. 1--5.

\bibitem{8988246}
T.~{Shan}, X.~{Pan}, M.~{Li}, S.~{Xu}, and F.~{Yang}, ``Coding programmable
  metasurfaces based on deep learning techniques,'' \emph{IEEE J. Emerging and
  Selected Topics in Circuits and Systems}, vol.~10, no.~1, pp. 114--125, Mar.
  2020.

\bibitem{li2019machine}
L.~Li, H.~Ruan, C.~Liu, Y.~Li, Y.~Shuang, A.~Al{\`u}, C.-W. Qiu, and T.~J. Cui,
  ``Machine-learning reprogrammable metasurface imager,'' \emph{Nat. commun.},
  vol.~10, no.~1, pp. 1--8, Mar. 2019.

\bibitem{zhao}
J.~Zhao, X.~Yang, J.~Y. Dai, Q.~Cheng, X.~Li, N.~H. Qi, J.~C. Ke, G.~D. Bai,
  S.~Liu, S.~Jin, A.~Alu, and T.~J. Cui, ``Programmable time-domain
  digital-coding metasurface for non-linear harmonic manipulation and new
  wireless communication systems,'' \emph{National Science Review}, vol.~6,
  no.~2, pp. 231--238, Nov. 2019.

\bibitem{QPSK}
W.~{Tang}, X.~{Li}, J.~Y. {Dai}, S.~{Jin}, Y.~{Zeng}, Q.~{Cheng}, and T.~J.
  {Cui}, ``Wireless communications with programmable metasurface: {T}ransceiver
  design and experimental results,'' \emph{China Commun.}, vol.~16, no.~5, pp.
  46--61, May 2019.

\bibitem{zhang2}
L.~Zhang, X.~Q. Chen, S.~Liu, Q.~Zhang, J.~Zhao, J.~Y. Dai, G.~D. Bai, X.~Wan,
  Q.~Cheng, and G.~Castaldi, ``Space-time-coding digital metasurfaces,''
  \emph{Nat. Commun.}, vol.~9, no. 4334, pp. 1--11, Oct. 2018.

\bibitem{wang}
M.~Wang, H.~F. Ma, L.~W. Wu, S.~Sun, W.~X. Tang, and T.~J. Cui, ``Hybrid
  digital coding metasurface for independent control of propagating surface and
  spatial waves,'' \emph{Advanced Optical Materials}, vol.~7, no.~13, pp. 1--8,
  May 2019.

\bibitem{basar}
E.~Basar, ``Large intelligent surface-based index modulation: {A} new beyond
  {MIMO} paradigm for {6G},'' \emph{arXiv preprint, arXiv:1904.06704}, 2019.

\bibitem{guan}
X.~{Guan}, Q.~{Wu}, and R.~{Zhang}, ``Intelligent reflecting surface assisted
  secrecy communication: {I}s artificial noise helpful or not?'' \emph{IEEE
  Wireless Commun. Letters}, pp. 1--1, 2020.

\bibitem{yu}
X.~Yu, D.~Xu, and R.~Schober, ``Enabling secure wireless communications via
  intelligent reflecting surfaces,'' \emph{arXiv preprint, arXiv:1904.09573},
  2019.

\bibitem{shen}
H.~{Shen}, W.~{Xu}, S.~{Gong}, Z.~{He}, and C.~{Zhao}, ``Secrecy rate
  maximization for intelligent reflecting surface assisted multi-antenna
  communications,'' \emph{IEEE Commun. Letters}, vol.~23, no.~9, pp.
  1488--1492, Sep. 2019.

\bibitem{chen}
J.~{Chen}, Y.~{Liang}, Y.~{Pei}, and H.~{Guo}, ``Intelligent reflecting
  surface: {A} programmable wireless environment for physical layer security,''
  \emph{IEEE Access}, vol.~7, pp. 82\,599--82\,612, Jun. 2019.

\bibitem{wu}
Q.~Wu and R.~Zhang, ``Intelligent reflecting surface enhanced wireless network:
  {J}oint active and passive beamforming design,'' in \emph{Proc. IEEE Global
  Commun. Conf. (GLOBECOM)}, Abu Dhabi, Dec. 2018, pp. 1--6.

\bibitem{li}
L.~Li, H.~Liu, H.~Zhang, and W.~Xue, ``Efficient wireless power transfer system
  integrating with metasurface for biological applications,'' \emph{IEEE Trans.
  Ind. Electron.}, vol.~65, no.~4, pp. 3230--3239, Apr. 2017.

\bibitem{7354885}
M.~{Song}, P.~{Kapitanova}, I.~{Iorsh}, and P.~{Belov}, ``Metamaterials for
  wireless power transfer,'' in \emph{Proc. Days on Diffraction (DD)}, St.
  Petersburg, May 2015, pp. 1--4.

\bibitem{taha}
A.~Taha, M.~Alrabeiah, and A.~Alkhateeb, ``Enabling large intelligent surfaces
  with compressive sensing and deep learning,'' \emph{arXiv preprint,
  arXiv:1904.10136}, 2019.

\bibitem{del}
P.~del Hougne, M.~F. Imani, A.~V. Diebold, R.~Horstmeyer, and D.~R. Smith,
  ``Artificial neural network with physical dynamic metasurface layer for
  optimal sensing,'' \emph{arXiv preprint, arXiv:1906.10251}, 2019.

\bibitem{Huang}
C.~Huang, G.~C. Alexandropoulos, C.~Yuen, and M.~Debbah, ``Indoor signal
  focusing with deep learning designed reconfigurable intelligent surfaces,''
  in \emph{Proc. IEEE Int. Workshop on Signal Processing Advances in Wireless
  Commun. ({SPAWC}'19)}, Cannes, Jul. 2019, pp. 1--5.

\bibitem{8598647}
C.~{Kyrkou}, S.~{Timotheou}, P.~{Kolios}, T.~{Theocharides}, and
  C.~{Panayiotou}, ``Drones: {A}ugmenting our quality of life,'' \emph{IEEE
  Potentials}, vol.~38, no.~1, pp. 30--36, Jan. 2019.

\bibitem{gunes1}
M.~{Alzenad} and H.~{Yanikomeroglu}, ``Coverage and rate analysis for vertical
  heterogeneous networks {(VHetNets)},'' \emph{IEEE Trans. Wireless Commun.},
  vol.~18, no.~12, pp. 5643--5657, Dec. 2019.

\bibitem{gunes2}
I.~{Bor-Yaliniz}, M.~{Salem}, G.~{Senerath}, and H.~{Yanikomeroglu}, ``Is {5G}
  ready for drones: {A} look into contemporary and prospective wireless
  networks from a standardization perspective,'' \emph{IEEE Wireless Commun.},
  vol.~26, no.~1, pp. 18--27, Feb. 2019.

\bibitem{dronesO}
\BIBentryALTinterwordspacing
G.~Sachs, ``{DRONES} reporting for work,'' Mar. 2019. [Online]. Available:
  \url{https://www.goldmansachs.com/insights/technology-driving-innovation/drones/}
\BIBentrySTDinterwordspacing

\bibitem{8756714}
C.~{D'Andrea}, A.~{Garcia-Rodriguez}, G.~{Geraci}, L.~G. {Giordano}, and
  S.~{Buzzi}, ``Cell-free massive {MIMO} for {UAV} communications,'' in
  \emph{IEEE International Conference on Communications Workshops (ICC
  Workshops)}, Shanghai, 2019, pp. 1--6.

\bibitem{7827017}
H.~Q. {Ngo}, A.~{Ashikhmin}, H.~{Yang}, E.~G. {Larsson}, and T.~L. {Marzetta},
  ``Cell-free massive {MIMO} versus small cells,'' \emph{IEEE Trans. Wireless
  Commun.}, vol.~16, no.~3, pp. 1834--1850, Mar. 2017.

\bibitem{8952782}
C.~{D'Andrea}, A.~{Garcia-Rodriguez}, G.~{Geraci}, L.~G. {Giordano}, and
  S.~{Buzzi}, ``Analysis of {UAV} communications in cell-free massive {MIMO}
  systems,'' \emph{IEEE Open J. Commun. Soc.}, vol.~1, pp. 133--147, 2020.

\bibitem{8660516}
M.~{Mozaffari}, W.~{Saad}, M.~{Bennis}, Y.~{Nam}, and M.~{Debbah}, ``A tutorial
  on {UAV}s for wireless networks: {A}pplications, challenges, and open
  problems,'' \emph{IEEE Commun. Surveys Tuts.}, vol.~21, no.~3, pp.
  2334--2360, Third quarter 2019.

\bibitem{gunes3}
M.~{Alzenad}, A.~{El-Keyi}, F.~{Lagum}, and H.~{Yanikomeroglu}, ``{3-D}
  placement of an unmanned aerial vehicle base station {(UAV-BS)} for
  energy-efficient maximal coverage,'' \emph{IEEE Wireless Commun. Letters},
  vol.~6, no.~4, pp. 434--437, Aug. 2017.

\bibitem{7470933}
Y.~{Zeng}, R.~{Zhang}, and T.~J. {Lim}, ``Wireless communications with unmanned
  aerial vehicles: {O}pportunities and challenges,'' \emph{IEEE Commun. Mag.},
  vol.~54, no.~5, pp. 36--42, May 2016.

\bibitem{7463007}
S.~{Hayat}, E.~{Yanmaz}, and R.~{Muzaffar}, ``Survey on unmanned aerial vehicle
  networks for civil applications: {A} communications viewpoint,'' \emph{IEEE
  Commun. Surveys Tuts.}, vol.~18, no.~4, pp. 2624--2661, Fourth quarter 2016.

\bibitem{7306534}
U.~{Siddique}, H.~{Tabassum}, E.~{Hossain}, and D.~I. {Kim}, ``Wireless
  backhauling of 5{G} small cells: {C}hallenges and solution approaches,''
  \emph{IEEE Wireless Commun.}, vol.~22, no.~5, pp. 22--31, Oct. 2015.

\bibitem{8254715}
U.~{Challita} and W.~{Saad}, ``Network formation in the sky: {U}nmanned aerial
  vehicles for multi-hop wireless backhauling,'' in \emph{Proc. IEEE Global
  Commun. Conf. (GLOBECOM)}, Singapore, Dec. 2017, pp. 1--6.

\bibitem{8533634}
M.~{Mozaffari}, A.~T.~Z. Kasgari, W.~{Saad}, M.~{Bennis}, and M.~{Debbah},
  ``Beyond {5G} with {UAVs}: {F}oundations of a 3d wireless cellular network,''
  \emph{IEEE Trans. Wireless Commun.}, vol.~18, no.~1, pp. 357--372, Jan. 2019.

\bibitem{N1}
A.~A. {Al-Habob}, A.~{Ibrahim}, O.~A. {Dobre}, and A.~G. {Armada},
  ``Collision-free sequential task offloading for mobile edge computing,''
  \emph{IEEE Commun. Lett.}, vol.~24, no.~1, pp. 71--75, Jan. 2020.

\bibitem{N2}
A.~A. {Al-Habob} and O.~A. {Dobre}, ``Mobile edge computing and artificial
  intelligence: A mutually-beneficial relationship,'' \emph{IEEE ComSoc
  Techncial Committees Newsletter}, Nov. 2019.

\bibitem{7423671}
M.~{Gharibi}, R.~{Boutaba}, and S.~L. {Waslander}, ``Internet of drones,''
  \emph{IEEE Access}, vol.~4, pp. 1148--1162, Mar. 2016.

\bibitem{7545004}
N.~{Chen}, Y.~{Chen}, Y.~{You}, H.~{Ling}, P.~{Liang}, and R.~{Zimmermann},
  ``Dynamic urban surveillance video stream processing using fog computing,''
  in \emph{Proc. IEEE Int. Conf. on Multimedia Big Data (BigMM)}, Taipei, Apr.
  2016, pp. 105--112.

\bibitem{7964096}
A.~{Koubaa}, B.~{Qureshi}, M.~{Sriti}, Y.~{Javed}, and E.~{Tovar}, ``A
  service-oriented cloud-based management system for the
  {I}nternet-of-drones,'' in \emph{Proc. IEEE Int. Conf. on Autonomous Robot
  Systems and Competitions (ICARSC)}, Coimbra, Apr. 2017, pp. 329--335.

\bibitem{7888557}
Y.~{Zeng} and R.~{Zhang}, ``Energy-efficient {UAV} communication with
  trajectory optimization,'' \emph{IEEE Trans. Wireless Commun.}, vol.~16,
  no.~6, pp. 3747--3760, Mar. 2017.

\bibitem{8316986}
D.~{Yang}, Q.~{Wu}, Y.~{Zeng}, and R.~{Zhang}, ``Energy tradeoff in
  ground-to-{UAV} communication via trajectory design,'' \emph{IEEE Trans. Veh.
  Technol.}, vol.~67, no.~7, pp. 6721--6726, Jul. 2018.

\bibitem{8119562}
C.~{Zhan}, Y.~{Zeng}, and R.~{Zhang}, ``Energy-efficient data collection in
  {UAV} enabled wireless sensor network,'' \emph{IEEE Wireless Commun. Lett.},
  vol.~7, no.~3, pp. 328--331, Nov. 2018.

\bibitem{9075988}
W.~{Wang}, X.~{Li}, M.~{Zhang}, K.~{Cumanan}, D.~W.~K. {Ng}, G.~{Zhang},
  J.~{Tang}, and O.~A. {Dobre}, ``Energy-constrained {UAV}-assisted secure
  communications with position optimization and cooperative jamming,''
  \emph{IEEE Trans. Commun.}, vol.~xx, no.~nn, pp. 1--1, Apr. 2020.

\bibitem{8915758}
S.~{Yin}, Y.~{Zhao}, L.~{Li}, and F.~R. {Yu}, ``{UAV}-assisted cooperative
  communications with time-sharing information and power transfer,'' \emph{IEEE
  Trans. Veh. Technol.}, vol.~69, no.~2, pp. 1554--1567, Nov. 2020.

\bibitem{8876848}
M.~{Tatar Mamaghani} and Y.~{Hong}, ``On the performance of low-altitude
  {UAV}-enabled secure {AF} relaying with cooperative jamming and {SWIPT},''
  \emph{IEEE Access}, vol.~7, pp. 153\,060--153\,073, Jun. 2019.

\bibitem{8821282}
F.~{Huang}, J.~{Chen}, H.~{Wang}, G.~{Ding}, Y.~{Gong}, and Y.~{Yang},
  ``Multiple-{UAV}-assisted {SWIPT} in internet of things: User association and
  power allocation,'' \emph{IEEE Access}, vol.~7, pp. 124\,244--124\,255, Aug.
  2019.

\bibitem{8667026}
X.~{Sun}, W.~{Yang}, Y.~{Cai}, R.~{Ma}, and L.~{Tao}, ``Physical layer security
  in millimeter wave {SWIPT UAV}-based relay networks,'' \emph{IEEE Access},
  vol.~7, pp. 35\,851--35\,862, Mar. 2019.

\bibitem{8611204}
X.~{Sun}, W.~{Yang}, Y.~{Cai}, Z.~{Xiang}, and X.~{Tang}, ``Secure
  transmissions in millimeter wave {SWIPT UAV}-based relay networks,''
  \emph{IEEE Wireless Commun. Lett.}, vol.~8, no.~3, pp. 785--788, Jan. 2019.

\bibitem{8653332}
X.~{Hong}, P.~{Liu}, F.~{Zhou}, S.~{Guo}, and Z.~{Chu}, ``Resource allocation
  for secure {UAV}-assisted {SWIPT} systems,'' \emph{IEEE Access}, vol.~7, pp.
  24\,248--24\,257, Feb. 2019.

\bibitem{8581510}
M.~{Wazid}, A.~K. {Das}, N.~{Kumar}, A.~V. {Vasilakos}, and J.~J. P.~C.
  {Rodrigues}, ``Design and analysis of secure lightweight remote user
  authentication and key agreement scheme in {I}nternet of drones deployment,''
  \emph{IEEE Internet Things J.}, vol.~6, no.~2, pp. 3572--3584, Apr. 2019.

\bibitem{8691741}
S.~P. {Arteaga}, L.~A.~M. {Hernandez}, G.~S. {Perez}, A.~L.~S. {Orozco}, and
  L.~J.~G. {Villalba}, ``Analysis of the {GPS} spoofing vulnerability in the
  drone {3DR} solo,'' \emph{IEEE Access}, vol.~7, pp. 51\,782--51\,789, Apr.
  2019.

\bibitem{8486625}
Y.~{Chen} and L.~{Wang}, ``Privacy protection for {I}nternet of drones: {A}
  network coding approach,'' \emph{IEEE Internet Things J.}, vol.~6, no.~2, pp.
  1719--1730, Apr. 2019.

\bibitem{8795473}
S.~H. {Alsamhi}, O.~{Ma}, M.~S. {Ansari}, and F.~A. {Almalki}, ``Survey on
  collaborative smart drones and {I}nternet of things for improving smartness
  of smart cities,'' \emph{IEEE Access}, vol.~7, pp. 128\,125--128\,152, Aug.
  2019.

\bibitem{8255736}
D.~{Solomitckii}, M.~{Gapeyenko}, V.~{Semkin}, S.~{Andreev}, and
  Y.~{Koucheryavy}, ``Technologies for efficient amateur drone detection in
  {5G} millimeter-wave cellular infrastructure,'' \emph{IEEE Commun. Mag.},
  vol.~56, no.~1, pp. 43--50, Jan. 2018.

\bibitem{8255739}
C.~{Lin}, D.~{He}, N.~{Kumar}, K.~R. {Choo}, A.~{Vinel}, and X.~{Huang},
  ``Security and privacy for the {I}nternet of drones: {C}hallenges and
  solutions,'' \emph{IEEE Commun. Mag.}, vol.~56, no.~1, pp. 64--69, Jan. 2018.

\bibitem{8337900}
I.~{Guvenc}, F.~{Koohifar}, S.~{Singh}, M.~L. {Sichitiu}, and D.~{Matolak},
  ``Detection, tracking, and interdiction for amateur drones,'' \emph{IEEE
  Commun. Mag.}, vol.~56, no.~4, pp. 75--81, Apr. 2018.

\bibitem{8337908}
V.~{Sharma}, D.~N.~K. {Jayakody}, I.~{You}, R.~{Kumar}, and J.~{Li}, ``Secure
  and efficient context-aware localization of drones in urban scenarios,''
  \emph{IEEE Commun. Mag.}, vol.~56, no.~4, pp. 120--128, Apr. 2018.

\bibitem{8337902}
X.~{Yue}, Y.~{Liu}, J.~{Wang}, H.~{Song}, and H.~{Cao}, ``Software defined
  radio and wireless acoustic networking for amateur drone surveillance,''
  \emph{IEEE Commun. Mag.}, vol.~56, no.~4, pp. 90--97, Apr. 2018.

\bibitem{8325268}
J.~H. {Cheon}, K.~{Han}, S.~{Hong}, H.~J. {Kim}, J.~{Kim}, S.~{Kim}, H.~{Seo},
  H.~{Shim}, and Y.~{Song}, ``Toward a secure drone system: {F}lying with
  real-time homomorphic authenticated encryption,'' \emph{IEEE Access}, vol.~6,
  pp. 24\,325--24\,339, Mar. 2018.

\bibitem{8675384}
A.~{Fotouhi}, H.~{Qiang}, M.~{Ding}, M.~{Hassan}, L.~G. {Giordano},
  A.~{Garcia-Rodriguez}, and J.~{Yuan}, ``Survey on {UAV} cellular
  communications: {P}ractical aspects, standardization advancements,
  regulation, and security challenges,'' \emph{IEEE Commun. Surveys Tuts.},
  vol.~21, no.~4, pp. 3417--3442, Fourth quarter 2019.

\bibitem{8301585}
L.~{Wang}, B.~{Hu}, and S.~{Chen}, ``Energy efficient placement of a drone base
  station for minimum required transmit power,'' \emph{IEEE Wireless Commun.
  Lett.}, vol.~PP, pp. 1--1, Feb. 2018, early access.

\bibitem{8255733}
T.~{Long}, M.~{Ozger}, O.~{Cetinkaya}, and O.~B. {Akan}, ``Energy neutral
  internet of drones,'' \emph{IEEE Commun. Mag.}, vol.~56, no.~1, pp. 22--28,
  Jan. 2018.

\bibitem{8701666}
P.~{Yang}, X.~{Cao}, X.~{Xi}, W.~{Du}, Z.~{Xiao}, and D.~{Wu},
  ``Three-dimensional continuous movement control of drone cells for
  energy-efficient communication coverage,'' \emph{IEEE Trans. Veh. Technol.},
  vol.~68, no.~7, pp. 6535--6546, Jul. 2019.

\bibitem{8653315}
A.~{Alsharoa}, H.~{Ghazzai}, A.~{Kadri}, and A.~E. {Kamal}, ``Spatial and
  temporal management of cellular {HetNets} with multiple solar powered
  drones,'' \emph{IEEE Trans. Mobile Comput.}, vol.~PP, pp. 1--1, Feb. 2019.

\bibitem{8708295}
J.~{Yao} and N.~{Ansari}, ``{QoS}-aware power control in {I}nternet of drones
  for data collection service,'' \emph{IEEE Trans. Veh. Technol.}, vol.~68,
  no.~7, pp. 6649--6656, Jul. 2019.

\bibitem{8613833}
M.~B. {Ghorbel}, D.~{Rodriguez-Duarte}, H.~{Ghazzai}, M.~J. {Hossain}, and
  H.~{Menouar}, ``Joint position and travel path optimization for energy
  efficient wireless data gathering using unmanned aerial vehicles,''
  \emph{IEEE Trans. Veh. Technol.}, vol.~68, no.~3, pp. 2165--2175, Mar. 2019.

\bibitem{8469090}
J.~M. {Arteaga}, S.~{Aldhaher}, G.~{Kkelis}, C.~{Kwan}, D.~C. {Yates}, and
  P.~D. {Mitcheson}, ``Dynamic capabilities of {Multi-MHz} inductive power
  transfer systems demonstrated with batteryless drones,'' \emph{IEEE Trans.
  Power Electron.}, vol.~34, no.~6, pp. 5093--5104, Jun. 2019.

\bibitem{8846189}
A.~M. {Jawad}, H.~M. {Jawad}, R.~{Nordin}, S.~K. {Gharghan}, N.~F. {Abdullah},
  and M.~J. {Abu-Alshaeer}, ``Wireless power transfer with magnetic resonator
  coupling and sleep/active strategy for a drone charging station in smart
  agriculture,'' \emph{IEEE Access}, vol.~7, pp. 139\,839--139\,851, Sep. 2019.

\bibitem{8253543}
W.~{Shi}, H.~{Zhou}, J.~{Li}, W.~{Xu}, N.~{Zhang}, and X.~{Shen}, ``Drone
  assisted vehicular networks: {A}rchitecture, challenges and opportunities,''
  \emph{IEEE Network}, vol.~32, no.~3, pp. 130--137, May 2018.

\bibitem{8316776}
S.~{Sekander}, H.~{Tabassum}, and E.~{Hossain}, ``Multi-tier drone architecture
  for {5G/B5G} cellular networks: {C}hallenges, trends, and prospects,''
  \emph{IEEE Commun. Mag.}, vol.~56, no.~3, pp. 96--103, Mar. 2018.

\bibitem{7744808}
I.~{Bor-Yaliniz} and H.~{Yanikomeroglu}, ``The new frontier in {RAN}
  heterogeneity: {M}ulti-tier drone-cells,'' \emph{IEEE Commun. Mag.}, vol.~54,
  no.~11, pp. 48--55, Nov. 2016.

\bibitem{8701689}
P.~{Horstrand}, R.~{Guerra}, A.~{Rodriguez}, M.~{Diaz}, S.~{Lopez}, and J.~F.
  {Lopez}, ``A {UAV} platform based on a hyperspectral sensor for image
  capturing and on-board processing,'' \emph{IEEE Access}, vol.~7, pp.
  66\,919--66\,938, Apr. 2019.

\bibitem{7900406}
M.~{Haris}, T.~{Watanabe}, L.~{Fan}, M.~R. {Widyanto}, and H.~{Nobuhara},
  ``Superresolution for {UAV} images via adaptive multiple sparse
  representation and its application to {3-D} reconstruction,'' \emph{IEEE
  Trans. Geosci. and Remote Sens.}, vol.~55, no.~7, pp. 4047--4058, Jul. 2017.

\bibitem{8418072}
N.~{Yamamoto} and N.~{Uchida}, ``Improvement of image processing for a
  collaborative security flight control system with multiple drones,'' in
  \emph{Proc. Int. Conf. on Advanced Information Networking and Applications
  Workshops (WAINA)}, Krakow, May 2018, pp. 199--202.

\bibitem{8755983}
A.~{Fouda}, A.~S. {Ibrahim}, I.~{Guvenc}, and M.~{Ghosh}, ``Interference
  management in {UAV}-assisted integrated access and backhaul cellular
  networks,'' \emph{IEEE Access}, vol.~7, pp. 104\,553--104\,566, Jul. 2019.

\bibitem{8555533}
Z.~{Zhang}, L.~{Li}, W.~{Liang}, X.~{Li}, A.~{Gao}, W.~{Chen}, and Z.~{Han},
  ``Downlink interference management in dense drone small cells networks using
  mean-field game theory,'' in \emph{Proc. Int. Conf. on Wireless
  Communications and Signal Processing (WCSP)}, Hangzhou, Oct. 2018, pp. 1--6.

\bibitem{8287974}
K.~{Yoshikawa}, S.~{Yamashita}, K.~{Yamamoto}, T.~{Nishio}, and M.~{Morikura},
  ``Resource allocation for {3D} drone networks sharing spectrum bands,'' in
  \emph{Proc. IEEE Veh. Tech. Conf. (VTC)}, Toronto, ON, Sep. 2017, pp. 1--5.

\bibitem{8851647}
S.~{Cheng}, Y.~{Chao}, L.~{Wang}, and A.~{Tsai}, ``Affinity propagation
  clustering for interference management in aerial small cells,'' in
  \emph{Proc. IEEE VTS Asia Pacific Wireless Commun. Symp. (APWCS)}, Singapore,
  Aug. 2019, pp. 1--5.

\bibitem{7756327}
C.~{Zhang} and W.~{Zhang}, ``Spectrum sharing for drone networks,'' \emph{IEEE
  J. Sel. Areas Commun.}, vol.~35, no.~1, pp. 136--144, Jan. 2017.

\bibitem{8654727}
U.~{Challita}, W.~{Saad}, and C.~{Bettstetter}, ``Interference management for
  cellular-connected {UAVs}: {A} deep reinforcement learning approach,''
  \emph{IEEE Trans. Wireless Commun.}, vol.~18, no.~4, pp. 2125--2140, Apr.
  2019.

\bibitem{8733128}
E.~{Belyaev} and S.~{Forchhammer}, ``An efficient storage of infrared video of
  drone inspections via iterative aerial map construction,'' \emph{IEEE Signal
  Proc. Lett.}, vol.~26, no.~8, pp. 1157--1161, Aug. 2019.

\bibitem{8734799}
V.~{Sharma}, I.~{You}, D.~N.~K. {Jayakody}, D.~G. {Reina}, and K.~R. {Choo},
  ``Neural-blockchain-based ultrareliable caching for edge-enabled {UAV}
  networks,'' \emph{IEEE Trans. Ind. Informat.}, vol.~15, no.~10, pp.
  5723--5736, Oct. 2019.

\bibitem{7120024}
P.~{Kamalinejad}, C.~{Mahapatra}, Z.~{Sheng}, S.~{Mirabbasi}, V.~C. {M. Leung},
  and Y.~L. {Guan}, ``Wireless energy harvesting for the {I}nternet of
  {T}hings,'' \emph{IEEE Commun. Mag.}, vol.~53, no.~6, pp. 102--108, Jun.
  2015.

\bibitem{8368232}
N.~{Van Huynh}, D.~T. {Hoang}, X.~{Lu}, D.~{Niyato}, P.~{Wang}, and D.~I.
  {Kim}, ``Ambient backscatter communications: {A} contemporary survey,''
  \emph{IEEE Commun. Surveys Tuts.}, vol.~20, no.~4, pp. 2889--2922, Fourth
  quarter 2018.

\bibitem{7984754}
D.~{Niyato}, D.~I. {Kim}, M.~{Maso}, and Z.~{Han}, ``Wireless powered
  communication networks: {R}esearch directions and technological approaches,''
  \emph{IEEE Wireless Commun.}, vol.~24, no.~6, pp. 88--97, Dec. 2017.

\bibitem{BS-EH}
D.~Niyato, E.~Hossain, D.~I. Kim, V.~Bhargava, and L.~Shafai,
  \emph{Wireless-Powered Communication Networks: {A}rchitectures, Protocols,
  and Applications}, 1st~ed.\hskip 1em plus 0.5em minus 0.4em\relax Cambridge,
  U.K: Cambridge Univ. Press, 2016.

\bibitem{RFID}
C.~H. Loo \emph{et~al.}, ``Chip impedance matching for {UHF} {RFID} tag antenna
  design,'' \emph{Progr. Electromagn. Res.}, vol.~81, pp. 359--370, Jan. 2008.

\bibitem{RFID2}
J.~Zhang, G.~Y. Tian, A.~M. Marindra, A.~I. Sunny, and A.~B. Zhao, ``A review
  of passive {RFID} tag antenna-based sensors and systems for structural health
  monitoring applications,'' \emph{Sensors}, vol.~17, no.~2, p. 265, Feb. 2017.

\bibitem{AL}
J.~Li, A.~Liu, G.~Shen, L.~Li, C.~Sun, and F.~Zhao, ``Retro-{VLC}: {E}nabling
  battery-free duplex visible light communication for mobile and {IoT}
  applications,'' in \emph{Proc. Int. Workshop Mobile Comput. Syst. Appl.},
  Santa Fe New Mexico, Feb. 2015.

\bibitem{AL2}
S.~{Shao}, A.~{Khreishah}, and H.~{Elgala}, ``Pixelated {VLC}-backscattering
  for self-charging indoor {IoT} devices,'' \emph{IEEE Photonics Technol.
  Lett.}, vol.~29, no.~2, pp. 177--180, Jan. 2017.

\bibitem{8491095}
R.~{Di Candia}, R.~{Jantti}, R.~{Duan}, J.~{Lietzen}, H.~{Khalifa}, and
  K.~{Ruttik}, ``Quantum backscatter communication: {A} new paradigm,'' in
  \emph{Proc. Int. Symp. on Wireless Commun. Systems (ISWCS)}, Lisbon, Aug.
  2018, pp. 1--6.

\bibitem{7432027}
F.~{Huo}, P.~{Mitran}, and G.~{Gong}, ``Analysis and validation of active
  eavesdropping attacks in passive {FHSS} {RFID} systems,'' \emph{IEEE Trans.
  Inf. Forensics Security}, vol.~11, no.~7, pp. 1528--1541, Jul. 2016.

\bibitem{8000668}
K.~{Fukuda}, J.~{Heidemann}, and A.~{Qadeer}, ``Detecting malicious activity
  with {DNS} backscatter over time,'' \emph{IEEE/ACM Trans. Netw.}, vol.~25,
  no.~5, pp. 3203--3218, Oct. 2017.

\bibitem{8681643}
Y.~{Zhang}, F.~{Gao}, L.~{Fan}, X.~{Lei}, and G.~K. {Karagiannidis}, ``Secure
  communications for multi-tag backscatter systems,'' \emph{IEEE Wireless
  Commun. Lett.}, vol.~8, no.~4, pp. 1146--1149, Aug. 2019.

\bibitem{6987335}
G.~{Yao}, J.~{Bi}, and A.~V. {Vasilakos}, ``Passive {IP} traceback:
  {D}isclosing the locations of {IP} spoofers from path backscatter,''
  \emph{IEEE Trans. Inf. Forensics Security}, vol.~10, no.~3, pp. 471--484,
  Mar. 2015.

\bibitem{7122464}
X.~{Wang}, Z.~{Su}, and G.~{Wang}, ``Relay selection for secure backscatter
  wireless communications,'' \emph{IET Electronics Lett.}, vol.~51, no.~12, pp.
  951--952, Nov. 2015.

\bibitem{7556997}
Q.~{Yang}, H.~{Wang}, Y.~{Zhang}, and Z.~{Han}, ``Physical layer security in
  {MIMO} backscatter wireless systems,'' \emph{IEEE Trans. Wireless Commun.},
  vol.~15, no.~11, pp. 7547--7560, Nov. 2016.

\bibitem{7867859}
H.~{Park}, J.~{Yu}, H.~{Roh}, and W.~{Lee}, ``{SCBF}: {E}xploiting a collision
  for authentication in backscatter networks,'' \emph{IEEE Commun. Lett.},
  vol.~21, no.~6, pp. 1413--1416, Jun. 2017.

\bibitem{6836141}
W.~{Saad}, X.~{Zhou}, Z.~{Han}, and H.~V. {Poor}, ``On the physical layer
  security of backscatter wireless systems,'' \emph{IEEE Trans. Wireless
  Commun.}, vol.~13, no.~6, pp. 3442--3451, Jun. 2014.

\bibitem{8219367}
S.~G. {Hong}, Y.~M. {Hwang}, S.~Y. {Lee}, Y.~{Shin}, D.~I. {Kim}, and J.~Y.
  {Kim}, ``Game-theoretic modeling of backscatter wireless sensor networks
  under smart interference,'' \emph{IEEE Commun. Lett.}, vol.~22, no.~4, pp.
  804--807, Apr. 2018.

\bibitem{7913737}
W.~{Liu}, K.~{Huang}, X.~{Zhou}, and S.~{Durrani}, ``Full-duplex backscatter
  interference networks based on time-hopping spread spectrum,'' \emph{IEEE
  Trans. Wireless Commun.}, vol.~16, no.~7, pp. 4361--4377, Jul. 2017.

\bibitem{8353418}
S.~{Ebrahimi-Asl}, M.~T. {Ahmad Ghasr}, and M.~{Zawodniok}, ``Cooperative
  interference control in neighboring passive scattering antennas,'' \emph{IEEE
  J. Radio Freq. Identif.}, vol.~2, no.~3, pp. 152--158, Sep. 2018.

\bibitem{6942226}
A.~{Bekkali}, S.~{Zou}, A.~{Kadri}, M.~{Crisp}, and R.~V. {Penty},
  ``Performance analysis of passive {UHF RFID} systems under cascaded fading
  channels and interference effects,'' \emph{IEEE Trans. Wireless Commun.},
  vol.~14, no.~3, pp. 1421--1433, Mar. 2015.

\bibitem{8666220}
A.~L. T.~B. {da Fonseca}, F.~L. {Cabrera}, and F.~R. {de Sousa}, ``{CMOS} fully
  integrated quasi-circulator with self-interference cancellation technique,''
  \emph{IET Electronics Lett.}, vol.~55, no.~6, pp. 329--331, Mar. 2019.

\bibitem{8302845}
W.~{Gong}, H.~{Liu}, J.~{Liu}, X.~{Fan}, K.~{Liu}, Q.~{Ma}, and X.~{Ji},
  ``Channel-aware rate adaptation for backscatter networks,'' \emph{IEEE/ACM
  Trans. Netw.}, vol.~26, no.~2, pp. 751--764, Apr. 2018.

\bibitem{8735851}
Q.~{Yao}, J.~{Ma}, R.~{Li}, X.~{Li}, J.~{Li}, and J.~{Liu}, ``Energy-aware
  {RFID} authentication in edge computing,'' \emph{IEEE Access}, vol.~7, pp.
  77\,964--77\,980, Jun. 2019.

\bibitem{8700258}
S.~H. {Kim} and D.~I. {Kim}, ``Traffic-aware backscatter communications in
  wireless-powered heterogeneous networks,'' \emph{IEEE Trans. Mobile Comput.},
  vol.~PP, pp. 1--1, Apr. 2019.

\bibitem{8501953}
L.~{Bariah}, S.~{Muhaidat}, and A.~{Al-Dweik}, ``Error probability analysis of
  non-orthogonal multiple access over {Nakagami}- $m$ fading channels,''
  \emph{IEEE Trans. Commun.}, vol.~67, no.~2, pp. 1586--1599, Feb. 2019.

\bibitem{8756283}
C.~{Zheng}, W.~{Cheng}, and H.~{Zhang}, ``Optimal resource allocation for
  two-user and single-{DF}-relay network with ambient backscatter,'' \emph{IEEE
  Access}, vol.~7, pp. 91\,375--91\,389, Jul. 2019.

\bibitem{8851217}
G.~{Yang}, X.~{Xu}, and Y.~{Liang}, ``Resource allocation in {NOMA}-enhanced
  backscatter communication networks for wireless powered {IoT},'' \emph{IEEE
  Wireless Commun. Lett.}, vol.~9, no.~1, pp. 117--120, Jan. 2020.

\bibitem{8847703}
C.~{Le} and D.~{Do}, ``Outage performance of backscatter {NOMA} relaying
  systems equipping with multiple antennas,'' \emph{IET Electronics Lett.},
  vol.~55, no.~19, pp. 1066--1067, Sep. 2019.

\bibitem{8636518}
Q.~{Zhang}, L.~{Zhang}, Y.~{Liang}, and P.~{Kam}, ``Backscatter-{NOMA}: A
  symbiotic system of cellular and {I}nternet-of-things networks,'' \emph{IEEE
  Access}, vol.~7, pp. 20\,000--20\,013, Feb. 2019.

\bibitem{8439079}
J.~{Guo}, X.~{Zhou}, S.~{Durrani}, and H.~{Yanikomeroglu}, ``Design of
  non-orthogonal multiple access enhanced backscatter communication,''
  \emph{IEEE Trans. Wireless Commun.}, vol.~17, no.~10, pp. 6837--6852, Oct.
  2018.

\bibitem{8412618}
X.~{Hui} and E.~C. {Kan}, ``Collaborative reader code division multiple access
  in the harmonic {RFID} system,'' \emph{IEEE J. Radio Freq. Identif.}, vol.~2,
  no.~2, pp. 86--92, Jun. 2018.

\bibitem{8017383}
X.~{Hui}, Y.~{Ma}, and E.~C. {Kan}, ``Code division multiple access in
  centimeter accuracy harmonic {RFID} locating system,'' \emph{IEEE J. Radio
  Freq. Identif.}, vol.~1, no.~1, pp. 51--58, Mar. 2017.

\bibitem{7995033}
C.~{Kang}, W.~{Lee}, Y.~{You}, and H.~{Song}, ``Signal detection scheme in
  ambient backscatter system with multiple antennas,'' \emph{IEEE Access},
  vol.~5, pp. 14\,543--14\,547, Jul. 2017.

\bibitem{8399824}
W.~{Liu}, Y.~{Liang}, Y.~{Li}, and B.~{Vucetic}, ``Backscatter multiplicative
  multiple-access systems: {F}undamental limits and practical design,''
  \emph{IEEE Trans. Wireless Commun.}, vol.~17, no.~9, pp. 5713--5728, Sep.
  2018.

\bibitem{8320359}
S.~{Ma}, G.~{Wang}, R.~{Fan}, and C.~{Tellambura}, ``Blind channel estimation
  for ambient backscatter communication systems,'' \emph{IEEE Commun. Lett.},
  vol.~22, no.~6, pp. 1296--1299, Jun. 2018.

\bibitem{8618337}
D.~{Mishra} and E.~G. {Larsson}, ``Optimal channel estimation for
  reciprocity-based backscattering with a full-duplex {MIMO} reader,''
  \emph{IEEE Trans. Signal Process.}, vol.~67, no.~6, pp. 1662--1677, Mar.
  2019.

\bibitem{8746230}
W.~{Zhao}, G.~{Wang}, S.~{Atapattu}, R.~{He}, and Y.~{Liang}, ``Channel
  estimation for ambient backscatter communication systems with massive-antenna
  reader,'' \emph{IEEE Trans. Veh. Technol.}, vol.~68, no.~8, pp. 8254--8258,
  Aug. 2019.

\bibitem{7982802}
A.~{Guerra}, F.~{Guidi}, D.~{Dardari}, A.~{Clemente}, and R.~{D'Errico}, ``A
  millimeter-wave indoor backscattering channel model for environment
  mapping,'' \emph{IEEE Trans. Antennas Propag.}, vol.~65, no.~9, pp.
  4935--4940, Sep. 2017.

\bibitem{8523801}
Y.~{Zhang}, F.~{Gao}, L.~{Fan}, X.~{Lei}, and G.~K. {Karagiannidis},
  ``Backscatter communications over correlated nakagami- $m$ fading channels,''
  \emph{IEEE Trans. Commun.}, vol.~67, no.~2, pp. 1693--1704, Feb. 2019.

\bibitem{7820135}
D.~{Darsena}, G.~{Gelli}, and F.~{Verde}, ``Modeling and performance analysis
  of wireless networks with ambient backscatter devices,'' \emph{IEEE Trans.
  Commun.}, vol.~65, no.~4, pp. 1797--1814, Apr. 2017.

\bibitem{8730429}
Y.~{Ye}, L.~{Shi}, R.~{Qingyang Hu}, and G.~{Lu}, ``Energy-efficient resource
  allocation for wirelessly powered backscatter communications,'' \emph{IEEE
  Commun. Lett.}, vol.~23, no.~8, pp. 1418--1422, Aug. 2019.

\bibitem{8476159}
G.~{Yang}, D.~{Yuan}, Y.~{Liang}, R.~{Zhang}, and V.~C.~M. {Leung}, ``Optimal
  resource allocation in full-duplex ambient backscatter communication networks
  for wireless-powered {IoT},'' \emph{IEEE Internet Things J.}, vol.~6, no.~2,
  pp. 2612--2625, Apr. 2019.

\bibitem{8700623}
S.~{Xiao}, H.~{Guo}, and Y.~{Liang}, ``Resource allocation for
  full-duplex-enabled cognitive backscatter networks,'' \emph{IEEE Trans.
  Wireless Commun.}, vol.~18, no.~6, pp. 3222--3235, Jun. 2019.

\bibitem{8360017}
S.~T. {Shah}, K.~W. {Choi}, T.~{Lee}, and M.~Y. {Chung}, ``Outage probability
  and throughput analysis of {SWIPT} enabled cognitive relay network with
  ambient backscatter,'' \emph{IEEE Internet Things J.}, vol.~5, no.~4, pp.
  3198--3208, Aug. 2018.

\bibitem{8116748}
T.~{Lin}, J.~{Bito}, J.~G.~D. {Hester}, J.~{Kimionis}, R.~A. {Bahr}, and M.~M.
  {Tentzeris}, ``On-body long-range wireless backscattering sensing system
  using inkjet-/{3-D}-printed flexible ambient {RF} energy harvesters capable
  of simultaneous {DC} and harmonics generation,'' \emph{IEEE Trans. Microw.
  Theory Techn.}, vol.~65, no.~12, pp. 5389--5400, Dec. 2017.

\bibitem{8340034}
N.~{Van Huynh}, D.~T. {Hoang}, D.~{Niyato}, P.~{Wang}, and D.~I. {Kim},
  ``Optimal time scheduling for wireless-powered backscatter communication
  networks,'' \emph{IEEE Wireless Commun. Lett.}, vol.~7, no.~5, pp. 820--823,
  Oct. 2018.

\bibitem{7981380}
S.~H. {Kim} and D.~I. {Kim}, ``Hybrid backscatter communication for
  wireless-powered heterogeneous networks,'' \emph{IEEE Trans. Wireless
  Commun.}, vol.~16, no.~10, pp. 6557--6570, Oct. 2017.

\bibitem{8413073}
J.~C. {Kwan} and A.~O. {Fapojuwo}, ``Sum-throughput maximization in wireless
  sensor networks with radio frequency energy harvesting and backscatter
  communication,'' \emph{IEEE Sensors J.}, vol.~18, no.~17, pp. 7325--7339,
  Sep. 2018.

\bibitem{8434224}
D.~{Li}, W.~{Peng}, and Y.~{Liang}, ``Hybrid ambient backscatter communication
  systems with harvest-then-transmit protocols,'' \emph{IEEE Access}, vol.~6,
  pp. 45\,288--45\,298, Aug. 2018.

\bibitem{8588295}
X.~{Liu}, Y.~{Gao}, and F.~{Hu}, ``Optimal time scheduling scheme for wireless
  powered ambient backscatter communications in {IoT} networks,'' \emph{IEEE
  Internet Things J.}, vol.~6, no.~2, pp. 2264--2272, Apr. 2019.

\bibitem{Berg}
D.~{Van Den Berg}, R.~{Glans}, D.~{De Koning}, F.~A. {Kuipers},
  J.~{Lugtenburg}, K.~{Polachan}, P.~T. {Venkata}, C.~{Singh}, B.~{Turkovic},
  and B.~{Van Wijk}, ``Challenges in haptic communications over the tactile
  {I}nternet,'' \emph{IEEE Access}, vol.~5, pp. 23\,502--23\,518, Oct. 2017.

\bibitem{8197491}
Y.~{Yuan}, ``Paving the road for virtual and augmented reality,'' \emph{IEEE
  Consum. Electron. Mag.}, vol.~7, no.~1, pp. 117--128, Jan. 2018.

\bibitem{8344795}
D.~{You}, B.~{Seo}, E.~{Jeong}, and D.~H. {Kim}, ``Internet of things ({IoT})
  for seamless virtual reality space: {C}hallenges and perspectives,''
  \emph{IEEE Access}, vol.~6, pp. 40\,439--40\,449, Apr. 2018.

\bibitem{7786938}
Y.~{Yuan}, ``Changing the world with virtual\/augmented reality technologies,''
  \emph{IEEE Consum. Electron. Mag.}, vol.~6, no.~1, pp. 40--41, Jan. 2017.

\bibitem{8689144}
F.~{El Jamiy} and R.~{Marsh}, ``Survey on depth perception in head mounted
  displays: {D}istance estimation in virtual reality, augmented reality, and
  mixed reality,'' \emph{IET Image Process.}, vol.~13, no.~5, pp. 707--712,
  Apr. 2019.

\bibitem{8026164}
F.~J. {Detmer}, J.~{Hettig}, D.~{Schindele}, M.~{Schostak}, and C.~{Hansen},
  ``Virtual and augmented reality systems for renal interventions: {A}
  systematic review,'' \emph{IEEE Reviews in Biomed. Engr.}, vol.~10, pp.
  78--94, Sep. 2017.

\bibitem{8733996}
D.~{Wang}, K.~{Ohnishi}, and W.~{Xu}, ``Multimodal haptic display for virtual
  reality: {A} survey,'' \emph{IEEE Trans. Ind. Electron.}, vol.~67, no.~1, pp.
  610--623, Jan. 2020.

\bibitem{8337839}
J.~{Tan}, G.~{Cheung}, and R.~{Ma}, ``360-degree virtual-reality cameras for
  the masses,'' \emph{IEEE MultiMedia}, vol.~25, no.~1, pp. 87--94, Jan. 2018.

\bibitem{7924239}
J.~{Yu}, ``A light-field journey to virtual reality,'' \emph{IEEE MultiMedia},
  vol.~24, no.~2, pp. 104--112, Apr. 2017.

\bibitem{8291482}
G.~M.~N. {Taira}, A.~C. {Sementille}, and S.~R.~R. {Sanches}, ``Influence of
  the camera viewpoint on augmented reality interaction,'' \emph{IEEE Latin
  America Trans.}, vol.~16, no.~1, pp. 260--264, Jan. 2018.

\bibitem{8019876}
B.~{Bach}, R.~{Sicat}, J.~{Beyer}, M.~{Cordeil}, and H.~{Pfister}, ``The
  hologram in my hand: {H}ow effective is interactive exploration of {3D}
  visualizations in immersive tangible augmented reality?'' \emph{IEEE Trans.
  Vis. Comput. Graphics}, vol.~24, no.~1, pp. 457--467, Jan. 2018.

\bibitem{8197481}
B.~K. {Wiederhold}, I.~T. {Miller}, and M.~D. {Wiederhold}, ``Using virtual
  reality to mobilize health care: {M}obile virtual reality technology for
  attenuation of anxiety and pain,'' \emph{IEEE Consum. Electron. Mag.},
  vol.~7, no.~1, pp. 106--109, Jan. 2018.

\bibitem{5665813}
J.~{Ryu}, M.~S. {Heo}, and H.~C. {Kim}, ``Development of portable breast
  self-examination device using enhanced tactile feedback,'' \emph{Electron.
  Lett.}, vol.~46, no.~25, pp. 1651--1653, Dec. 2010.

\bibitem{7707355}
K.~B. {Chen}, M.~E. {Sesto}, K.~{Ponto}, J.~{Leonard}, A.~{Mason},
  G.~{Vanderheiden}, J.~{Williams}, and R.~G. {Radwin}, ``Use of virtual
  reality feedback for patients with chronic neck pain and kinesiophobia,''
  \emph{IEEE Trans. Neural Syst. Rehabil. Eng.}, vol.~25, no.~8, pp.
  1240--1248, Aug. 2017.

\bibitem{7829437}
J.~{Orlosky}, Y.~{Itoh}, M.~{Ranchet}, K.~{Kiyokawa}, J.~{Morgan}, and
  H.~{Devos}, ``Emulation of physician tasks in eye-tracked virtual reality for
  remote diagnosis of neurodegenerative disease,'' \emph{IEEE Trans. Vis.
  Comput. Graphics}, vol.~23, no.~4, pp. 1302--1311, Apr. 2017.

\bibitem{8755882}
J.~{Torner}, S.~{Skouras}, J.~L. {Molinuevo}, J.~D. {Gispert}, and
  F.~{Alpiste}, ``Multipurpose virtual reality environment for biomedical and
  health applications,'' \emph{IEEE Trans. Neural Syst. Rehabil. Eng.},
  vol.~27, no.~8, pp. 1511--1520, Aug. 2019.

\bibitem{8673005}
D.~{Rojo}, J.~{Mayor}, J.~J.~G. {Rueda}, and L.~{Raya}, ``A virtual reality
  training application for adults with asperger's syndrome,'' \emph{IEEE Comput
  Graph Appl.}, vol.~39, no.~2, pp. 104--111, Mar. 2019.

\bibitem{8371236}
W.~{Si}, X.~{Liao}, Y.~{Qian}, and Q.~{Wang}, ``Mixed reality guided
  radiofrequency needle placement: {A} pilot study,'' \emph{IEEE Access},
  vol.~6, pp. 31\,493--31\,502, Jun. 2018.

\bibitem{7036100}
K.~{Murphy} and M.~{Darrah}, ``Haptics-based apps for middle school students
  with visual impairments,'' \emph{IEEE Trans. Haptics}, vol.~8, no.~3, pp.
  318--326, Jul. 2015.

\bibitem{8765378}
A.~A. {Ali}, G.~A. {Dafoulas}, and J.~C. {Augusto}, ``Collaborative educational
  environments incorporating mixed reality technologies: {A} systematic mapping
  study,'' \emph{IEEE Trans. Learn. Technol.}, vol.~12, no.~3, pp. 321--332,
  Jul. 2019.

\bibitem{7845618}
S.~{Matsutomo}, T.~{Manabe}, V.~{Cingoski}, and S.~{Noguchi}, ``A computer
  aided education system based on augmented reality by immersion to {3-D}
  magnetic field,'' \emph{IEEE Trans. Magn.}, vol.~53, no.~6, pp. 1--4, Jun.
  2017.

\bibitem{7891552}
N.~{Drljevic}, L.~H. {Wong}, and I.~{Boticki}, ``Where does my augmented
  reality learning experience ({ARLE}) belong? {A} student and teacher
  perspective to positioning {ARLEs},'' \emph{IEEE Trans. Learn. Technol.},
  vol.~10, no.~4, pp. 419--435, Oct. 2017.

\bibitem{7123626}
M.~{Ibanez}, A.~{Di-Serio}, D.~{Villaran-Molina}, and C.~{Delgado-Kloos},
  ``Support for augmented reality simulation systems: {T}he effects of
  scaffolding on learning outcomes and behavior patterns,'' \emph{IEEE Trans.
  Learn. Technol.}, vol.~9, no.~1, pp. 46--56, Jan. 2016.

\bibitem{8377984}
Y.~{Sheng}, ``Scalable intelligence-enabled networking with traffic engineering
  in {5G} scenarios for future audio-visual-tactile {I}nternet,'' \emph{IEEE
  Access}, vol.~6, pp. 30\,378--30\,391, Jun. 2018.

\bibitem{8624569}
A.~{Benloucif}, A.~{Nguyen}, C.~{Sentouh}, and J.~{Popieul}, ``Cooperative
  trajectory planning for haptic shared control between driver and automation
  in highway driving,'' \emph{IEEE Trans. Ind. Electron.}, vol.~66, no.~12, pp.
  9846--9857, Dec. 2019.

\bibitem{7995874}
M.~{Corno}, L.~{D'Avico}, G.~{Panzani}, and S.~M. {Savaresi}, ``A haptic-based,
  safety-oriented, braking assistance system for road bicycles,'' in
  \emph{Proc. IEEE Intelligent Vehicles Symp. (IV)}, Los Angeles, CA, Jun.
  2017, pp. 1189--1194.

\bibitem{5648353}
M.~{Mulder}, D.~A. {Abbink}, M.~M. {van Paassen}, and M.~{Mulder}, ``Design of
  a haptic gas pedal for active car-following support,'' \emph{IEEE Trans.
  Intell. Transp. Syst.}, vol.~12, no.~1, pp. 268--279, Mar. 2011.

\bibitem{8718538}
P.~{Mekikis}, K.~{Ramantas}, L.~{Sanabria-Russo}, J.~{Serra},
  A.~{Antonopoulos}, D.~{Pubill}, E.~{Kartsakli}, and C.~{Verikoukis},
  ``{NFV}-enabled experimental platform for {5G} tactile {I}nternet support in
  industrial environments,'' \emph{IEEE Trans. Ind. Informat.}, vol.~16, no.~3,
  pp. 1--1, Mar. 2020.

\bibitem{8542940}
A.~{Aijaz} and M.~{Sooriyabandara}, ``The tactile {I}nternet for industries:
  {A} review,'' \emph{Proc. IEEE}, vol. 107, no.~2, pp. 414--435, Feb. 2019.

\bibitem{8281493}
O.~{Blanco-Novoa}, T.~M. {Fernandez-Carames}, P.~{Fraga-Lamas}, and M.~A.
  {Vilar-Montesinos}, ``A practical evaluation of commercial industrial
  augmented reality systems in an industry 4.0 shipyard,'' \emph{IEEE Access},
  vol.~6, pp. 8201--8218, Feb. 2018.

\bibitem{7962162}
J.~M. {Jacinto-Villegas}, M.~{Satler}, A.~{Filippeschi}, M.~{Bergamasco},
  M.~{Ragaglia}, A.~{Argiolas}, M.~{Niccolini}, and C.~A. {Avizzano}, ``A novel
  wearable haptic controller for teleoperating robotic platforms,'' \emph{IEEE
  Robot. Autom. Lett.}, vol.~2, no.~4, pp. 2072--2079, Oct. 2017.

\bibitem{7980645}
M.~{Dohler}, T.~{Mahmoodi}, M.~A. {Lema}, M.~{Condoluci}, F.~{Sardis},
  K.~{Antonakoglou}, and H.~{Aghvami}, ``Internet of skills, where robotics
  meets {AI}, {5G} and the tactile {I}nternet,'' in \emph{Proc. European Conf.
  on Networks and Communications (EuCNC)}, Oulu, Jun. 2017, pp. 1--5.

\bibitem{8736514}
A.~{Ebrahimzadeh}, M.~{Chowdhury}, and M.~{Maier}, ``Human-agent-robot task
  coordination in {FiWi}-based tactile {I}nternet infrastructures using
  context- and self-awareness,'' \emph{IEEE Trans. Netw. Service Manag.},
  vol.~16, no.~3, pp. 1127--1142, Sep. 2019.

\bibitem{8063891}
M.~{Chowdhury} and M.~{Maier}, ``Collaborative computing for advanced tactile
  {I}nternet human-to-robot ({H2R}) communications in integrated {FiWi}
  multirobot infrastructures,'' \emph{IEEE Internet Things J.}, vol.~4, no.~6,
  pp. 2142--2158, Dec. 2017.

\bibitem{8556371}
S.~{Haddadin}, L.~{Johannsmeier}, and F.~{Diaz Ledezma}, ``Tactile robots as a
  central embodiment of the tactile {I}nternet,'' \emph{Proc. IEEE}, vol. 107,
  no.~2, pp. 471--487, Feb. 2019.

\bibitem{8491368}
C.~{Reardon}, H.~{Zhang}, R.~{Wright}, and L.~E. {Parker}, ``Robots can teach
  students with intellectual disabilities: {E}ducational benefits of using
  robotic and augmented reality applications,'' \emph{IEEE Robot. Autom. Mag.},
  vol.~26, no.~2, pp. 79--93, Jun. 2019.

\bibitem{8241709}
F.~{Brizzi}, L.~{Peppoloni}, A.~{Graziano}, E.~D. {Stefano}, C.~A. {Avizzano},
  and E.~{Ruffaldi}, ``Effects of augmented reality on the performance of
  teleoperated industrial assembly tasks in a robotic embodiment,'' \emph{IEEE
  Trans. Human-Mach. Syst.}, vol.~48, no.~2, pp. 197--206, Apr. 2018.

\bibitem{7306533}
Z.~{Gao}, L.~{Dai}, D.~{Mi}, Z.~{Wang}, M.~A. {Imran}, and M.~Z. {Shakir},
  ``Mmwave massive-{MIMO}-based wireless backhaul for the {5G} ultra-dense
  network,'' \emph{IEEE Wireless Commun.}, vol.~22, no.~5, pp. 13--21, 2015.

\bibitem{6600706}
S.~{Hur}, T.~{Kim}, D.~J. {Love}, J.~V. {Krogmeier}, T.~A. {Thomas}, and
  A.~{Ghosh}, ``Millimeter wave beamforming for wireless backhaul and access in
  small cell networks,'' \emph{IEEE Trans. Commun.}, vol.~61, no.~10, pp.
  4391--4403, Oct. 2013.

\bibitem{7445132}
K.~{Venugopal} and R.~W. {Heath}, ``Millimeter wave networked wearables in
  dense indoor environments,'' \emph{IEEE Access}, vol.~4, pp. 1205--1221, Mar.
  2016.

\bibitem{8663550}
Z.~{Chen}, X.~{Ma}, B.~{Zhang}, Y.~{Zhang}, Z.~{Niu}, N.~{Kuang}, W.~{Chen},
  L.~{Li}, and S.~{Li}, ``A survey on {T}erahertz communications,'' \emph{China
  Commun.}, vol.~16, no.~2, pp. 1--35, Feb. 2019.

\bibitem{THz3}
T.~Kurner and S.~Priebe, ``Towards {THz} communications - status in research,
  standardization and regulation,'' \emph{J. of Infrared, Millimeter, and
  Terahertz Waves}, vol.~35, no.~1, pp. 53--62, Aug. 2014.

\bibitem{THZC}
S.~Mollahasani and E.~Onur, ``Evaluation of {T}erahertz channel in data
  centers,'' in \emph{Proc. IEEE/IFIP Network Operations and Management Symp.},
  Istanbul, 2016, pp. 727--730.

\bibitem{THZC2}
B.~Peng and T.~Kurner, ``A stochastic channel model for future wireless {THz}
  data centers,'' in \emph{Proc. Int. Symp. on Wireless Commun. Systems},
  Brussels, 2015, pp. 741--745.

\bibitem{THz4}
H.~Sarieddeen, N.~Saeed, T.~Y. Al-Naffouri, and M.-S. Alouini, ``Next
  generation {T}rahertz communications: {A} rendezvous of sensing, imaging and
  localization,'' \emph{arXiv preprint, arXiv:1909.10462v1}, 2019.

\bibitem{THz5}
I.~Akyildiz, J.~Jornet, and C.~Hana, ``Terahertz band: {N}ext frontier for
  wireless communications,'' \emph{Physical Commun.}, vol.~12, pp. 16--32, Sep.
  2014.

\bibitem{gamma}
\BIBentryALTinterwordspacing
D.~M. Barbu. (2015) The effects of radiation on the eye in industrial
  environments. [Online]. Available: \url{https://imt.uoradea.ro/auo.fmte/}
\BIBentrySTDinterwordspacing

\bibitem{6852088}
T.~{Yamazato}, I.~{Takai}, H.~{Okada}, T.~{Fujii}, T.~{Yendo}, S.~{Arai},
  M.~{Andoh}, T.~{Harada}, K.~{Yasutomi}, K.~{Kagawa}, and S.~{Kawahito},
  ``Image-sensor-based visible light communication for automotive
  applications,'' \emph{IEEE Commun. Mag.}, vol.~52, no.~7, pp. 88--97, Jul.
  2014.

\bibitem{5722675}
R.~{Perez-Jimenez}, J.~{Rufo}, C.~{Quintana}, J.~{Rabadan}, and F.~J.
  {Lopez-Hernandez}, ``Visible light communication systems for passenger
  in-flight data networking,'' in \emph{Proc. IEEE International Conference on
  Consumer Electronics (ICCE)}, Las Vegas, NV, Jan. 2011, pp. 445--446.

\bibitem{6924006}
D.~{Iturralde}, C.~{Azurdia-Meza}, N.~{Krommenacker}, I.~{Soto},
  Z.~{Ghassemlooy}, and N.~{Becerra}, ``A new location system for an
  underground mining environment using visible light communications,'' in
  \emph{Proc. 9th International Symposium on Communication Systems, Networks
  Digital Sign (CSNDSP)}, Manchester, Jul. 2014, pp. 1165--1169.

\bibitem{med}
N.~Lawrentschuk and D.~Bolton, ``Mobile phone interference with medical
  equipment and its clinical relevance: A systematic review,'' \emph{Med. J.
  Aust.}, vol. 181, no.~3, pp. 154--159, Aug. 2004.

\bibitem{7151783}
D.~R. {Dhatchayeny}, A.~{Sewaiwar}, S.~V. {Tiwari}, and Y.~H. {Chung},
  ``Experimental biomedical {EEG} signal transmission using {VLC},'' \emph{IEEE
  Sensors J.}, vol.~15, no.~10, pp. 5386--5387, Oct. 2015.

\bibitem{song}
M.~Song, K.~Baryshnikova, A.~Markvart, P.~Belov, E.~Nenasheva, C.~Simovski, and
  P.~Kapitanova, ``Smart table based on a metasurface for wireless power
  transfer,'' \emph{Phys. Rev. Appl.}, vol.~11, no.~5, pp. 1--9, May 2019.

\bibitem{tian}
X.~Tian, P.~M. Lee, Y.~J. Tan, T.~L. Wu, H.~Yao, M.~Zhang, Z.~Li, K.~A. Ng,
  B.~C. Tee, and J.~S. Ho, ``Wireless body sensor networks based on
  metamaterial textiles,'' \emph{Nat. Electron.}, pp. 243--251, Jun. 2019.

\bibitem{droneApp}
M.~Hassanalian and A.~Abdelkefi, ``Classifications, applications, and design
  challenges of drones: {A} review,'' \emph{Prog. Aerosp. Sci.}, vol.~91, pp.
  99--131, May 2017.

\bibitem{Amazon}
\BIBentryALTinterwordspacing
D.~Gross. (2013) Amazon's drone delivery: {H}ow would it work? [Online].
  Available:
  \url{http://www.cnn.com/2013/12/02/tech/innovation/amazon-drones-questions/}
\BIBentrySTDinterwordspacing

\bibitem{marine}
L.~Koh and S.~Wich, ``Dawn of drone ecology: {L}ow-cost autonomous aerial
  vehicles for conservation,'' \emph{Trop. Conserv. Sci.}, vol.~5, pp.
  121--132, Jul. 2012.

\bibitem{bss}
A.~Gudipati, D.~Perry, L.~E. Li, and S.~Katti, ``{SoftRAN}: Software defined
  radio access network,'' in \emph{Proc. the second ACM SIGCOMM workshop on Hot
  topics in software defined networking}, Hong Kong, Aug. 2013, pp. 25--30.

\bibitem{8387218}
A.~A. {Boulogeorgos}, A.~{Alexiou}, T.~{Merkle}, C.~{Schubert}, R.~{Elschner},
  A.~{Katsiotis}, P.~{Stavrianos}, D.~{Kritharidis}, P.~{Chartsias},
  J.~{Kokkoniemi}, M.~{Juntti}, J.~{Lehtomaki}, A.~{Teixeira}, and
  F.~{Rodrigues}, ``Terahertz technologies to deliver optical network quality
  of experience in wireless systems beyond {5G},'' \emph{IEEE Commun. Mag.},
  vol.~56, no.~6, pp. 144--151, Jun. 2018.

\bibitem{6998944}
C.~{Han}, A.~O. {Bicen}, and I.~F. {Akyildiz}, ``Multi-ray channel modeling and
  wideband characterization for wireless communications in the {T}erahertz
  band,'' \emph{IEEE Trans. Wireless Commun.}, vol.~14, no.~5, pp. 2402--2412,
  May 2015.

\bibitem{8622294}
M.~{Obeed}, A.~M. {Salhab}, M.~{Alouini}, and S.~A. {Zummo}, ``Survey on
  physical layer security in optical wireless communication systems,'' in
  \emph{Seventh International Conference on Communications and Networking
  (ComNet)}, Hammamet, Nov. 2018, pp. 1--5.

\bibitem{2020arXiv200207583N}
S.~{Naser}, L.~{Bariah}, W.~{Jaafar}, S.~{Muhaidat}, P.~C. {Sofotasios},
  M.~{Al-Qutayri}, and O.~A. {Dobre}, ``Optical rate-splitting multiple access
  for visible light communications,'' \emph{arXiv preprint, arXiv:2002.07583},
  Feb. 2020.

\bibitem{schober}
M.~{Najafi}, B.~{Schmauss}, and R.~{Schober}, ``Intelligent reconfigurable
  reflecting surfaces for free space optical communications,'' \emph{arXiv
  preprint, arXiv:2005.04499}, May 2020.

\bibitem{9057633}
L.~{Yang}, W.~{Guo}, and I.~S. {Ansari}, ``Mixed dual-hop {FSO}-{RF}
  communication systems through reconfigurable intelligent surface,''
  \emph{IEEE Commun. Lett.}, pp. 1--1, 2020.

\bibitem{2020arXiv200105715W}
H.~{Wang}, Z.~{Zhang}, B.~{Zhu}, J.~{Dang}, L.~{Wu}, L.~{Wang}, K.~{Zhang}, and
  Y.~{Zhang}, ``Performance of wireless optical communication with
  reconfigurable intelligent surfaces and random obstacles,'' \emph{arXiv
  preprints, arXiv:2001.05715}, Jan. 2020.

\bibitem{yang2016}
H.~Yang, X.~Cao, F.~Yang, J.~Gao, S.~Xu, M.~Li, X.~Chen, Y.~Zhao, Y.~Zheng, and
  S.~Li, ``A programmable metasurface with dynamic polarization, scattering and
  focusing control,'' \emph{Sci. Rep.}, vol.~6, pp. 1--11, Oct. 2016.

\bibitem{7465674}
R.~J. {Hall}, ``An {I}nternet of drones,'' \emph{IEEE Internet Comput.},
  vol.~20, no.~3, pp. 68--73, May 2016.

\bibitem{8288376}
W.~{Khawaja}, O.~{Ozdemir}, and I.~{Guvenc}, ``{UAV} air-to-ground channel
  characterization for {mmWave} systems,'' in \emph{Proc. IEEE Veh. Tech. Conf.
  (VTC)}, Toronto, ON, Sep. 2017, pp. 1--5.

\bibitem{7407385}
D.~W. {Matolak} and R.~{Sun}, ``Air-ground channel characterization for
  unmanned aircraft systems-{P}art {I}: {M}ethods, measurements, and models for
  over-water settings,'' \emph{IEEE Trans. Veh. Technol.}, vol.~66, no.~1, pp.
  26--44, Jan. 2017.

\bibitem{7842372}
W.~{Khawaja}, I.~{Guvenc}, and D.~{Matolak}, ``{UWB} channel sounding and
  modeling for {UAV} air-to-ground propagation channels,'' in \emph{Proc. IEEE
  Global Commun. Conf. (GLOBECOM)}, Washington, DC, Dec. 2016, pp. 1--7.

\bibitem{8353365}
Z.~{Kaleem} and M.~H. {Rehmani}, ``Amateur drone monitoring: {S}tate-of-the-art
  architectures, key enabling technologies, and future research directions,''
  \emph{IEEE Wireless Commun.}, vol.~25, no.~2, pp. 150--159, Apr. 2018.

\bibitem{AL3}
V.~Liu, V.~Talla, and S.~Gollakota, ``Enabling instantaneous feedback with
  full-duplex backscatter,'' in \emph{Proc. Int. Conf. Mobile Comput. Netw.},
  Maui Hawaii, Sep. 2014, pp. 67--78.

\bibitem{Aijaz}
A.~{Aijaz}, M.~{Dohler}, A.~H. {Aghvami}, V.~{Friderikos}, and M.~{Frodigh},
  ``Realizing the tactile {I}nternet: {H}aptic communications over next
  generation 5{G} cellular networks,'' \emph{IEEE Wireless Commun.}, vol.~24,
  no.~2, pp. 82--89, Apr. 2017.

\bibitem{Antonakoglou}
K.~{Antonakoglou}, X.~{Xu}, E.~{Steinbach}, T.~{Mahmoodi}, and M.~{Dohler},
  ``Toward haptic communications over the 5{G} tactile {I}nternet,'' \emph{IEEE
  Commun. Surveys Tuts.}, vol.~20, no.~4, pp. 3034--3059, Fourth quarter 2018.

\end{thebibliography}

\EOD
\end{document}